\documentclass[aps,prd,10pt,nofootinbib,twocolumn,eqsecnum,showpacs,showkeys,superscriptaddress,preprintnumbers,altaffilletter]{revtex4-1}
\usepackage[T1]{fontenc}
\usepackage[utf8]{inputenc}
\RequirePackage{graphicx}
\usepackage{amsmath}
\usepackage{amssymb}
\usepackage{booktabs}
\usepackage{soul}

\usepackage{hyperref}
\hypersetup{
    colorlinks=true,
    linkcolor=blue,
    filecolor=magenta,
    urlcolor=cyan,
    citecolor=cyan
}

\newcommand{\simgt}{\lower.5ex\hbox{$\; \buildrel > \over \sim \;$}}
\newcommand{\simlt}{\lower.5ex\hbox{$\; \buildrel < \over \sim \;$}}

\begin{document}

\title{Testing non-local gravity by   clusters of galaxies}

\author{Filippo Bouchè}
\email{filippo.bouche@gmail.com}
\affiliation{Dipartimento di Fisica ``E. Pancini'', Universit\`a  di Napoli  ``Federico II", Via Cinthia 21, I-80126, Napoli, Italy}
\author{Salvatore Capozziello}
\email{capozziello@na.infn.it}
\affiliation{Dipartimento di Fisica ``E. Pancini'', Universit\`a  di Napoli  ``Federico II", Via Cinthia 21, I-80126, Napoli, Italy}
\affiliation{DipaScuola Superiore Meridionale, Largo S. Marcellino 10, I-80138, Napoli, Italy}
\affiliation{Istituto Nazionale di Fisica Nucleare, Sez. di Napoli,  Via Cinthia 21, I-80126, Napoli, Italy}
\author{V. Salzano}
\email{vincenzo.salzano@usz.edu.pl}
\affiliation{Institute of Physics, University of Szczecin, Wielkopolska 15, 70-451 Szczecin, Poland}
\author{Keiichi Umetsu}
\email{keiichi@asiaa.sinica.edu.tw}
\affiliation{Academia Sinica Institute of Astronomy and Astrophysics (ASIAA), No.~1, Section~4, Roosevelt Road, Taipei 10617, Taiwan}

\date{\today}

\begin{abstract}
Extended theories of gravity have been extensively investigated during the last thirty years, aiming at fixing  infrared and ultraviolet shortcomings of General Relativity and of the associated $\Lambda$CDM cosmological model. Recently, non-local theories of gravity have drawn increasing attention due to their  potential to ameliorate both the ultraviolet and infrared behavior of gravitational interaction. In particular, Integral Kernel theories of Gravity provide a viable mechanism to explain the late time cosmic acceleration so as to avoid the introduction of any form of unknown dark energy. On the other hand, these models represent a natural link towards quantum gravity. Here, we study a scalar-tensor equivalent model of General Relativity corrected with non-local terms, where corrections are selected by the existence of  Noether symmetries.   After performing the weak field limit and generalizing the results to extended mass distributions, we analyse the non-local model at galaxy cluster scales, by comparing the theoretical predictions with gravitational lensing observations from the CLASH program. We obtain agreement with data at the same level of statistical significance as General Relativity. We also provide  constraints for the Navarro--Frenk--White parameters and lower bounds for the non-local length scales. The results are finally compared with those from the literature. 

\end{abstract}

\maketitle

\section{Introduction}

In the last decades, the growing availability of an increasingly larger amount of astrophysical and cosmological data has supposedly led us into a so-called ``precision cosmology'' era. The effective model that best fits most of the collected observations is the $\Lambda$CDM model, which is based on General Relativity (GR) and on the introduction of two exotic fluids, the cold dark matter (DM) and the dark energy (DE). They should represent the $\sim27\%$ and $\sim68\%$ of the matter--energy content of the present-day Universe, respectively \cite{Planck:2018vyg,eBOSS:2020yzd,DES:2021wwk}, and being responsible for the dynamical features of the Universe at all scales. Despite being the best model to explain the collected data, the $\Lambda$CDM paradigm is plagued by several problems, both experimental and theoretical \cite{Bull:2015stt}. On the one hand we have a huge variety of solutions for DM but we face the complete lack of any detection of any viable particle candidate for DM at fundamental scales \cite{Bertone:2004pz}. On the other hand, we also have an humongous number of possibilities to describe DE, but also many theoretical shortcomings regarding its nature and behaviour, such as the discrepancy ($\sim$120 orders of magnitude) between the observed value of the cosmological constant and the vacuum energy density calculated via QFT. Other big issues are the presence of singularities in the theory and the completely unsatisfactory description of gravity at quantum level.

All these problems led to the idea of developing  theories of gravity ``beyond GR'', aiming at fixing its infrared (IR) and ultraviolet (UV) shortcomings. Several proposal for extended theories of gravity (ETGs) have been made during last thirty years \cite{Clifton:2011jh,Ishak:2018his}. Some of these are based on  modifications of the geometrical content of the theory, namely the Hilbert-Einstein Lagrangian; some others are based on  modifications of the matter content, for example by adding  extra scalar fields. Often these two kind of modifications are introduced together. The most famous and extensively investigated ETGs are $f(R)$ theories \cite{Capozziello:2002rd,RevModPhys.82.451, Capozziello:2012ie, Clifton:2011jh, CAPOZZIELLO2011167, ReviewDeFelice}, where the Hilbert-Einstein Lagrangian is replaced by a general function of the Ricci scalar $R$, and scalar-tensor theories \cite{Nojiri:2010wj, Oikonomou,  doi:10.1142/S021827180600942X, RevModPhys.75.559, Euclid, Huterer_2017}, in which one or more scalar fields are  minimally or non-minimally coupled to gravity. These two classes of theories can be made equivalent, since $f(R)$ theories can be recovered by scalar-tensor ones with some specific changes of variables and viceversa.

Other ETGs approaches replace the Ricci scalar $R$ with the torsion scalar $T$, instead \cite{Cai:2015emx}. In fact, it is possible to build the so called Teleparallel Equivalent of General Relativity (TEGR) by giving up the Equivalence Principle and replacing the Ricci scalar $R$ with the Torsion scalar $T$. TEGR, firstly introduced by Einstein himself \cite{unzicker2005translation}, and GR are dynamically equivalent. However, the equivalence does not hold for the teleparallel equivalent of $f(R)$, i.e. $f(T)$ theories. The latter leads to second-order field equations, while the field equations derived by $f(R)$ gravity, in metric formalism,  are of  fourth-order \cite{Bamba}. 

An interesting approach is the introduction of non-local terms \cite{Capozziello:ReviewNonLocal, Nojiri:2010wj}. Non-locality is one of the main   feature of  quantum theory, thanks to the Heisenberg Principle,  and it automatically arises in quantum field theory (QFT) when one-loop effective actions are considered. Instead, GR is a classical theory, hence local by definition. In order to merge the gravitational interaction with the quantum formalism, it is thus possible to implement an effective approach by adding non-local operators to the gravitational Lagrangian. The procedure can be considered as an  effective approach to link gravitation and QFT.

In general, two main different families of non-local theories of gravity can be considered. The first ones are the Infinite Derivatives Theories of Gravity (IDGs), which present, at short range, non-locality caused by terms with entire analytic transcendental functions of a differential operator. The most commonly used terms are the exponential functions of the covariant d'Alembert operator, $\Box = g_{\mu\nu} \nabla^{\mu} \nabla^{\nu}$. IDGs have been  introduced to solve one of the main problems of $f(R)$ theories, namely the lack of unitarity caused by bad ghosts. Moreover, this class of non-local terms weakens gravitational attraction at short length scale. Hence, a natural solution to UV shortcomings of GR is provided \cite{Buoninfante2018, Modesto:2013ioa, PhysRevD.87.083507, Conroy:2014eja}.
    
Then, we also have Integral Kernel Theories of Gravity (IKGs), inspired by quantum corrections obtained in QFT on curved spacetime. In IKGs, the Lagrangian is extended by adding transcendental functions of the fields ($R$, $T$, $G$, etc.), which can be always represented by the integral kernels of differential operators, such as
\begin{equation}\label{1.1}
        \Box^{-1} \eta(x) \equiv \int d^{4}x' G(x,x') \eta(x')
\end{equation}
Due to its integral nature, this class of terms gives rise to long range non-localities, which can fix the IR shortcomings of GR \cite{doi:10.1142/S0217732315400039, Deser:2007jk, Capriolo1,Capriolo2}. Interesting IKG models have been proposed in \cite{RRmodel,RTmodel}. Subsequent comparisons with cosmological observations \cite{Dirian:2016puz,RRtest2018,Nesseris:2014mea,Acunzo} showed agreement with data at a level statistically equivalent to $\Lambda$CDM and an intriguing reduction of the Hubble tension.

In this paper, we consider a metric IKG proposed  in \cite{Deser:2007jk} to explain the late-time cosmic acceleration. Non-locality is introduced by a general function of the inverse d'Alembert operator, Eq.~(\ref{1.1}), the form of which can be found by applying the so-called Noether Symmetry Approach as discussed in \cite{Capozziello:ReviewNonLocal}. Furthermore, we consider the local representation of the theory, firstly introduced in \cite{Nojiri:2007uq}. Two auxiliary scalar fields emerge as Lagrange multipliers, so that the new scalar-tensor theory is made equivalent to the original non-local theory. The Newtonian limit is finally performed and  two Newtonian potentials, $\phi$ and $\psi$, are derived. These point-mass potentials can be generalized to extended spherically symmetric mass distributions which can model gravitational structures like galaxy clusters. Then, the non-local model can be compared to observational data. In other words, signatures of non-locality can emerge from large-scale structure.

Here, we perform an analysis similar to \cite{Salzano:2017qac, Salzano:2016udu, Laudato:2021mnm} using a sample of 19 high-mass galaxy clusters, for which strong- and weak-lensing data sets from deep \textit{Hubble Space Telescope} (\textit{HST}) and ground-based wide-field observations are available. All data products used in this study are obtained from the Cluster Lensing and Supernova survey with Hubble (CLASH) program \cite{CLASH2012}.
The aim of the paper is to constrain the free parameters of the non-local theory and compare the results to those obtained in \cite{Dialektopoulos:2018iph,Borka:2021omc}, where the same theory has been analyzed using the orbits of S2 stars around the Galactic center. 

The paper is organised as follow: in section \ref{Section 2} we present the non-local model and its local scalar-tensor representation. Then, we sketch the Noether Symmetry Approach to derive the form of the non-local terms related to the existence of conserved quantities. Moreover, we perform the weak-field  limit in order to find the Newtonian potentials, $\phi$ and $\psi$, and we generalize the results to extended spherically symmetric systems. Finally, our choice for the mass density profile is presented and an overview of the gravitational lensing theory, used to derive the theoretical prediction, is introduced. In Sec. \ref{Section 3},  the data sets of the CLASH program that we have used in our analysis is presented. Then, we introduce the main points of the statistical analysis we have performed. Furthermore, in Sec. \ref{Section 4}, the results of the statistical analysis are discussed and a comparison between them and previous results in the literature is made. In Sec. \ref{Section 5}, the conclusions are drawn.

\section{The Model}\label{Section 2}


The model  we are going to  study in this paper was firstly introduced in \cite{Deser:2007jk}. Here, the Hilbert-Einstein action is extended by adding an extra non-local term characterized by a general function of the inverse d'Alembert operator
\begin{equation}\label{2.1}
    S \, = \, \frac{1}{16 \pi G} \int d^{4}x \, \sqrt{-g} \, \Big[ R \big( 1 + f(\Box^{-1}R) \big) \Big]
\end{equation}
The function $f$ is called \textit{distortion function} and GR is immediately restored when $f$ is set to zero. Notice that the theory defined by Eq.~(\ref{2.1}) can be seen as a special case of the Generalized non-local Teleparallel Gravity (GNTG) proposed in \cite{Bahamonde:2017sdo}, from which the above model is recovered by setting the coupling constants equal to $1$ and $-1$, respectively.

The model in Eq.~(\ref{2.1}) explains well the current cosmic acceleration. In fact, the non-local terms here introduced allow a delayed response to the transition from radiation to matter dominance, and then avoid fine tunings often introduced to address late time acceleration \cite{Deser:2007jk}. Furthermore, using a proper piecewise-defined distortion function, the non-local model may lead to the unification of the early-time inflation with late-time acceleration \cite{Nojiri:2007uq}. 

A local representation of the non-local model Eq.~(\ref{2.1}) has been proposed in \cite{Nojiri:2007uq}, where an equivalent scalar-tensor theory is built by introducing two auxiliary scalar fields, $\eta$ and $\xi$, so that the action is rewritten as follows
\begin{equation}\label{2.2}
\begin{aligned}
    S \, &= \, \frac{1}{16 \pi G} \int d^{4}x \, \sqrt{-g} \, \Big[ R \big( 1 + f(\eta) \big) + \xi \big( \Box\eta - R \big) \Big] \, = \\
    &= \, \frac{1}{16 \pi G} \int d^{4}x \, \sqrt{-g} \, \Big[ R \big( 1 + f(\eta) - \xi \big) - \nabla^{\mu}\xi \, \nabla_{\mu}\eta \Big]
\end{aligned}
\end{equation}
where  the term $\xi\Box\eta$ has been integrated by parts and  the boundary term is set to zero. A general procedure to build a local representation of non-local theories is presented in \cite{Capozziello:ReviewNonLocal}. Varying the action in Eq.~(\ref{2.2}) with respect to $\xi$ and $\eta$, we obtain the two  Klein-Gordon equations:
\begin{equation}\label{2.3}
    \Box \eta = R \;\; \Rightarrow \;\; \eta = \Box^{-1}R
\end{equation}
\begin{equation}\label{2.4}
    \Box \xi = -R \, \frac{df(\eta)}{d\eta}
\end{equation}
where the auxiliary fields behave as Lagrange multipliers.
Eq.~(\ref{2.3}) is a constraint such that substituting it into (\ref{2.2}) one recovers (\ref{2.1}). Instead, equation (\ref{2.4}) is a non-trivial equation for the dynamics of the scalar field $\xi$. Furthermore, the variation of the action in Eq.~(\ref{2.2}) plus the matter action, with respect to the metric tensor $g_{\mu\nu}$, yields to the following gravitational field equation:
\begin{equation}\label{2.5}
\begin{aligned}
    G_{\mu\nu} \, &= \, \frac{1}{1+f(\eta)-\xi} \, \Big[ \big(8\pi G \big)^{2} \, T^{(m)}_{\mu\nu} + \nabla_{\mu}\xi \, \nabla_{\nu}\eta \,\, + \\
    &+ \big( \nabla_{\mu}\nabla_{\nu} - g_{\mu\nu} \, \Box \big) \big( f(\eta) - \xi \big) - \frac{1}{2} g_{\mu\nu} \nabla^{\alpha}\xi \, \nabla_{\alpha}\eta \Big]
\end{aligned}
\end{equation}
A non-trivial form for the distortion function has been reconstructed in \cite{WoodardDistFun} in order to reproduce the $\Lambda$CDM expansion. The model provided by substituting such form of $f$ in Eq.~(\ref{2.1}) has been analysed in its localized version in \cite{NersisyanConstraints2017}, using Redshift-Space Distortions (RSD) data: a slightly lower value of $\sigma_{8}$ is derived and, consequently, a better agreement with data results with respect to the $\Lambda$CDM model. Moreover, in \cite{Park2018},  the author has shown that the same results can be obtained in the original non-local version in \cite{WoodardDistFun}, as long as the initial conditions are set the same. Further analysis have been performed in \cite{AmendolaIdrianConstraints}, using Cosmic Microwave Background (CMB), Type Ia supernovae (SNIa) and RSD observations: a deficit of growth of linear structures and a higher lensing power as compared to $\Lambda$CDM is derived, so that a significant shift for the parameter $\sigma_{8}$ and $\tau_{re}$ results. It follows a CMB-RSD tension and ``weak'' evidence for the $\Lambda$CDM model. The same authors also highlight a modification in the propagation equations for Gravitational Waves (GWs), which provides a powerful way for testing deviations to GR with future third generation GW interferometers. See also \cite{Capriolo1, Capriolo2}.

A drawback that could arise in the non-local theories of gravitation is the lack of a screening mechanism, which is necessary to avoid any undesired effect at Solar System scale. In \cite{DeserScreening}, it is argued that inside
bound objects $\Box^{-1}R$ acquires positive value, whereas it is negative in cosmology. Thus, as one is free to choose the distortion function, one can set it so that it vanishes for positive values of $\Box^{-1}R$, i.e. $f(\Box^{-1}R) \sim \theta(\Box^{-1}R)$ where $\theta$ is the Heaviside step function. It follows a perfect screening mechanism which allows the  model to reduce to GR at the Solar System scales. However, in \cite{LunarLaserRanging},  has been shown that the value of $\Box^{-1}R$ is actually negative also at Solar System scale, therefore this procedure cannot be applied. As a consequence, the above model would present a time dependence of the effective Newton constant in the small scale limit and it would thus be ruled out by Lunar Laser Ranging (LLR) observations. This conclusion, drawn in \cite{LunarLaserRanging}, seems to be too strong, since it is still not clear how a Friedman-Lema\^itre-Robertson-Walker (FLRW) background quantities behave when evaluated from cosmological scales down to Solar System ones, where the system decouples from the Hubble flow. In fact, a full non-linear time- and scale-dependent solution around a non-linear structure would be necessary.

However, some proposals exist in this direction and the so-called Vainshtein mechanism \cite{Vainshtein:1972sx} can be considered the paradigm to realize the screening. The main problem is that GR is very well-tested at Solar System scales and then any extension requires fine constraints to be physically viable. For example, $f(R)$ gravity requires an accurate \textit{chameleon mechanism} to give realistic models ranging from Solar System up to cosmology (see \cite{Capozziello:2007eu} for a discussion). Basically, any screening mechanism require a scalar field coupled to matter, and mediating a “fifth-force” which might span from Solar System up to cosmological scales. For high density, this force has to be suppressed, so that no deviation from GR should  emerge. For lower densities, the modification to GR become effective with some observational signature. The screening can be accomplished in several ways: for example, a weak coupling between the field and matter in regions of high density can be considered. This coupling should induce a weak fifth-force. In this situation, the field can acquire a large mass in high density environments, being short-ranged and undetectable. In lower density regions, it should be light and long-ranged, as in the case of chameleon fields \cite{Capozziello:2007eu}. Finally, the field  may change the kinetic contribution in the effective Lagrangian, with first or second order derivatives becoming important in a certain range, as in the Vainshtein case. In \cite{Salzano:2017qac}, it has been discussed for clusters of galaxies in presence of galileon fields. In the case we are going to discuss here, non-local terms can be "localized" and they result in an effective scalar field depending on the scale \cite{Acunzo}. This feature can naturally give rise to some screening mechanism solving the above reported problems. We will discuss this topic in detail elsewhere.

\subsection{The Noether Symmetry Approach}

The action in  Eq.~(\ref{2.2}) as well as the two non-trivial field equations Eqs.~(\ref{2.4}) and (\ref{2.5}), all depend on the specific form of the distortion function $f(\Box^{-1}R)$. It is thus possible to use the Noether Symmetry Approach \cite{Capozziello:1996bi} to select a form of the distortion function such that the theory is invariant under point transformations. This method has been extensively used in the literature \cite{Capozziello:1996bi, Dialektopoulos:2018qoe, sym10070233, Bahamonde2019, PhysRevD.95.064031, Bahamonde:2017sdo}, aiming to study modified theories of gravity based on general functions, such as $f(R)$, $f(T)$, etc. Such theories have to be checked against data in order to be constrained
and then obtain physically reliable models. However, it is possible to constraint modified-gravity theories using a theoretical approach rather than a phenomenological one. Specifically, Noether symmetries can be used as a geometric criterion to choose among different models. Moreover, the presence of symmetries imply conserved quantities that, in many cases, have a physical meaning and allow to reduce dynamics and find exact solutions \cite{Capozziello:1996bi}.

The Noether Symmetry Approach works as follows: 
\begin{itemize}
    \item one first selects a class of background space-time metrics, which, in our case, is spherically symmetric, and one therefore writes the metric;
    \item then, one substitutes the metric into the Lagrangian density Eq.~(\ref{2.1}) and, after integrating out all the total derivative terms, it is possible to  obtain a point-like canonical Lagrangian;
    \item one derives the Noether vector $X$, i.e. the infinitesimal generator of  point transformations;
    \item  it is  then possible to apply the Noether symmetry condition  \cite{Dialektopoulos:2018qoe}
    \begin{equation}\label{2.6}
        X^{[1]} \mathcal{L} + \mathcal{L} \bigg( \frac{d\xi^{t}}{dt} + \frac{d\xi^{r}}{dr} \bigg) \, = \, \frac{dh^{t}}{dt} + \frac{dh^{r}}{dr}
    \end{equation}
    where $X^{[1]}$ is the first prolongation of $X$, $\xi^{t}$ and $\xi^{r}$ are the coefficients of the Noether vector and $h^{t}$ and $h^{r}$ are two arbitrary functions depending on time and  generalized coordinates. Expanding the condition Eq.~(\ref{2.6}), one finds a system of  equations with 9 unknown variables, which yields to two possible models that are invariant under point transformations
    \begin{equation}\label{2.7}
        f(\eta) \, = \, c_{4} + c_{3} \,, \eta
    \end{equation}
    \begin{equation}\label{2.8}
        f(\eta) \, = \, c_{4} + \frac{c_{5}}{c_{1}} \, e^{c_{1}\eta}\,.
    \end{equation}
\end{itemize}
See \cite{Capozziello:ReviewNonLocal} for details.
An overview of the general method is given in \cite{Dialektopoulos:2018qoe}, while the explicit calculation of the Noether Symmetry Approach applied to the  model \eqref{2.1} is performed in \cite{Dialektopoulos:2018iph}. It is interesting to note that the two forms Eq.~(\ref{2.7}) and (\ref{2.8}) of the distortion function correspond to those phenomenologically introduced in \cite{Nojiri2017, Wetterich1998}.

Hereinafter,  we consider the exponential form Eq.~(\ref{2.8}) for the distortion function and we set all the integration constants to one. Thus, we have
\begin{equation}\label{2.9}
    f(\eta) \, = \, 1 + e^{\eta}\,.
\end{equation}
As reported in \cite{Modesto1}, this form of coupling is particularly relevant to achieve the so called "super-renormalizability" for effective theories of gravity. Here, it is selected thanks to the existence of related Noether symmetries. 

\subsection{The Newtonian limit}

Now, let us perform the weak field limit for  the above non-local model, in order to derive the Newtonian potentials which will be used to match observations. The gravitational field for a static and  spherically symmetric metric is described by
\begin{equation}\label{2.10}
    ds^{2} \, = \, A(r) \, dt^{2} - B(r) \, dr^{2} - r^{2} \, d\Omega^{2}\,.
\end{equation}
 Notice that the Birkhoff theorem is not guaranteed in non-local gravity, but we expect that static and spherically symmetric solutions are a good approximation when the Newtonian limit is performed. Not even the existence of the solution $B(r) = 1/A(r)$ is guaranteed in modified theories of gravity \cite{Diego}, so it cannot be chosen \textit{a priori} in Eq.~(\ref{2.10}).

It is well known, from GR, that expanding $g_{00}$ up to $v^{2} \sim \mathcal{O}(2)$, one obtains the Newtonian potential $\phi$ for time-like particles. However, for theories beyond GR the Post-Newtonian (PN) limit is necessary, i.e.
\begin{equation}\label{2.11}
    g_{00} \sim \mathcal{O}(6), \quad g_{0i} \sim \mathcal{O}(5), \quad g_{ij} \sim \mathcal{O}(4)\, .
\end{equation}
When higher-order corrections are taken into account, two length scales arise, which are related to the scalar degrees of freedom we introduced to localize the theory. The expansion of the metric components as well as the scalar fields therefore gives
\begin{equation}\label{2.12}
    A(r) \, = \, 1 + \frac{1}{c^2} \phi^{(2)}(r) + \frac{1}{c^4} \phi^{(4)}(r) + \frac{1}{c^6} \phi^{(6)}(r) + \mathcal{O}(8)
\end{equation}
\begin{equation}\label{2.13}
    B(r) \, = \, 1 + \frac{1}{c^2} \psi^{(2)}(r) + \frac{1}{c^4} \psi^{(4)}(r) + \mathcal{O}(6)
\end{equation}
\begin{equation}\label{2.14}
    \eta(r) \, = \, \eta_{0} + \frac{1}{c^2} \eta^{(2)}(r) + \frac{1}{c^4} \eta^{(4)}(r) + \frac{1}{c^6} \eta^{(6)}(r) + \mathcal{O}(8)
\end{equation}
\begin{equation}\label{2.15}
    \xi(r) \, = \, \xi_{0} + \frac{1}{c^2} \xi^{(2)}(r) + \frac{1}{c^4} \xi^{(4)}(r) + \frac{1}{c^6} \xi^{(6)}(r) + \mathcal{O}(8)
\end{equation}
Combining these expansions together with the 00- and 11- components of Eq.~(\ref{2.5}) and with the equations for the scalar fields, Eqs.~(\ref{2.3}) and (\ref{2.4}), one obtains \cite{Dialektopoulos:2018iph}
\begin{equation}\label{2.16}
\begin{aligned}
     A(r) \, &= \, 1 - \frac{2GM\eta_{c}}{c^{2}r} + \frac{G^{2}M^{2}}{c^{4}r^{2}} \Bigg[ \frac{14}{9}\eta_{c}^{2} + \frac{18 r_{\xi} - 11 r_{\eta}}{6 r_{\eta} r_{\xi}} \, r \Bigg] + \\
     &- \frac{G^{3}M^{3}}{c^{6}r^{3}} \Bigg[ \frac{50 r_{\xi} - 7 r_{\eta}}{12 r_{\eta} r_{\xi}} \, \eta_{c} \, r + \frac{16}{27} \eta_{c}^{3} - \frac{2 r_{\xi}^{2} - r_{\eta}^{2}}{r_{\eta}^{2} r_{\xi}^{2}} \, r^{2} \Bigg]
\end{aligned}
\end{equation}
\begin{equation}\label{2.17}
     B(r) \, = \, 1 + \frac{2GM\eta_{c}}{3 c^{2}r} + \frac{G^{2}M^{2}}{c^{4}r^{2}} \Bigg[ \frac{2}{9}\eta_{c}^{2} + \frac{3 r_{\eta} - 2 r_{\xi}}{2 r_{\eta} r_{\xi}} \, r \Bigg]
\end{equation}
\begin{equation}\label{2.18}
\begin{aligned}
     \eta(r) \, &= \, \frac{4GM\eta_{c}}{3 c^{2}r} - \frac{G^{2}M^{2}}{c^{4}r^{2}} \Bigg[ \frac{11 r_{\eta} + 6 r_{\xi}}{6 r_{\eta} r_{\xi}} \, r - \frac{2}{9}\eta_{c}^{2} \Bigg] + \\
     &- \frac{G^{3}M^{3}}{c^{6}r^{3}} \Bigg[ \frac{r^{2}}{r_{\eta}^{2}} - \frac{25 r_{\eta} - 14 r_{\xi}}{12 r_{\eta} r_{\xi}} \, \eta_{c} \, r - \frac{4}{81} \eta_{c}^{3} \Bigg]
\end{aligned}
\end{equation}
\begin{equation}\label{2.19}
\begin{aligned}
     \xi(r) \, &= \, 1 + \frac{G^{2}M^{2}}{c^{4}r^{2}} \Bigg[ \frac{2}{3}\eta_{c}^{2} - \frac{13 r_{\eta} - 6 r_{\xi}}{6 r_{\eta} r_{\xi}} \, r \Bigg] + \frac{G^{3}M^{3}}{c^{6}r^{3}} \, \cdot \\
     &\cdot \Bigg[ \frac{20}{27} \eta_{c}^{3} - \frac{r_{\eta}^{2} - r_{\xi}^{2}}{r_{\eta}^{2} r_{\xi}^{2}} \, r^{2} - \frac{131 r_{\eta} + 6 r_{\xi}}{36 r_{\eta} r_{\xi}} \, \eta_{c} \, r \Bigg]
\end{aligned}
\end{equation}
where $\eta_{c}$ is a dimensionless constant, which can be set to 1 in order to recover the Newtonian limit for $\phi$. The two parameters $r_{\eta}$ and $r_{\xi}$, which arise in the higher-order terms $\mathcal{O}(4)$ and $\mathcal{O}(6)$, are the length scales related to the the scalar degrees of freedom and thus to the non-localities. These length scales are the free parameters of the theory that we want to constrain by observations.

\subsection{Extended spherically symmetric systems}\label{sec:potentials}

In order to confront the theory against data, it is necessary to generalize the results in Eqs.~(\ref{2.16})~-~(\ref{2.17}) to extended mass distributions. First of all, from Eqs.~(\ref{2.16})~-~(\ref{2.17}) we derive the point-mass expressions for the gravitational $(\phi)$ and metric $(\psi)$ potentials, which are, respectively:
\begin{align}\label{phi_point}
\phi(r) &= \phi_{0}(r) + \phi_{1}(r) + \phi_{2}(r)\,, \\
\phi_{0}(r) &=  -\frac{GM}{r}\,, \nonumber \\
\phi_{1}(r) &= \frac{G^{2}M^{2}}{2c^{2}r^{2}} \bigg[ \frac{14}{9} + \left(\frac{3} {r_{\eta}} - \frac{11}{6 r_{\xi}}\right) r \bigg]\,, \nonumber \\
\phi_{2}(r) &= \frac{G^{3}M^{3}}{2c^{4}r^{3}} \bigg[\frac{7 r_{\eta}-50r_{\xi}}{12r_{\eta}r_{\xi}} r -\frac{16}{27} + \frac{2r_{\xi}^{2}-r_{\eta}^{2}}{r_{\eta}^{2}r_{\xi}^2}r^2\bigg]\,; \nonumber
\end{align}
and
\begin{align}\label{psi_point}
\psi(r) &= \psi_{0}(r) + \psi_{1}(r)\,, \\
\psi_{0}(r) &= -\frac{GM}{3r} \nonumber\,, \\
\psi_{1}(r) &= -\frac{G^{2}M^{2}}{2c^{2}r^{2}} \bigg[ \frac{2}{9} + \left(\frac{3}{2 r_{\xi}} - \frac{1}{r_{\eta}} \right) r \bigg]\,. \nonumber
\end{align}
Before  proceeding, it is worth having  a look at the orders of magnitude of each of the above contributions, to forecast their weight in the analysis and thus estimate which kind of constraints (if any) it is possible to  put on the theory. A rough estimation, using the typical masses and radii of the clusters of galaxies (respectively, $M=10^{15}$ $M_{\odot}$ and $r=3$ Mpc), tells us that:
\begin{equation}\label{2.20}
    \mathcal{O}\left(\frac{G M}{r}\right)\sim 10^{-27} \; \mathrm{kpc^{2} \, s^{-2}}
\end{equation}
\begin{equation}\label{2.21}
    \mathcal{O}\left(\frac{G^2 M^2}{2 c^2 r^2}\right)\sim 10^{-32} \; \mathrm{kpc^{2} \, s^{-2}}
\end{equation}
\begin{equation}\label{2.22}
    \mathcal{O}\left(\frac{G^3 M^3}{2 c^4 r^3}\right) \sim 10^{-37} \; \mathrm{kpc^{2} \, s^{-2}}
\end{equation}
It immediately follows that the terms (\ref{2.22}) are completely negligible in our analysis, and we will not consider them further. However we can also make two additional and more important considerations about these qualitative results.

The first one is that, in order to expect a significant contribution from $\phi_1$ and $\psi_1$ with respect to the standard terms, $\phi_0$ and $\psi_0$, we should have $r_{\eta} \sim r_{\xi}\sim 10^{-5}r$.

The second consideration is probably even more decisive: in case the non-local corrections to the potential were not large enough, we would have here a theory which \textit{does not} reduce to GR. In fact, it is straightforward to see that in that case we would have $\psi \sim \phi/3$. If such a scenario would be able to fit the data (and if yes, with which implications) is something we are going to check carefully in the following analysis.

The generalization to extended system is based on performing the following integrals (using spherical coordinates)
\begin{equation}\label{ext_Phi0}
\Phi_{0}(r) = \int_{0}^{\infty} \rho(x') x'^2 dx'\int_{0}^{\pi} \sin \theta' d\theta'
\int_{0}^{2\pi} d\phi' \phi_0(r)\,, 
\end{equation}
with $r$ defined as:
\begin{equation}
r = |\boldsymbol{x} - \boldsymbol{x'}| = \big(x^{2}+ x'^{2} - 2\,x\,x'\,\mathrm{cos}\theta \big)^{1/2}\, ,
\end{equation}
where $\boldsymbol{x}$ is the vector position of the point in the space where we want to calculate the potential, and $\boldsymbol{x'}$ is the vector position connected to the mass distribution (source of the gravitational potential). Note how
the integration over the radial coordinate has to be performed between $0$ and $\infty$ because Newton's theorems are not guaranteed in non-local gravity, so that we cannot apply the Gauss theorem and the effects of external shell of matter cannot be neglected\footnote{As discussed in \cite{Cardo}, the fact that the Gauss theorem could be violated in Extended Gravity is not a problem because conservation laws are guaranteed by the Bianchi identities.}.

While the term $\Psi_0(r)$ is derived as in Eq.~(\ref{ext_Phi0}), $\Phi_1(r)$ and $\Psi_1(r)$ need an intermediate step, as they involve the term $M^2$. Since
the mass element can be written as
\begin{equation}\label{eq:dM}
dM = \rho(x') \, x'^{2} \, dx' \, \sin \theta \, d\theta \, d\phi\, ,
\end{equation}
and considering that $dM^2 = 2 M dM$, with
\begin{equation}\label{2.27}
M(r) = \, 4\pi \int_{0}^{r} dx' \, x'^{2} \, \rho(x')\, 
\end{equation}
we get:
\begin{align}\label{ext_Phi1}
\Phi_{1}(r) &= 2 \int_{0}^{\infty} \rho(x') x'^2 dx'\int_{0}^{x'} \rho(x'') x''^2 dx'' \\ 
&\int_{0}^{\pi} \sin \theta' d\theta' \int_{0}^{\pi} \sin \theta'' d\theta'' \int_{0}^{2\pi} d\phi' \int_{0}^{2\pi} d\phi'' \phi_1(r)\,.  \nonumber
\end{align}
The same results holds for $\Psi_{1}(r)$.

\subsection{The mass density profile}

To compute the integrals of the extended potentials, it is necessary to make a choice for the mass density profile describing the mass distribution in galaxy clusters. In our analysis, the matter distribution in galaxy clusters is considered to be dominated by DM. 
We have chosen to describe the cluster mass distribution with a spherically symmetric Navarro--Frenk--White (NFW) mass density profile \cite{NFW1996},
\begin{equation}\label{eq:NFW}
\rho_\mathrm{NFW}(r) =  \frac{\rho_{s}}{\frac{r}{r_{s}} \Big( 1 + \frac{r}{r_{s}} \Big)^{2}}
\end{equation}
where $\rho_{s}$ and $r_{s}$ are the characteristic halo density and radius, respectively. Here, the choice of the NFW density profile is motivated by cosmological $N$-body simulations \cite{NFW1996,Child2018} and observational results based on gravitational lensing \cite{Umetsu:2015baa} (see Section~\ref{Section 3}), both in GR context.

It is useful to express the NFW parameters in terms of the overdensity radius $r_\Delta$ and the dimensionless concentration parameter $c_\Delta$ as
\begin{equation}\label{eq:cr_delta}
\rho_{s} = \frac{\Delta}{3} \rho_{c} \frac{c_{\Delta}^{3}}{\mathrm{ln}(1+ c_{\Delta}) - \frac{c_{\Delta}}{1 + c_{\Delta}}},
\end{equation}
\begin{equation}\label{2.31}
r_{s} = \frac{r_{\Delta}}{c_{\Delta}},
\end{equation}
where $r_{\Delta}$ is the spherical radius within which the mean interior density equals $\Delta$ times the critical density $\rho_c$ of the Universe at the cluster redshift. Instead of working with $r_{\Delta}$, it is more common to use $M_{\Delta}$, i.e., the total mass contained within the overdensity radius $r_{\Delta}$,
\begin{equation}\label{eq:M_delta}
M_{\Delta} = \frac{4}{3} \pi r_{\Delta}^{3} \Delta \rho_{c} = 4 \pi \rho_{s} r_{s}^{3} \bigg[ \mathrm{ln}(1+ c_{\Delta}) - \frac{c_{\Delta}}{1 + c_{\Delta}} \bigg]\, .
\end{equation}
In the literature, the typical choice of the overdensity characterizing the halo mass is $\Delta = 200$, while higher overdensities, such as $\Delta = 500$ and $2500$, are used to characterize the properties of halos in their inner regions. In our analysis, we set $\Delta=200$. Thus, the free NFW parameters that we have used for our analysis are $\{ M_{200}, c_{200} \}$.

Once the choice for the mass density profile has been made, it is finally possible to compute analytically the integrals for the extended potentials from Eqs.~(\ref{ext_Phi0}) and (\ref{ext_Phi1}), which we omit to write here explicitly just for the sake of clarity. 

\subsection{Gravitational lensing}

All the previous machinery is fundamental to calculate the quantity which we will then compare with observational data which, as explained in next section, are based on gravitational lensing event analysis from the clusters of galaxies we have considered.

The general configuration for a gravitational lensing system \cite{LensingReview,Umetsu2020} has a foreground object (the lens) between the observer and the background source of light. The angular diameter distances between the observer and the lens and between the observer and the source are denoted by $D_{L}$ and $D_{S}$, respectively, while the angular diameter distance between the lens and the source is $D_{LS}$. The angular diameter distance as a function of redshift is defined as:
\begin{equation}\label{eq:dA}
D_{A}(z) = \frac{c}{1+z} \int_{0}^{z} \frac{dz'}{H(z')}\, .
\end{equation}
In a $\Lambda$CDM scenario, the Hubble function $H(z)$ is given by the first Freedman equation,
\begin{equation}\label{2.41}
    H(z) = H_{0} \sqrt{\Omega_{m} (1+z)^{3} + \Omega_{k} (1+z)^{2} + \Omega_{\Lambda}}\, .
\end{equation}
In our analysis, we assume as our fiducial background cosmology the one from the latest release of the \textit{Planck} survey \cite{Planck:2018vyg}, with $\Omega_m = 0.308$, $H_0 = 67.89$ and $\Omega_k=0$ (from which $\Omega_{\Lambda} = 1-\Omega_m$). It is important to note here that we are implicitly assuming that the non-local model we are analyzing behaves on cosmological scales as this chosen $\Lambda$CDM one, at least effectively. Any cosmological implication and analysis is out of the scopes of this paper, however it is worth noticing that, according to the Deser and Woodard model \cite{Deser:2007jk} which we are essentially adopting here, using non-local term is a natural way to address  dark energy behaviour and recover the observed late universe (see also \cite{Capozziello:ReviewNonLocal}. In this perspective, assuming $\Lambda$CDM is in agreement with this result.

It is generally verified that the distances $D_{L}$ and $D_{LS}$ are much
larger than the physical extension of the lens, so that the latter can be approximated as a two-dimensional system (``thin-lens'' approximation). In such a configuration the relation among the angular position of the source $(\vec{\theta}_{s})$ and the angular position of the image $(\vec{\theta})$ is given by the lens equation:
\begin{equation}\label{eq:lens_eq}
\vec{\beta} = \vec{\theta} - \frac{D_{LS}}{D_S}\, \hat{\vec{\alpha}}(\vec{\theta})\,,
\end{equation}
where $\vec{\hat{\alpha}}$ is the deflection angle, which, in GR, is defined as:
\begin{equation}\label{eq:defl_angle}
\hat{\vec{\alpha}} =  \frac{2}{c^{2}} \int_{-\infty}^{+\infty} \vec{\nabla}_{\perp} \Phi \mathrm{d}z\, ,
\end{equation}
with $\vec{\nabla}_{\perp}$ being the two-dimensional gradient operator perpendicular to the light propagation and $z$ the line-of-sight direction. The deflection angle is then directly related to the quantity which we will use to constrain our model, the convergence 
\begin{equation}\label{eq:convergence_GR1}
\kappa (R) \equiv  \frac{1}{c^{2}} \frac{D_{LS}D_{L}}{D_{S}} \int_{-\infty}^{+\infty} \nabla^{2}_{r} \Phi(R,z)\mathrm{d}z
\end{equation}
where: $R$ is the two-dimensional projected radius in the lens plane\footnote{It is clear from the context that this $R$ is a radius which must not to be confused with the above curvature scalar.}; $r=\sqrt{R^2+z^2}$ is the three-dimensional radius; $\nabla^{2}_{r}$ is the Laplacian operator
in spherical coordinates; and $c$ is the speed of light. If we make use of the Poisson equation 
\begin{equation}\label{2.48}
\nabla^{2}_{r}  \Phi = 4\pi G \rho(r)\,
\end{equation}
we then obtain the final expression for the convergence
\begin{equation}\label{eq:convergence_GR2}
\kappa (R) = \int_{-\infty}^{+\infty} 
\frac{4 \pi G}{c^{2}} \frac{D_{LS}D_{L}}{D_{S}}
\rho(R,z)\mathrm{d}z \equiv \frac{\Sigma}{\Sigma_{cr}}\,
\end{equation}
where $\Sigma(R)$ is the surface density of the lens, defined as
\begin{equation}
\Sigma(R) = \int_{\infty}^{+\infty} \rho(R,z) \mathrm{d}z\, ,
\end{equation}
and $\Sigma_{c}$ is the critical surface density for gravitational lensing,
\begin{equation}\label{2.47}
    \Sigma_{c} \, \equiv \, \frac{c^{2}}{4\pi G} \, \frac{D_{s}}{D_{ls}D_{l}}\, .
\end{equation}
If we want to generalize all the results which we have obtained so far,  valid in GR, we must express the convergence in its most general form as
\begin{equation}\label{eq:convergence_All}
\kappa (R) \equiv  \frac{1}{c^{2}} \frac{D_{LS}D_{L}}{D_{S}} \int_{-\infty}^{+\infty} \nabla^{2}_{r} \left( \frac{\Phi(R,z) + \Psi(R,z)}{2}\right)\mathrm{d}z\, 
\end{equation}
where the potentials $\Phi$ and $\Psi$ which we are going to use are those defined in Sec.~\ref{sec:potentials}.

\section{The data}\label{Section 3}

The data sets we  use for our analysis are obtained from the CLASH program \cite{CLASH2012}.  One of the main goals of the CLASH program was to precisely determine the mass density profiles of 25 high-mass galaxy clusters using deep multi-band \textit{HST} imaging, in combination with wide-field weak-lensing observations \cite{Umetsu2014}. The CLASH sample contains 20 hot X-ray clusters ($>5$~keV) with nearly concentric X-ray isophotes and a well-defined X-ray peak located close to the brightest cluster galaxy (BCG). Notice that no lensing preselection was used to avoid a biased sample towards intrinsically concentrated clusters and those systems whose major axis is preferentially aligned with the line of sight. Cosmological hydrodynamical simulations suggest that the CLASH X-ray-selected subsample is mostly
composed of relaxed systems ($\sim 70\%$)
and largely free of such orientation bias \cite{Meneghetti:2014xna}. Additionally, the CLASH sample has five clusters selected by their exceptional lensing strength so as to magnify galaxies at high redshift. These clusters often turn out to be dynamically disturbed systems \cite{Umetsu2020}.

The CLASH sample spans nearly an order of magnitude in mass ($5\simlt M_{200}/10^{14}M_\odot\simlt 30$) and covers a wide redshift range ($0.18 < z < 0.90$ with a median redshift of $z_\mathrm{med} = 0.40$). For each of the 25 clusters, \textit{HST} weak- and strong-lensing data products are available in the central regions \cite{Zitrin2015}. For 20 of them (16 X-ray-selected and 4 lensing-selected clusters), radial convergence profiles were reconstructed \cite{Umetsu:2015baa} from the combination of ground-based weak-lensing shear and magnification data and \textit{HST} lensing data.

In our analysis, we focus on 15 X-ray-selected and 4 lensing-selected clusters taken from the CLASH subsample analyzed in \cite{Umetsu:2015baa}. Here, one of the X-ray-selected clusters (RXJ1532) has been discarded because no multiple image systems have been identified in the cluster \cite{Zitrin2015} and thus the mass reconstruction is based only on the wide-field weak-lensing data \cite{Umetsu:2015baa}. Our analysis sample spans a redshift range of $0.187 \le z \le 0.686$, with a median redshift of $z_\mathrm{med}= 0.352$. The typical resolution limit of the mass reconstruction, set by the \textit{HST} lensing data, is $10$~arcseconds, which corresponds to $\approx 35 h^{-1}h$~kpc at $z_\mathrm{med}$. Note that, according to \cite{Meneghetti:2014xna}, about half of the selected sample clusters are expected to be unrelaxed.

It was found in \cite{Umetsu:2015baa} that the ensemble-averaged surface mass density $\Sigma(R)$ of the CLASH X-ray-selected subsample is best described by the NFW profile, when GR is considered. The NFW model describes well the DM distribution in clusters, which is the dominant component over the whole cluster scale. The cluster baryons, such as the X-ray-emitting hot gas and BCGs, are sensitive to non-gravitational and local astrophysical phenomena. Accordingly, hydrostatic total mass estimates, based on the X-ray probe, are highly influenced by the dynamical and physical conditions of the cluster. In contrast, gravitational lensing can provide a direct probe of the projected mass distribution in galaxy clusters.

\subsection{Statistical analysis}

The aim of this work is to constrain the parameters of the non-local model (\ref{2.1}) and the parameters describing the NFW profile for each individual cluster. Thus, the vector of our free theoretical parameters is $\boldsymbol{\theta} = \{c_{200}, \, M_{200}, \, r_{\eta}, \, r_{\xi} \}$. The $\chi^{2}$ function for each cluster used in the analysis is defined as
\begin{equation}\label{3.1}
    \chi^{2} = \big( \boldsymbol{\kappa^{theo}}(\boldsymbol{\theta}) - \boldsymbol{\kappa^{obs}} \big) \cdot \mathbf{C}^{-1} \cdot \big( \boldsymbol{\kappa^{theo}}(\boldsymbol{\theta}) - \boldsymbol{\kappa^{obs}} \big)\, ,
\end{equation}
where $\boldsymbol{\kappa^{obs}}$ is the data vector containing the observed convergence values. The data vector consists of $15$ data elements, each representing the value of $\kappa$ measured in each radial bin. The vector $\boldsymbol{\kappa^{theo}}(\boldsymbol{\theta})$ contains theoretical predictions for the model convergence calculated from Eq.~(\ref{eq:convergence_All}). Finally, $\mathbf{C}$ is the covariance error matrix constructed by \cite{Umetsu:2015baa}.

The $\chi^{2}$ function is then minimized using our own Monte Carlo Markov Chain (MCMC) code written in $\texttt{Wolfram Mathematica}$. The convergence of the chains has been checked according to the method proposed in \cite{10.1111/j.1365-2966.2004.08464.x}. Full convergence has been reached for 18 out of 19 sample clusters, while MACSJ0717 shows a pathological behaviour. Such behaviour is due to a peculiar shape of the $\chi^{2}$ function, with two different minima, which is shared by all the samples, but is especially pronounced in MACSJ0717. This shape is directly related to the degeneracy of the geometrical and the matter effects, as will clearly emerge from the results of our analysis.

In order to assess the validity of our non-local model against the standard GR case, we calculated the Bayesian evidence $\mathcal{E}$, so as to provide a statistical meaningful comparison tool between the two theories. We have calculated  the Bayesian evidence using the algorithm proposed in \cite{Mukherjee_2006}. As this algorithm is stochastic, in order to reduce the statistical noise we run it $\sim 100$ times, obtaining a distribution of values from which we extract the best value of $\mathcal{E}$, as the median of the distribution, and the corresponding error.

Since the Bayesian evidence depends on the choice of the priors \cite{Nesseris_2013}, we have always used uninformative flat priors on the parameters. For each cluster, we have assumed flat positive priors on the NFW parameters, i.e., $c_{200}>0$ and $M_{200}>0$ (note that the choice of  NFW priors is different from that of \cite{Umetsu:2015baa}, who used flat logarithmic priors on $M_{200}$ and $c_{200}$), while for the non-local parameters we use flat logarithmic priors, $-5<\mathrm{log}\,r_{\eta,\xi}<5$.

\begin{figure*}[t!]
\centering
\includegraphics[width=8.5cm]{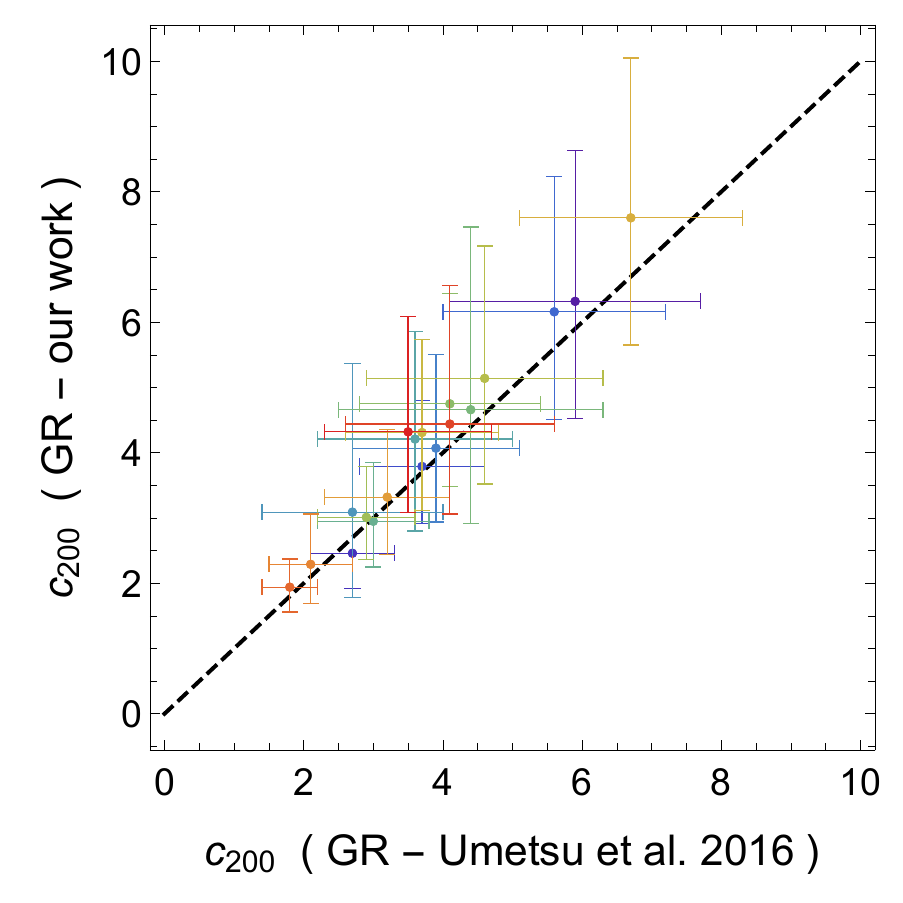}~~~
\includegraphics[width=8.5cm]{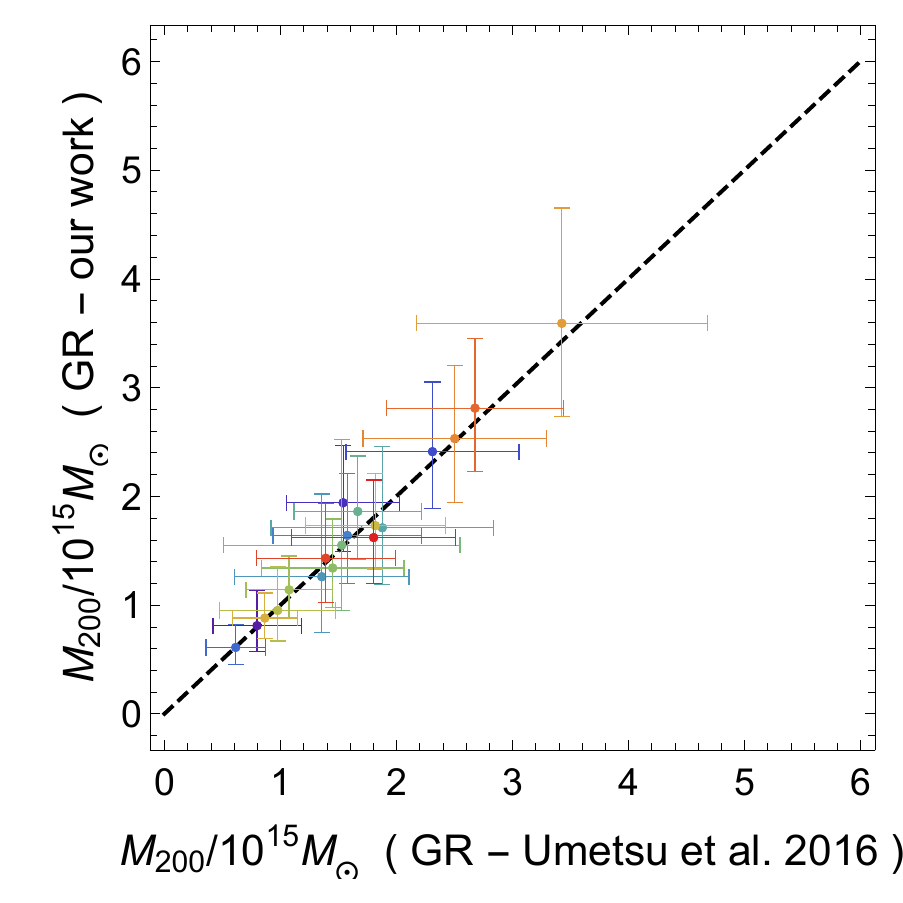}
\caption{Concentrations and masses of CLASH clusters: comparison between our results and those from \citep{Umetsu:2015baa} in the GR scenario.}\label{fig:compare_Umetsu}
\end{figure*}

Finally, the Bayes factors $\mathcal{B}$ are computed for each sample cluster. The Bayes factor is defined as the ratio of evidence factors of two models 
\begin{equation}\label{3.4}
    \mathcal{B}^{\,i}_{j} = \frac{\mathcal{E}(M_{i})}{\mathcal{E}(M_{j})}
\end{equation}
where $M_{j}$ is the reference model, which, in this case,  is GR. The model selection is then performed by using the so-called Jeffreys scale \cite{JeffreysScale}, which states that: for $\ln \mathcal{B}^{\,i}_{j}<0$ there is evidence against the model $M_{i}$, i.e. the reference model is statistically favored; for $0<\ln\mathcal{B}^{\,i}_{j} <1$ there is no significant evidence in favor of the model $M_{i}$; for $1<\ln\mathcal{B}^{\,i}_{j} <2.5$ the evidence is substantial; for $2.5< \ln\mathcal{B}^{\,i}_{j} <5$ the evidence is strong; for $\ln\mathcal{B}^{\,i}_{j} >5$ the evidence is decisive.

\section{Results and Discussion}\label{Section 4}

\begin{figure*}[t!]
\centering
\includegraphics[width=8.5cm]{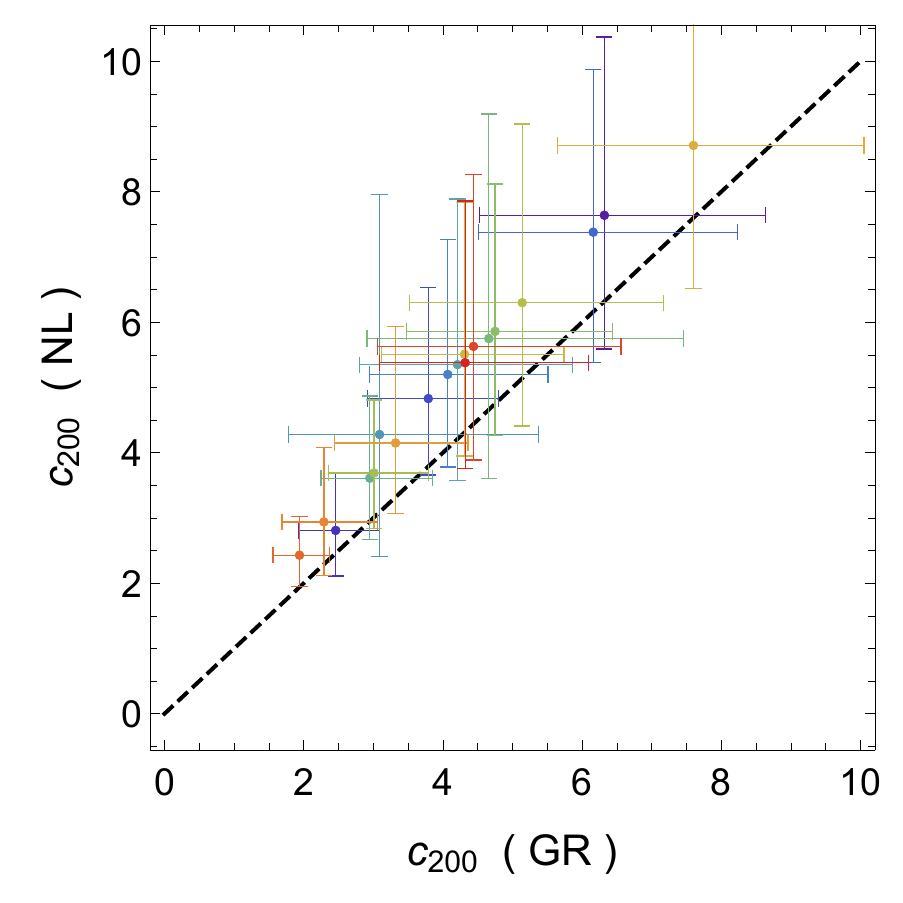}~~~
\includegraphics[width=8.5cm]{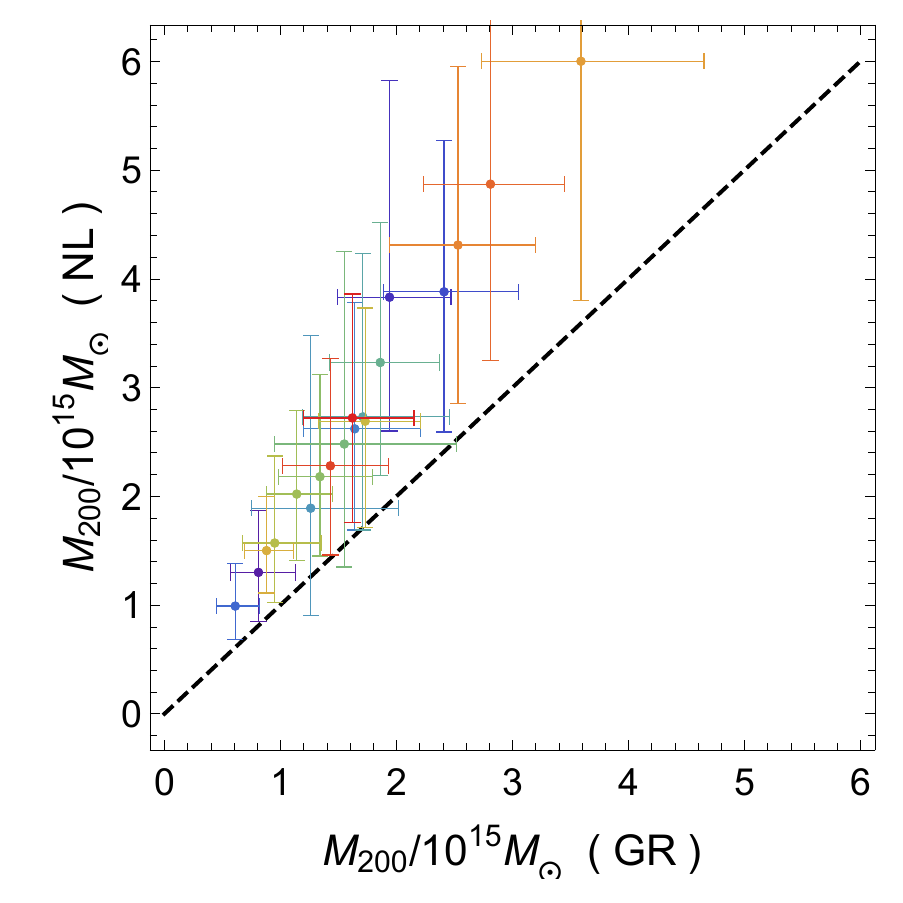}
\caption{Concentrations and masses of CLASH clusters: comparison between the GR and the nonlocal scenario from our analysis.}\label{fig:compare_GR_NL}
\end{figure*}

First of all, we have performed the fits in the classical GR scenario (second column of Table \ref{tab:results}), which will be our reference model. We can thereby cross-check our modelling and statistical analysis algorithm with results from the literature for the same sample. In Fig.~\ref{fig:compare_Umetsu},  we compare our results with \cite{Umetsu:2015baa}, the original work were the lensing data we are using  were first obtained and presented. Note that a direct comparison is possible because the same NFW parametrization, $\{c_{200},M_{200}\}$, has been used, with the same mass modelling, i.e. no further components (gas, galaxies) have been used. 
The cross-check shows an excellent agreement, with all the nineteen estimates of both NFW parameters, $c_{200}$ and $M_{200}$, which agree within the $1\sigma$ level. 

\begin{figure*}[h]
\centering
\includegraphics[width=7.cm]{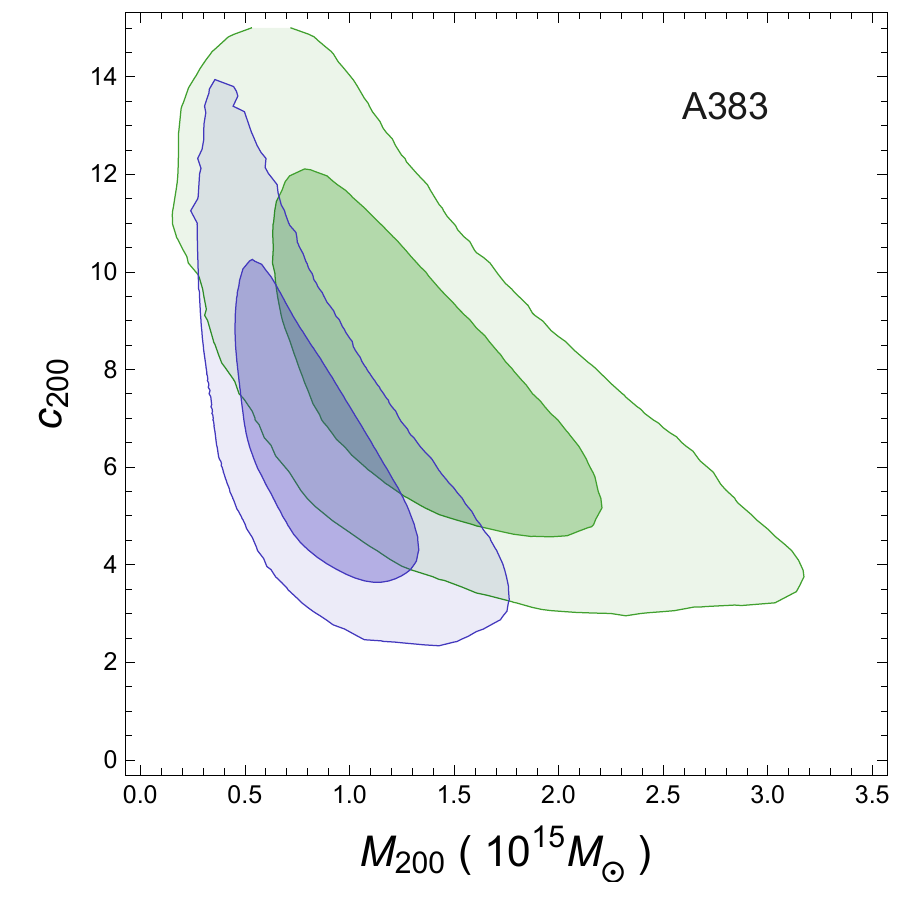}~~~
\includegraphics[width=7.cm]{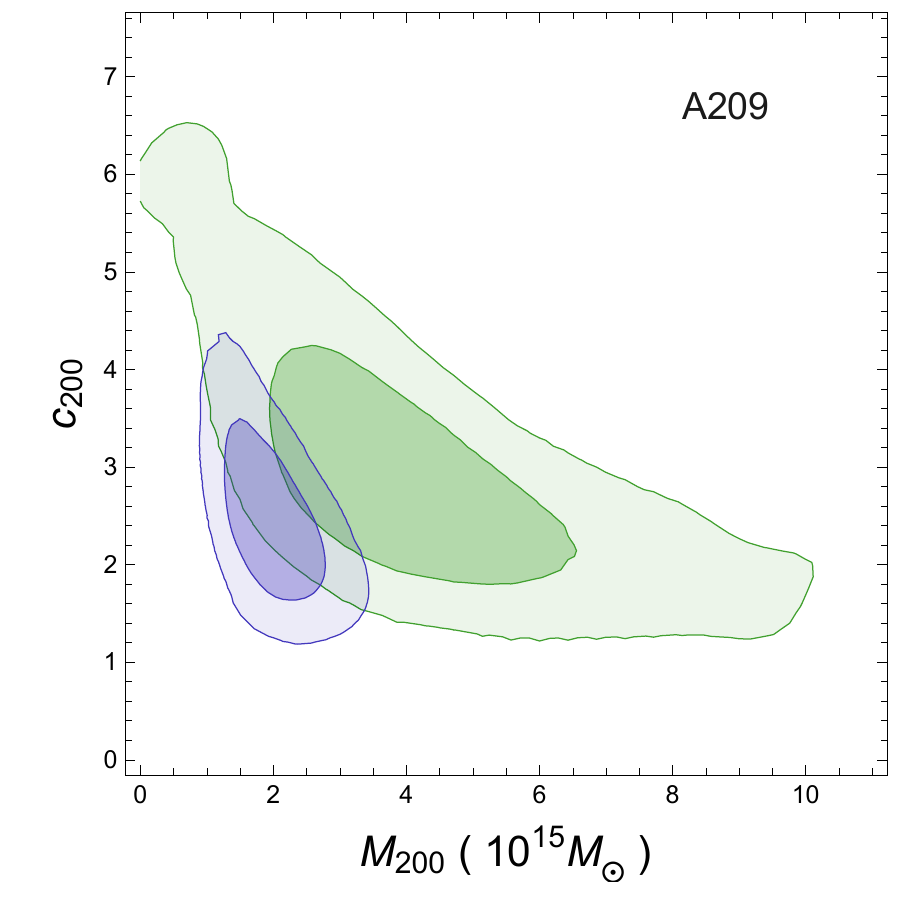}\\
~~~\\
\includegraphics[width=7.cm]{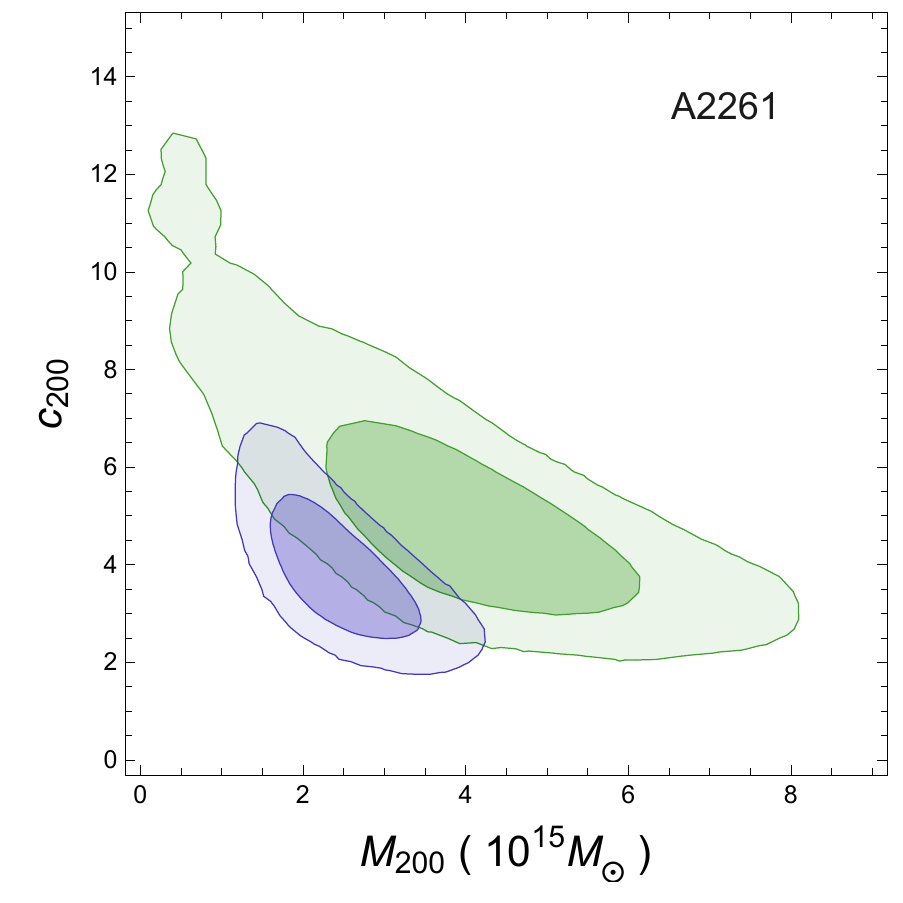}~~~
\includegraphics[width=7.cm]{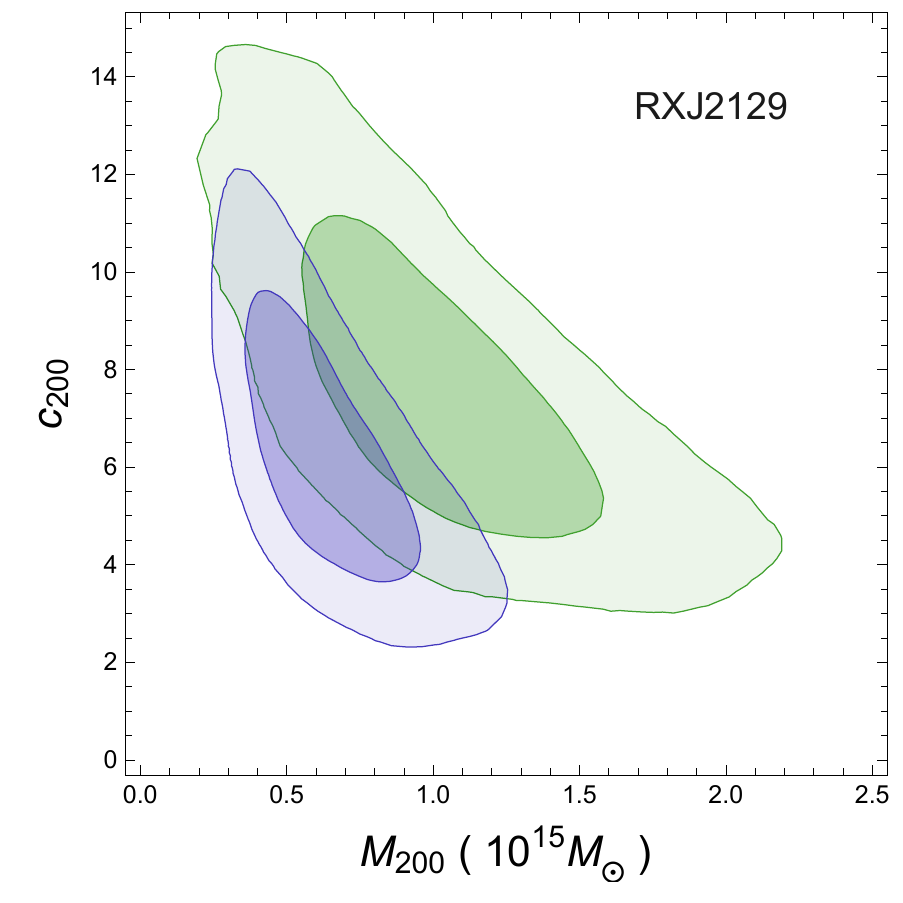}\\
~~~\\
\includegraphics[width=7.cm]{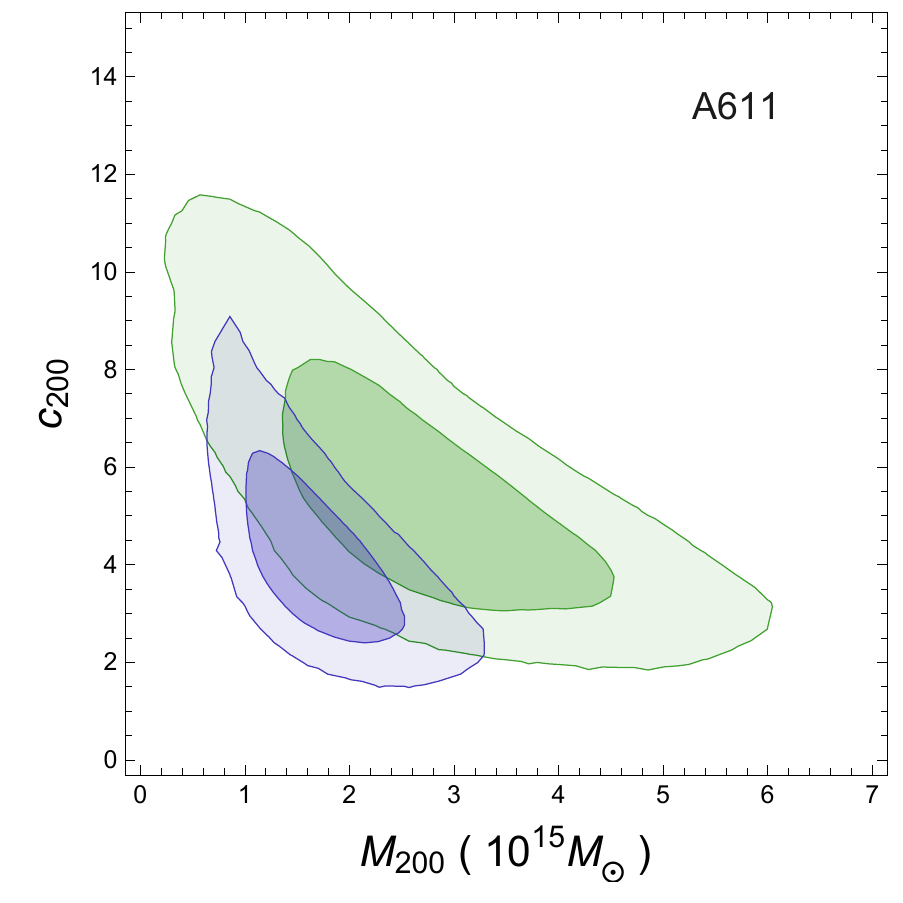}~~~
\includegraphics[width=7.cm]{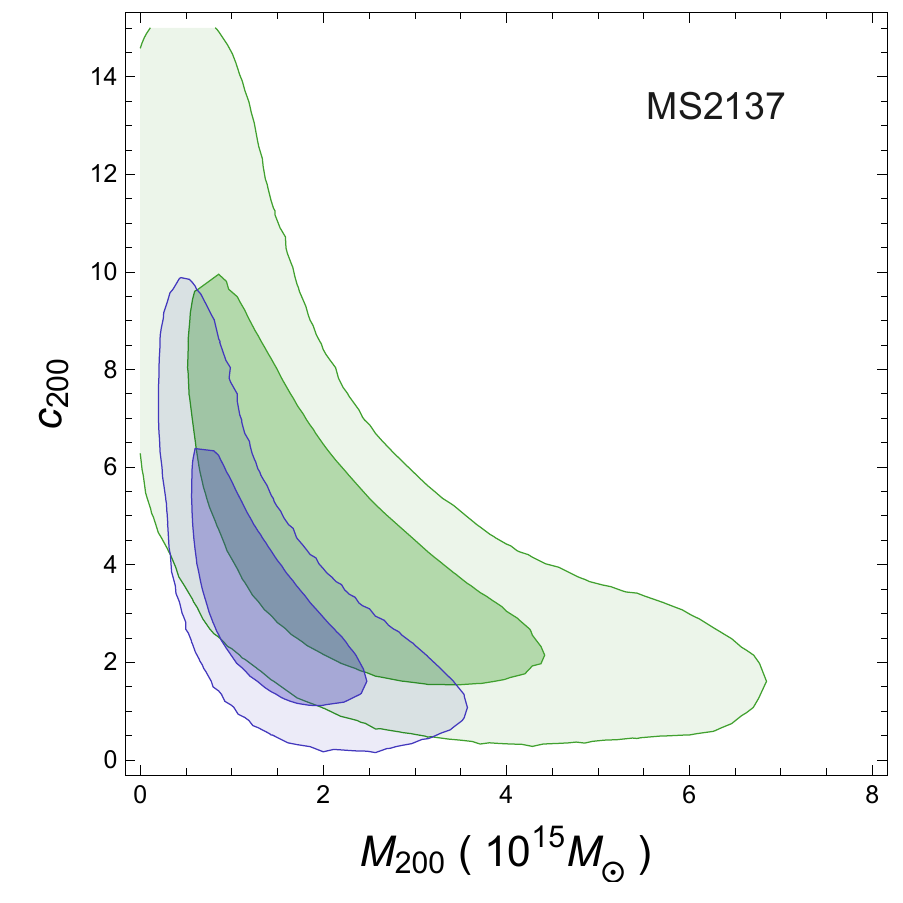}
\caption{Contour Plot: blue - GR; green - nonlocal.}\label{fig:comparisong_cM_1}
\end{figure*}

\begin{figure*}[h]
\centering
\includegraphics[width=7.cm]{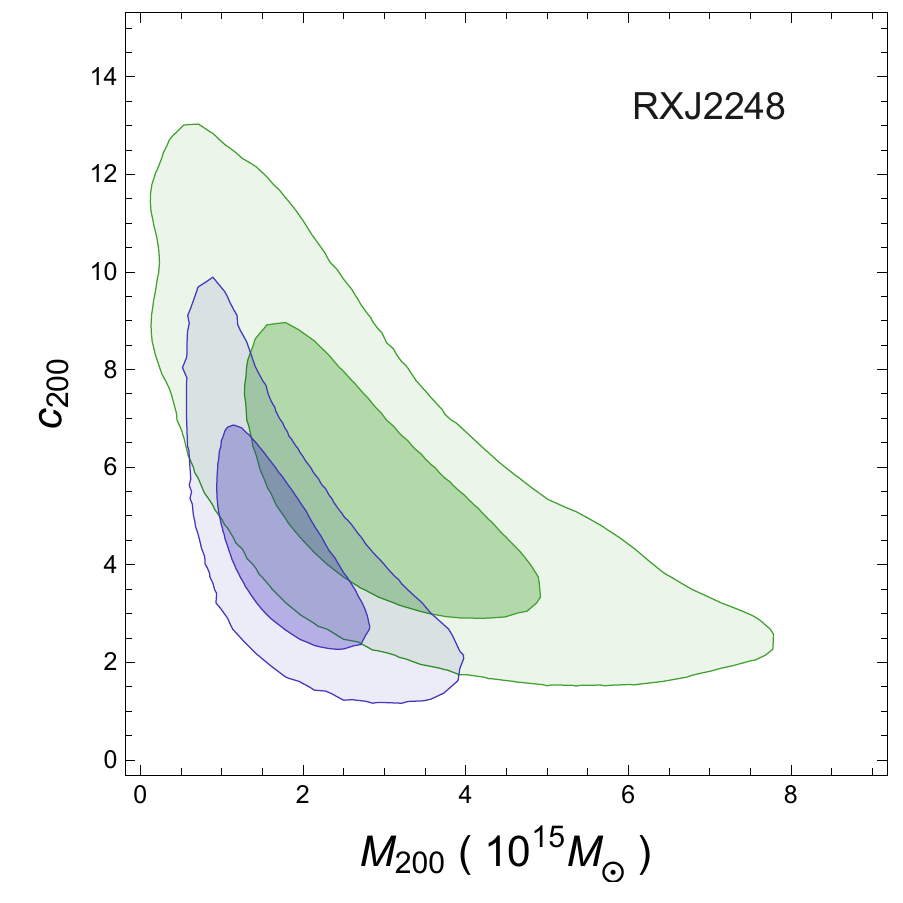}~~~
\includegraphics[width=7.cm]{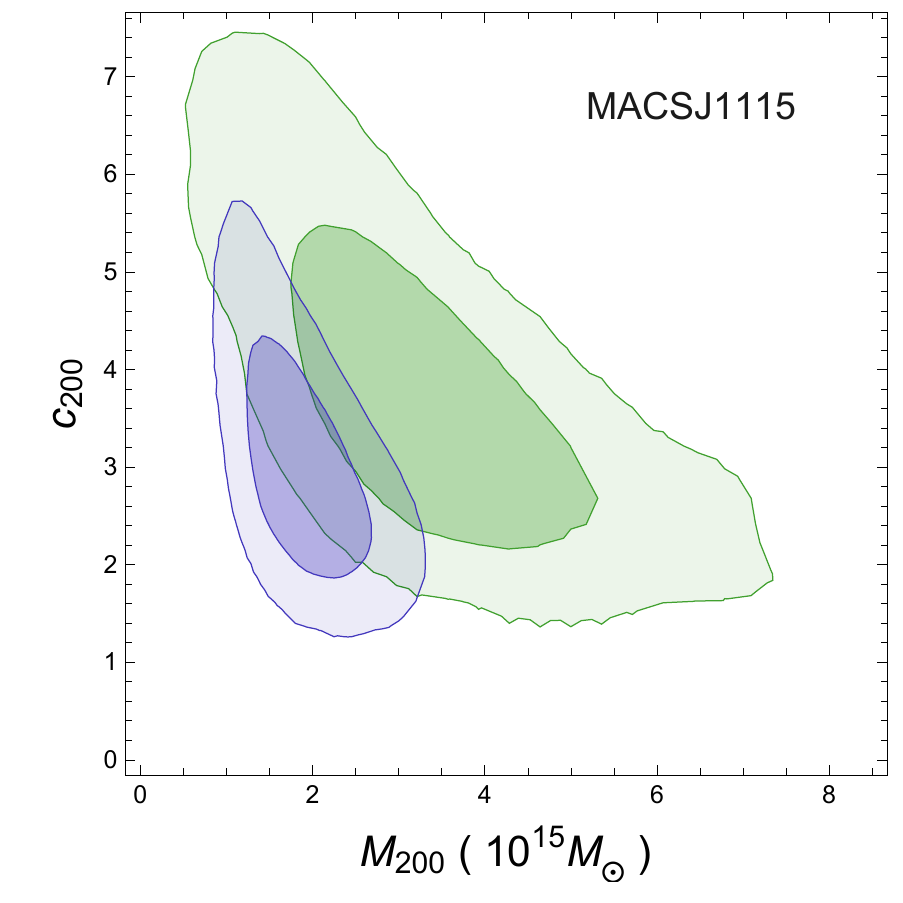}\\
~~~\\
\includegraphics[width=7.cm]{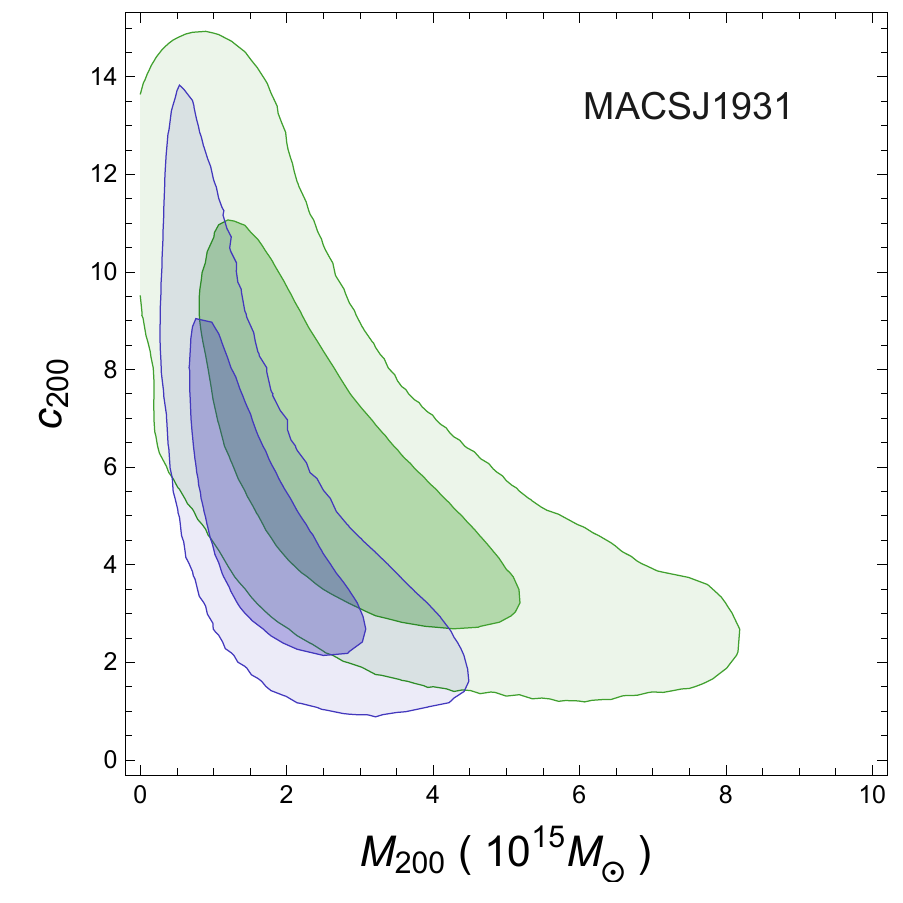}~~~
\includegraphics[width=7.cm]{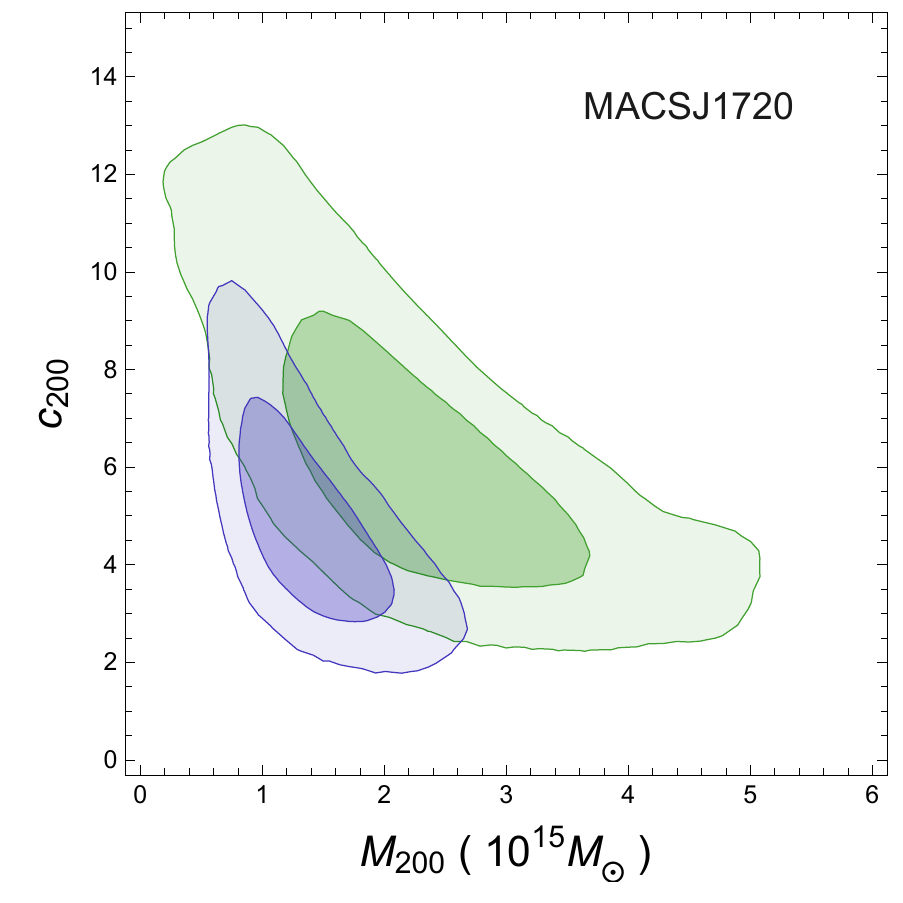}\\
~~~\\
\includegraphics[width=7.cm]{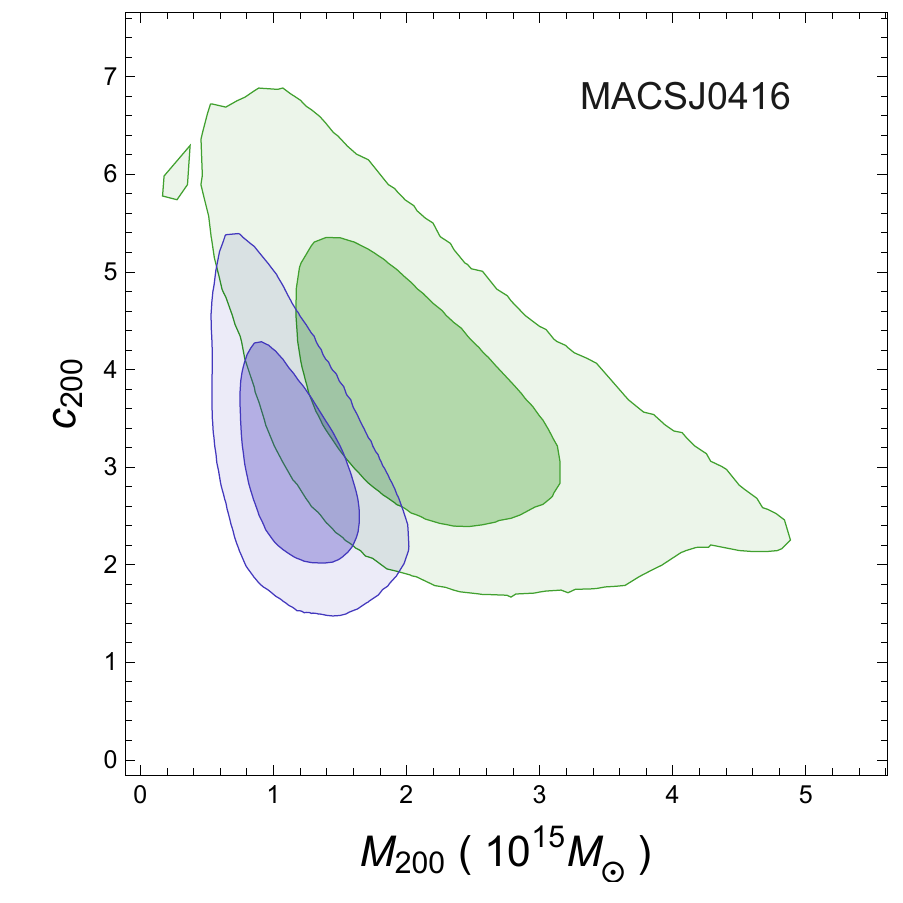}~~~
\includegraphics[width=7.cm]{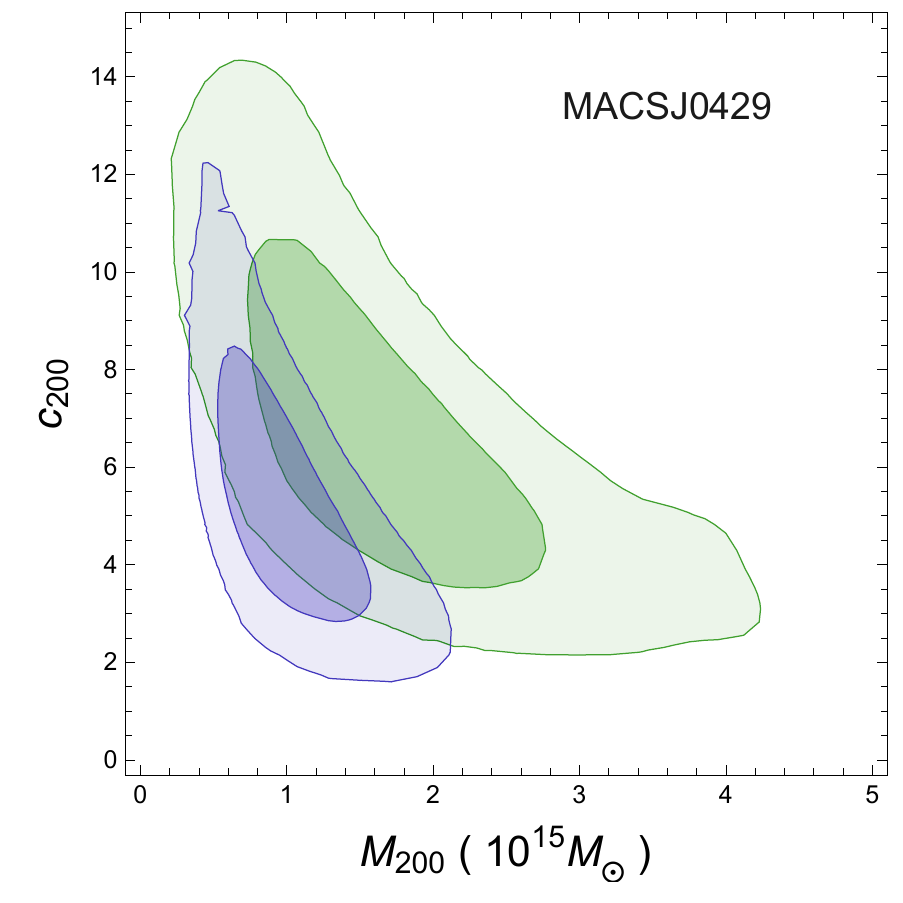}
\caption{Contour Plot: blue - GR; green - nonlocal.}\label{fig:comparisong_cM_2}
\end{figure*}

\begin{figure*}[h]
\centering
\includegraphics[width=7.cm]{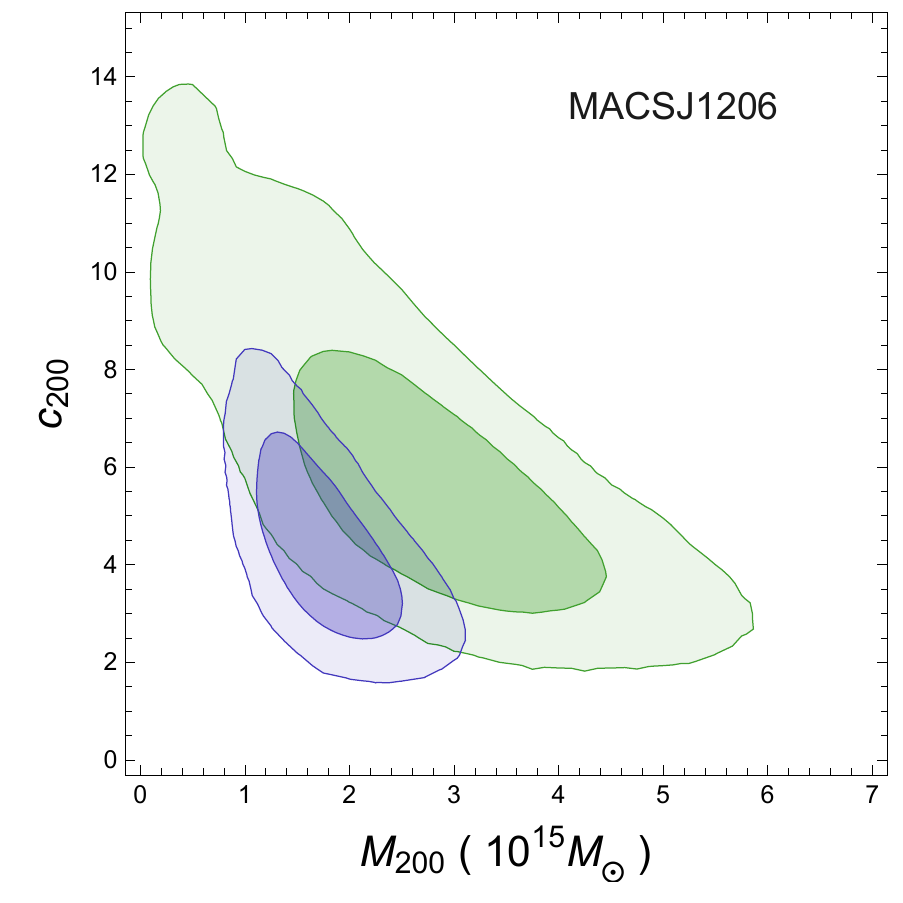}~~~
\includegraphics[width=7.cm]{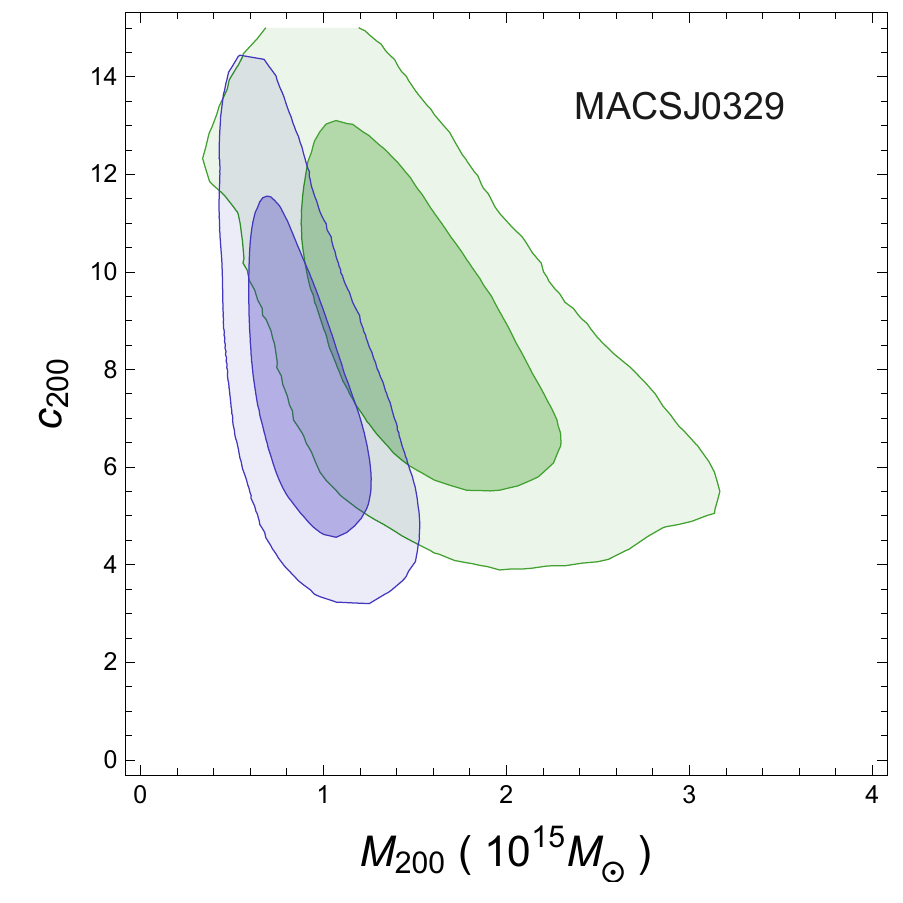}\\
~~~\\
\includegraphics[width=7.cm]{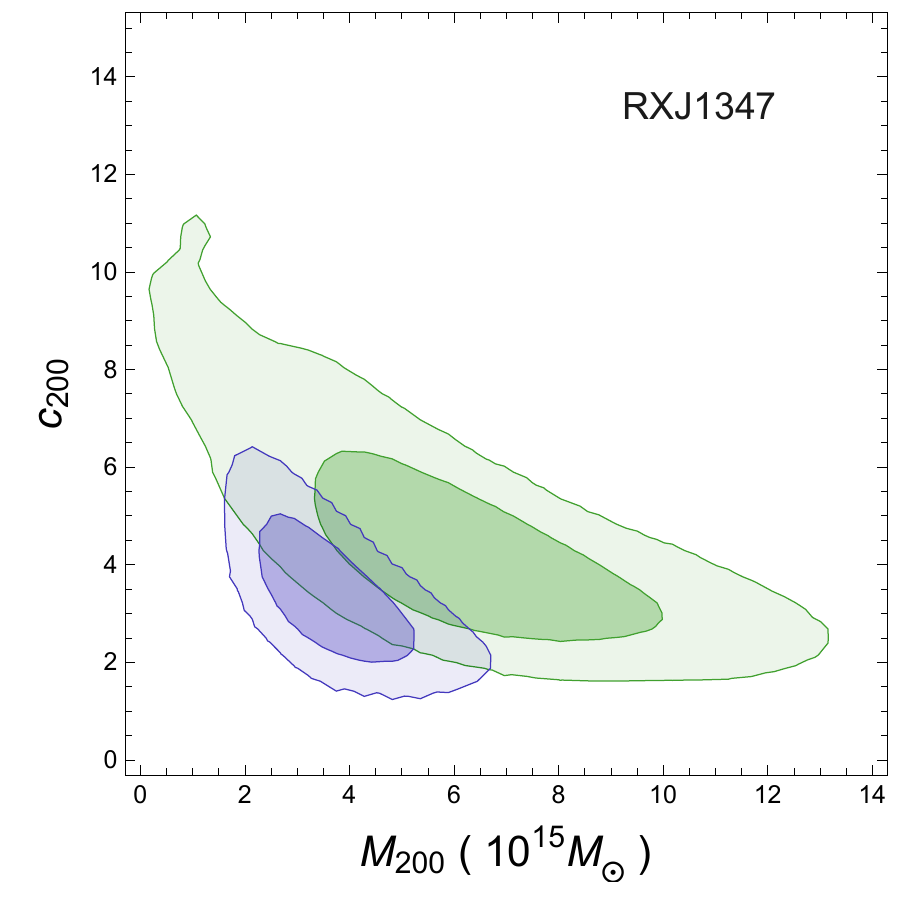}~~~
\includegraphics[width=7.cm]{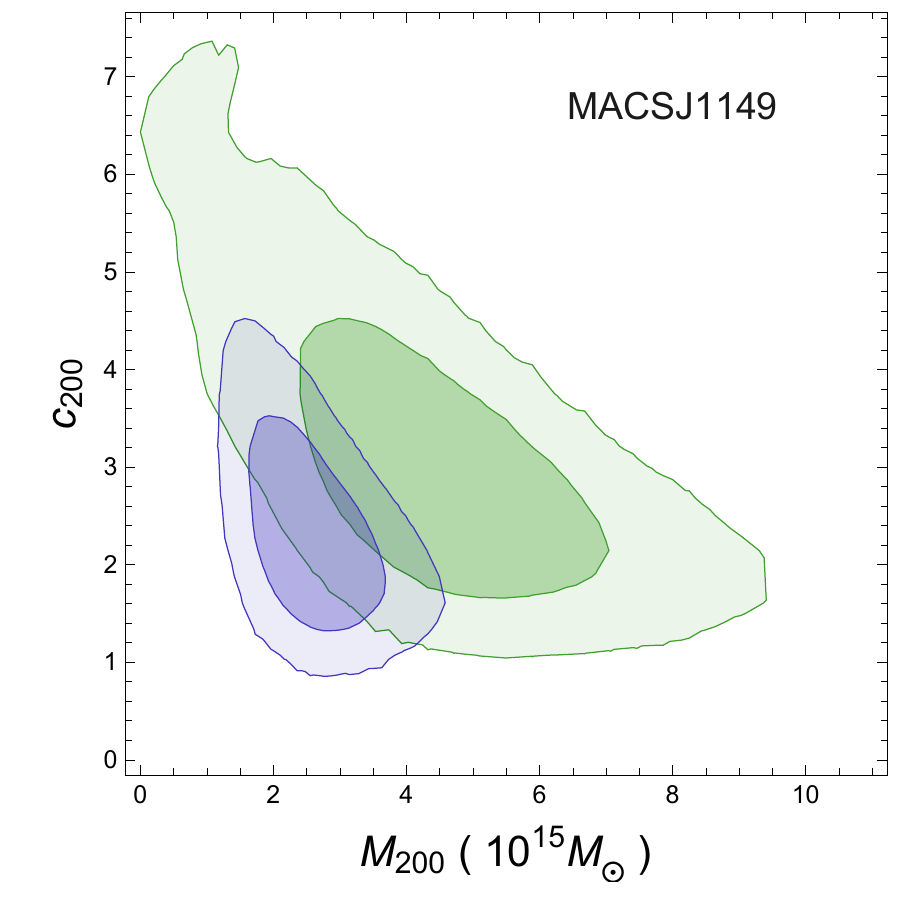}\\
~~~\\
\includegraphics[width=7.cm]{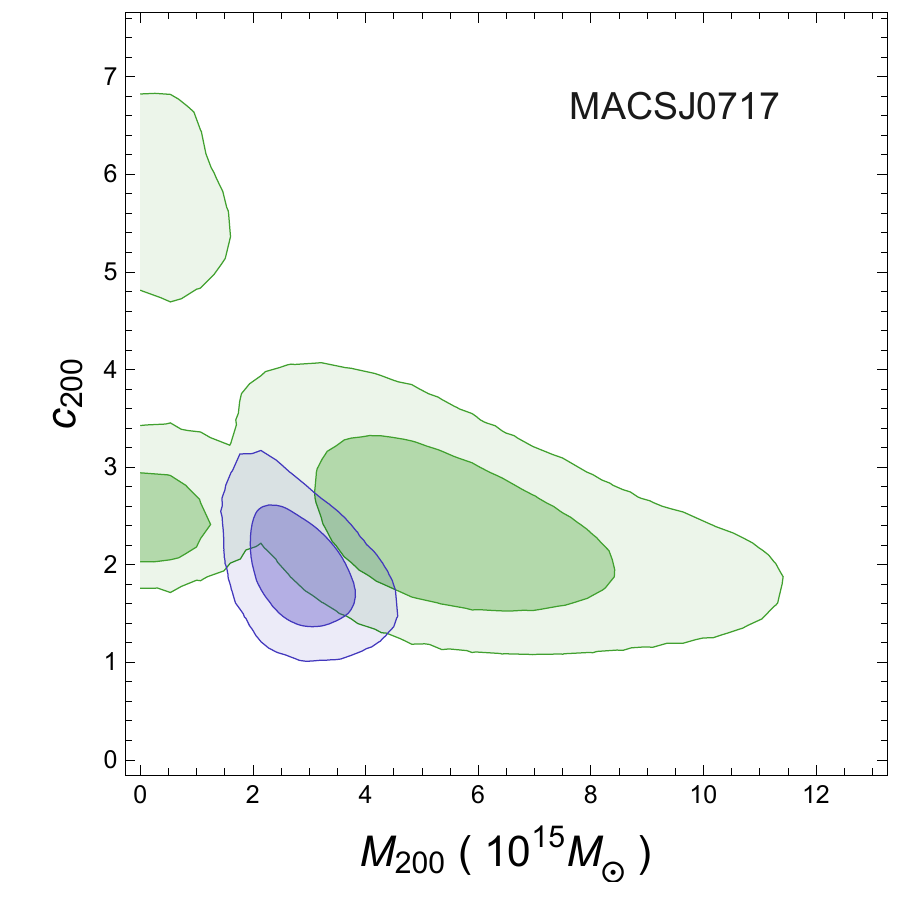}~~~
\includegraphics[width=7.cm]{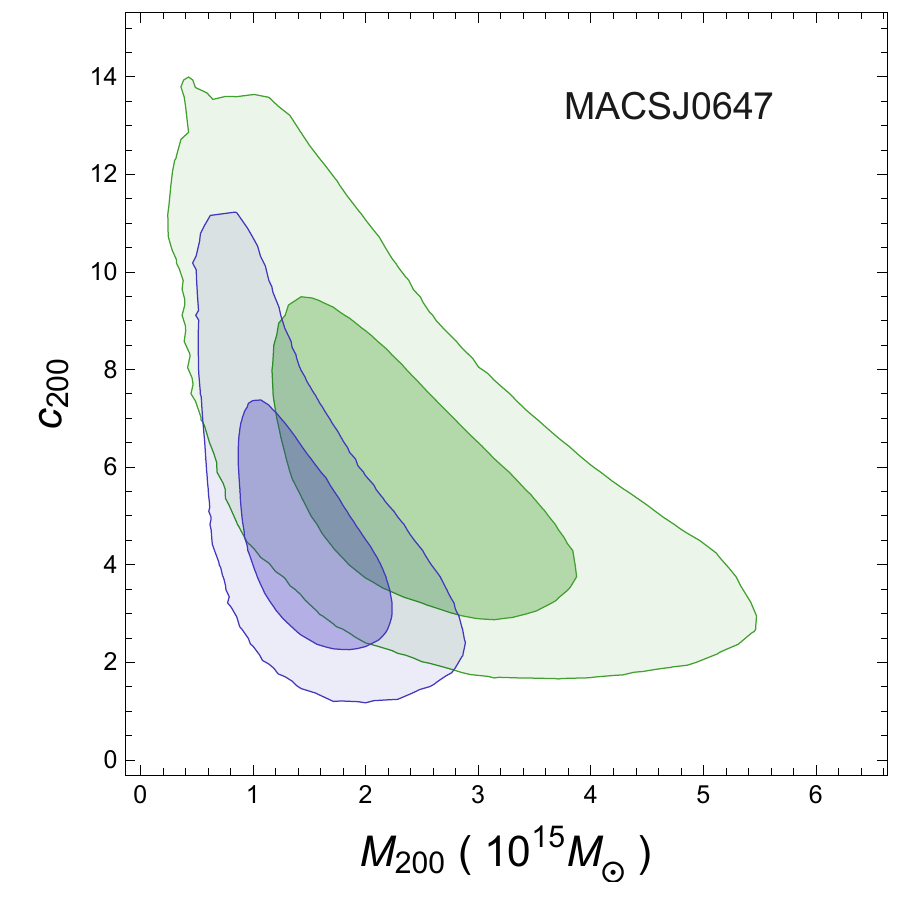}
\caption{Contour Plot: blue - GR; green - nonlocal.}\label{fig:comparisong_cM_3}
\end{figure*}

\begin{figure*}[h]
\centering
\includegraphics[width=7.cm]{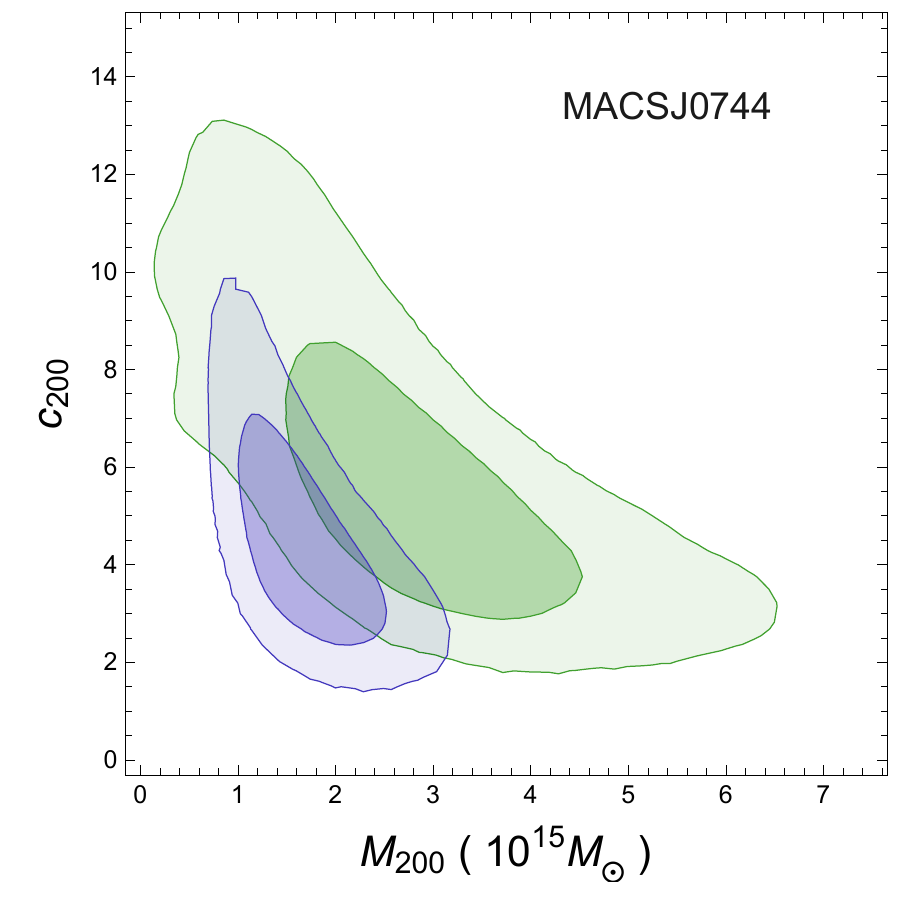}
\caption{Contour Plot: blue - GR; green - nonlocal.}\label{fig:comparisong_cM_4}
\end{figure*}

Concerning the non-local model,  results are shown in the third column of Table~\ref{tab:results}. The NFW parameters are well constrained by the lensing data.
From Fig.~\ref{fig:compare_GR_NL} and Figs.~\ref{fig:comparisong_cM_1}--\ref{fig:comparisong_cM_4}, one can see that the estimates of  concentration parameter $c_{200}$ show no significant differences between GR and our non-local model, being all consistent with each other at the $1\sigma$ level, although the non-local model tends to show higher concentrations than GR. On the other hand, this trend is much clearer in the cluster mass $M_{200}$, whose non-local estimates clearly deviate from the GR ones.

Such increased estimates of $M_{200}$ are very likely related to the form of the metric potentials and the role/weight of the correction terms. As explained above, if the non-local corrections are small and the $\Phi_0$ and $\Psi_0$ terms are the dominant contributions, the theory does not reduce to GR, because the potential $\Psi_0$ only takes into account $1/3$ of the contribution which would be expected. Thus, in the non-local model, the cluster mass (mostly, and slightly for $c_{200}$) must be increased to compensate the missing contribution from $\Psi_0$ to the convergence. From Figs.~\ref{fig:comparisong_cM_1}--\ref{fig:comparisong_cM_4}, it immediately appears that the $1\sigma$- and $2\sigma$-regions shift toward higher masses, except for two cases (A2261 and MACSJ0717) where the $1\sigma$ contours do not overlap. 

\begin{figure*}[h]
\centering
\includegraphics[width=8.5cm]{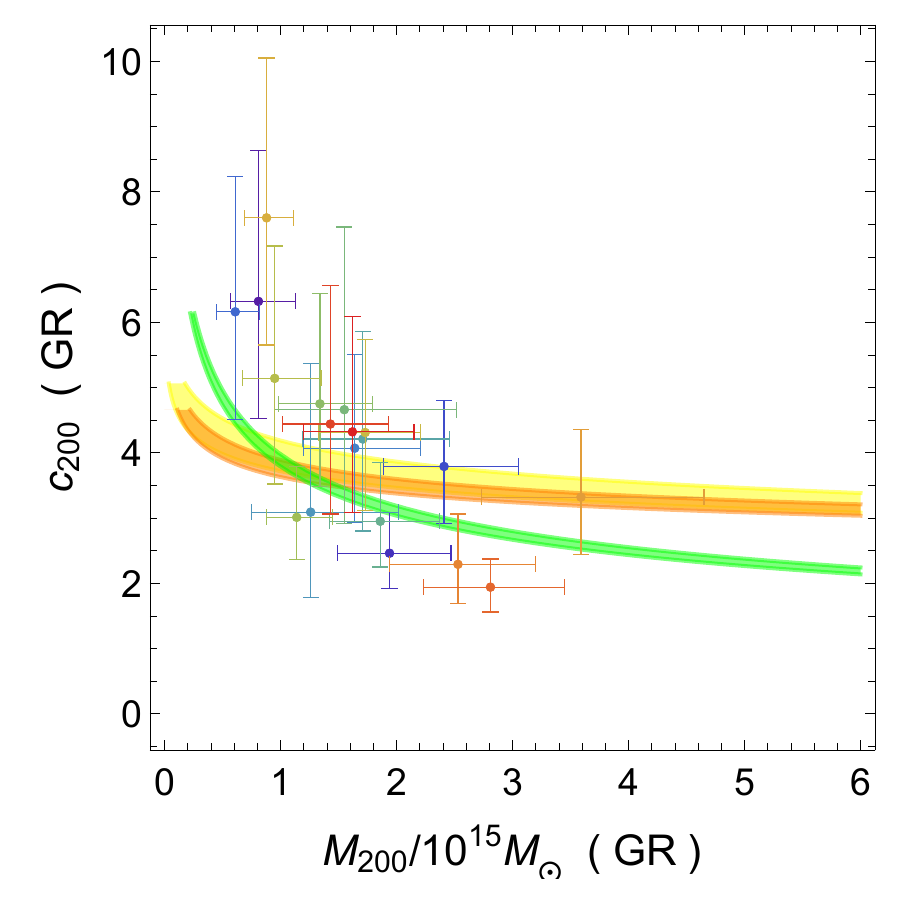}~~~
\includegraphics[width=8.5cm]{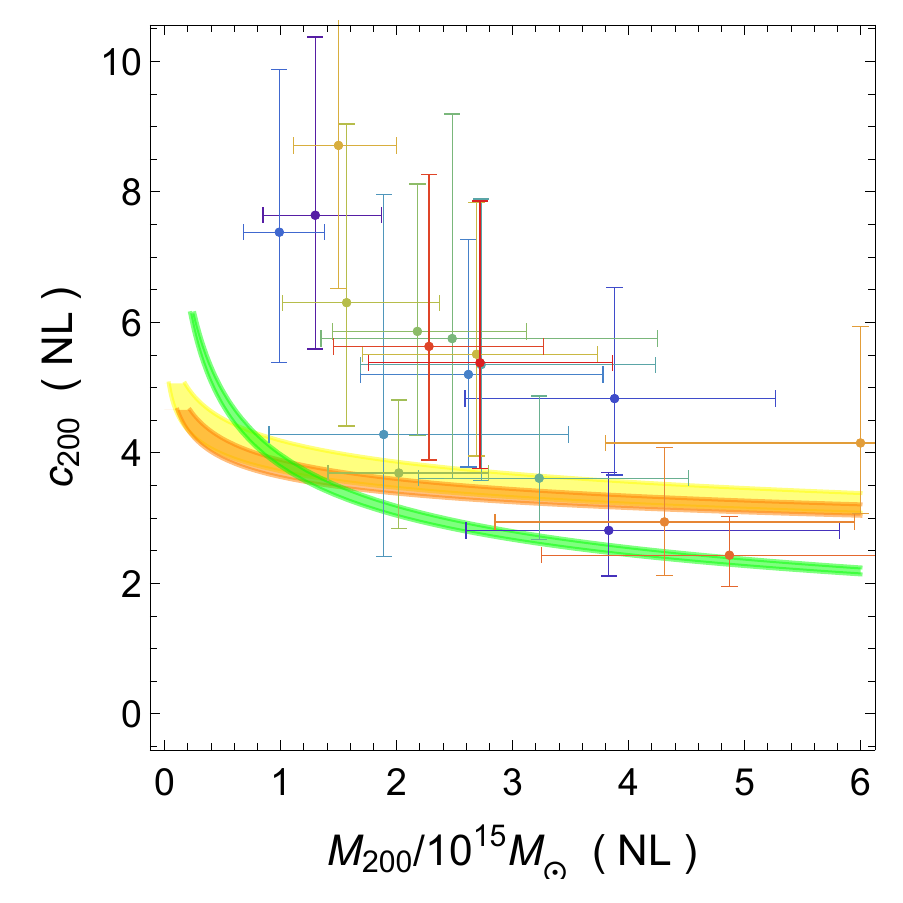}
\caption{Concentrations and masses of CLASH clusters: comparison between our work and $c$--$M$ relations from the literature. Yellow band: $c$--$M$ relation from \citep{Correa:2015dva}; orange band: $c$--$M$ relation from \citep{Diemer:2018vmz}; green band: $c$--$M$ relation for CLASH clusters derived by \cite{Merten2015}. The upper and lower limits of the colored bands correspond to the minimum and maximum redshift values of CLASH clusters.}\label{fig:comparing_cM_literature}
\end{figure*}

As a further check, in Fig.~\ref{fig:comparing_cM_literature} we compare the\\ $(c_{200},M_{200})$ constraints obtained in our work for GR and for the non-local model with $c$--$M$ relations from the literature. As expected, the points representing the $c$--$M$ relation for each cluster scatter due to the different physical properties of the samples. However, for GR, the region spanned by the clusters agrees with the bands representing the relation obtained in the literature. Instead, for the non-local model, the same region shifts towards higher concentrations and masses, hence it does not coincide with expected results, especially with the green band derived in \cite{Merten2015} using the same sample of our work. The gap is around 2$\sigma$, thus the non-local model is just slightly statistically disfavoured respect to GR and cannot be discarded.

When we consider   the non-local length scales $r_{\eta}$ and $r_{\xi}$, the statistics does not appear as much regular as in the case of the NFW parameters. The MCMC sampling does not identify a clear minimum in the $\chi^2$ function; instead, the $\chi^2$ exhibits a shallow valley in the parameter space. This is due to the fact that the non-local corrections are too small to be effectively detected given the observational uncertainties. We find that the MCMC samples are mostly concentrated in the region $-1<\mathrm{log}\,r_{\eta \, , \, \xi}<5$, and only lower limits can be extracted (see column 3 of Table~\ref{tab:results}).

Lower limits on the typical non-local scales  can be set with the present data. They are $r_{\eta}>4\times 10^{-5} - 7\times 10^{-2}\, \mathrm{kpc}$ and
$r_{\xi}>2\times 10^{-5} - 3\times 10^{-2}\, \mathrm{kpc}$,
 so that the magnitudes of the non-local corrections $\phi_{1}$ and $\psi_{1}$ in Eq.~(\ref{phi_point}~-~\ref{psi_point}) are
$\phi_{1} \sim 10^{-28} - 10^{-25}\; \mathrm{kpc^{2} \, s^{-2}}$ and $\psi_{1} \sim 10^{-27} - 10^{-24} \; \mathrm{kpc^{2} \, s^{-2}}$.
Thus, such terms might be dominant or of comparable magnitude with respect to the zeroth-order terms, $\phi_{0}$ and $\psi_{0}$, when the non-local parameters approach their lower bounds. This result would appear to conflict with the increased estimates of $c_{200}$ and $M_{200}$ that we obtained for the non-local model. 

However, a deeper insight of the MCMC results shows that it naturally emerges from the degeneracy of the geometrical and mass effects. In fact, given that the theory cannot be reduced to GR, and that $\Psi_{NL}$ can only account for $\Psi_{GR}/3$, there exist only two possibilities to fit the observations: an increase of estimated cluster mass or a strong contribution of the non-local corrections, $\Phi_{1}(r)$ and $\Psi_{1}(r)$, to the lensing potential. Both options are statistically viable, leading to an agreement with data at a level statistically equivalent to GR. 

This degeneracy of geometrical and matter effects corresponds to a peculiar shape of the $\chi^{2}$ function, which presents two minima. Such behaviour is extremely noticeable in MACSJ0717, whose Markov chain does not remain in a specific region and consequently does not converge to  particular values of the parameters. In particular, for MACSJ0717, the minimum corresponding to low values of the non-local length scales (i.e. strong non-local correction to the gravitational potential) is remarkably deep and the associated estimates of $M_{200}$ are extremely low. It follows that the non-local model seems to be able to fit the observations thanks to its non-local geometrical contributions to the lensing potential, with a very good statistical viability. Moreover, the fact that the corresponding $M_{200}$ estimates are very low means that little matter contribution is necessary to fit data, so that it would open to the possibility to fit the gravitational lensing observations without dark matter. Further studies will be necessary to investigate in detail this possibility. 

On the other hand, the $\chi^{2}$ function of our non-local model shows a low sensibility with respect to the non-local parameters, $r_{\eta}$ and $r_{\xi}$, so that the $\chi^{2}$ minimum corresponding to their lower limits results in an unstable equilibrium point. In contrast, the $\chi^2$ minimum corresponding to increased values of the NFW parameters turns out to be a stable equilibrium point and, as a result, the MCMC sampling is concentrated in this specific area. 

\begin{figure*}
\centering
\includegraphics[width=8.5cm]{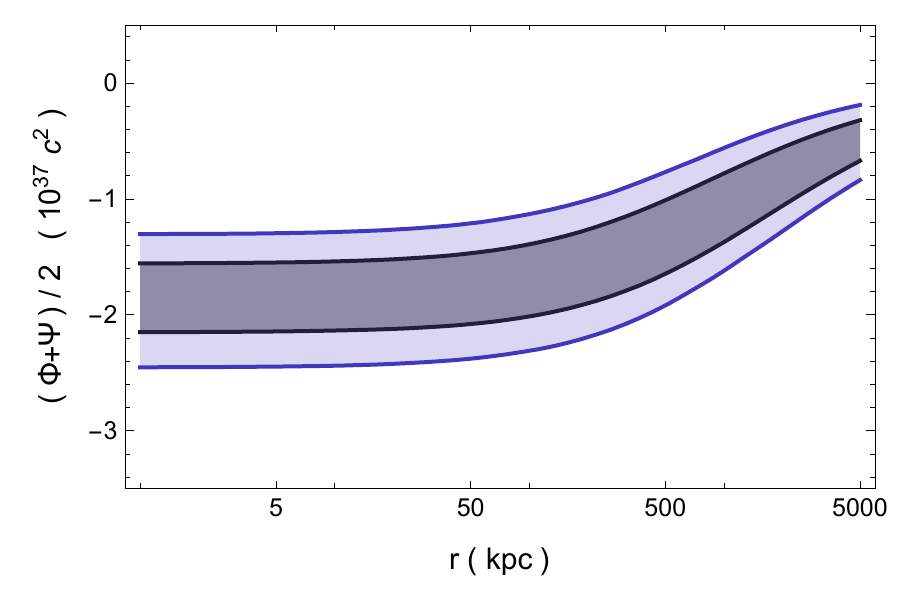}~~
\includegraphics[width=8.5cm]{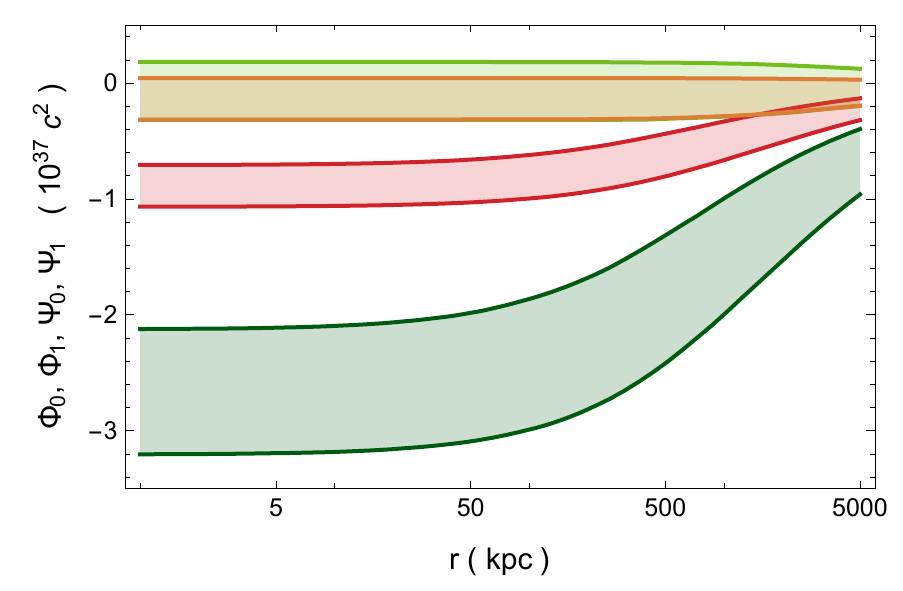}
\caption{Gravitational lensing potential (dimensionless) contributions for A611. \textit{Left panel}: total gravitational potential for GR (black) and nonlocal (blue).  \textit{Right panel}: $\Phi(r)$ (green) and $\Psi(r)$ (red) contributions; zeroth-order terms are in dark colors, first-order terms are in light color}\label{fig:lensingpotentialA611}
\end{figure*}

This behaviour is dominant in all the sample clusters except for MACSJ0717. Such an example is illustrated in Figure \ref{fig:lensingpotentialA611}. The upper panel shows an overlap of the non-local lensing potential with that of GR, demonstrating that the two theories are able to fit the data at similar levels of significance.  In the lower panels, the two contributions to the non-local lensing potential, $\Phi(r)$ and $\Psi(r)$, are presented separately and each of them is decomposed into its zeroth order and first order terms. One can immediately see that $\Phi_{1}(r)$ is completely subdominant with respect to the zeroth order term, which is greater (more negative) than the GR potential due to the increased mass estimate. Thus, the resulting effective potential is dominated by the contribution that corresponds to the ``2/3 of GR'' minimum. It follows the ostensibly contradictory results: the posterior constraints on the NFW parameters are dominated by regions of high posterior probability, while the lower bounds of the non-local length scales are related to the minimum corresponding to the fully non-local regime. Both solutions can fit data at similar levels of statistical significance, as a result of the matter-geometry degeneracy.


We may now compare our results with those of \cite{Dialektopoulos:2018iph}, which are obtained by numerically simulating the S2 star orbits around our Galactic centre. It immediately appears that the two results for $r_{\eta}$ are not consistent: in \cite{Dialektopoulos:2018iph} the authors obtain a constraint of $10^{7}$ km $< r_{\eta} < 10^{10}$ km, while our lower bound is $r_{\eta} > 10^{14}$km. Instead, the two estimates for the parameter $r_{\xi}$ are consistent with each other, because our lower bound $r_{\xi} > 10^{12}$ km is within that found in \cite{Dialektopoulos:2018iph}, $r_{\xi} > 10^{6}$ km. These results are well expected. Our analysis is performed at  galaxy cluster scales, while \cite{Dialektopoulos:2018iph} is at galactic scales. Even though the theoretical parameters should be independent of the test scale, it is difficult to reach a constraint of the AU order if the resolution limit of the galaxy clusters data is of the order of $\sim 10$~kpc. Notice that the results for $r_{\xi}$ are consistent, probably due to the fact that this parameter is associated with the scalar field which is not dynamical, but only plays an auxiliary role to localize the original Lagrangian in  Eq.~(\ref{2.1}). 

Finally, the computed Bayes factors are presented in the fourth column of Table \ref{tab:results}. Using the Jeffreys scale, we can see that there is no evidence in favour of the non-local model with respect to GR. For ten samples we have $0<\ln\mathcal{B}^{\,i}_{j} <1$, for eight samples $\ln\mathcal{B}^{\,i}_{j}$ is slightly negative and for one sample $\ln\mathcal{B}^{\,i}_{j}>1$. However, in both negative and ``substantial'' cases the logarithm of the Bayes factor is consistent with $[0,1]$ within few $\sigma$.

\section{Summary and Conclusions}\label{Section 5}
Since the non-local  theory provides a viable mechanism to explain the late time cosmic acceleration with no need to introduce any form of dark energy, it is of great interest to perform further tests of this model also to address dark matter issues. In fact, its features and the physical consequences of non-locality have to be further analysed both on astrophysical and cosmological scales. 

In this work, we have performed a completely new test of the non-local model, using gravitational lensing data at galaxy cluster scales. The theory provides two effective ways to fit the observations at the same level of statistical significance as GR. On the one hand, when the high-value limit of the non-local parameters is reached, the non-local model reduces to a GR-like theory, being able to fit the data at the cost of increased cluster mass. In this scenario, the non-local theory mimics GR at galaxy cluster scales, affecting only the estimated cluster mass. We cross-checked our results by comparing them with $c_{200}-M_{200}$ relations from the literature, finding that the non-local model is slightly disfavoured with respect to GR. However, further comparisons with the constraints from different probes are necessary. 

On the other hand, when the low-value limit of the non-local length scales is reached, the non-local corrections to the lensing potential become larger and comparable to, or even dominant over, the standard zeroth-order terms. In such a scenario, the non-local scenario may be able to mimic GR, neither affecting the mass estimates nor the statistical viability of the model. Moreover, when the non-local contributions becomes completely dominant, the non-local theory seems to be able to fit the lensing observations with extremely low cluster mass. Consequently, it emerges an intriguing possibility to fit data with no dark matter. In order to investigate such a scenario, further studies should be performed, taking into account the baryonic contributions from the hot gas and stellar components in galaxy clusters.

{\renewcommand{\tabcolsep}{1.mm}
{\renewcommand{\arraystretch}{2.}
\begin{table*}[h]\label{Table 1}
\begin{minipage}{\textwidth}
\centering
\caption{CLASH clusters ordered by redshift. For each cluster, we provide $1\sigma$ constraints on each parameter in the top line, and the minimum values in the $\chi^2$ function in bottom one. Units: cluster radii are in kpc. Lensing-selected CLASH clusters are indicated by stars.}\label{tab:results}
\resizebox*{\textwidth}{!}{
\begin{tabular}{c|cc|cccc|cc}
\hline
\hline
 & \multicolumn{2}{c|}{GR} & \multicolumn{4}{c|}{Nonlocal} & & \\
 name 
 & $c_{200}$ & $M_{200}$   
 & $c_{200}$ & $M_{200}$ & $\log r_{\eta}$  & $\log r_{\xi}$ & $\mathcal{B}^{NL}_{GR}$& $\ln \mathcal{B}^{NL}_{GR}$\\
 & & $(10^{15}\,\mathrm{M}_{\odot})$ & & $(10^{15}\,\mathrm{M}_{\odot})$ & $\mathrm{(kpc)}$ & $\mathrm{(kpc)}$ & & \\
\hline
\hline
A383 & 
$6.32^{+2.31}_{-1.79}$ & $0.81^{+0.32}_{-0.24}$ & $7.64^{+2.73}_{-2.05}$ & $1.30^{+0.57}_{-0.45}$ & $>-2.00$ &  $>-2.90$ & $1.05^{+0.03}_{-0.02}$ & $0.05^{+0.03}_{-0.02}$ \\
A209 & 
$2.46^{+0.63}_{-0.54}$ & $1.94^{+0.53}_{-0.45}$ & $2.81^{+0.91}_{-0.72}$ & $3.83^{+1.99}_{-1.25}$ & $>-1.46$ &  $>-1.52$ & $1.29^{+0.04}_{-0.03}$ & $0.26^{+0.03}_{-0.02}$ \\
A2261 & 
$3.79^{+1.01}_{-0.87}$ & $2.41^{+0.64}_{-0.52}$ & $4.83^{+1.70}_{-1.17}$ & $3.88^{+1.39}_{-1.29}$ & $>-1.61$ & $>-2.82$ & $0.97^{+0.02}_{-0.02}$ & $-0.03^{+0.02}_{-0.02}$ \\
RXJ2129 & 
$6.16^{+2.07}_{-1.65}$ & $0.61^{+0.21}_{-0.16}$ &  $7.38^{+2.49}_{-2.00}$ & $0.99^{+0.39}_{-0.31}$ & $>-2.34$ & $>-3.47$ & $0.99^{+0.02}_{-0.02}$ & $-0.01^{+0.02}_{-0.02}$ \\
A611 & 
$4.07^{+1.44}_{-1.13}$ & $1.64^{+0.57}_{-0.44}$ & 
$5.20^{+2.07}_{-1.42}$ & $2.62^{+1.16}_{-0.93}$ & $>-1.82$ & $>-3.21$ & $1.04^{+0.03}_{-0.02}$ & $0.04^{+0.03}_{-0.02}$ \\
MS2137 & 
$3.09^{+2.28}_{-1.31}$ & $1.26^{+0.76}_{-0.51}$ &  $4.28^{+3.68}_{-1.87}$ & $1.89^{+1.59}_{-0.99}$ & $>-1.74$ & $>-2.23$ & $1.23^{+0.02}_{-0.02}$ & $0.21^{+0.02}_{-0.02}$ \\
RXJ2248  & 
$4.21^{+1.65}_{-1.41}$ & $1.71^{+0.75}_{-0.52}$ & $5.35^{+2.54}_{-1.77}$ & $2.73^{+1.50}_{-1.04}$ & $>-1.56$ & $>-2.68$ & $1.09^{+0.03}_{-0.02}$ & $0.08^{+0.03}_{-0.02}$ \\
MACSJ1115  & 
$2.95^{+0.90}_{-0.70}$ & $1.86^{+0.51}_{-0.44}$ &  $3.61^{+1.26}_{-0.94}$ & $3.23^{+1.29}_{-1.04}$ & $>-1.93$ & $>2.90$ & $0.97^{+0.02}_{-0.02}$ & $-0.03^{+0.02}_{-0.02}$ \\
MACSJ1931  & 
$4.66^{+2.80}_{-1.75}$ & $1.55^{+0.97}_{-0.60}$ & $5.75^{+3.44}_{-2.14}$ & $2.48^{+1.77}_{-1.13}$ & $>-2.28$ & $>-3.25$ & $1.03^{+0.02}_{-0.02}$ & $0.03^{+0.02}_{-0.02}$ \\
MACSJ1720 & 
$4.75^{+1.69}_{-1.27}$ & $1.34^{+0.45}_{-0.36}$ &
$5.86^{+2.26}_{-1.59}$ & $2.18^{+0.94}_{-0.73}$ & $>-1.74$ & $>-3.80$ & $0.98^{+0.02}_{-0.02}$ & $-0.03^{+0.02}_{-0.02}$ \\
MACSJ0416* & 
$3.01^{+0.78}_{-0.65}$ & $1.14^{+0.31}_{-0.26}$ &
$3.69^{+1.12}_{-0.85}$ & $2.02^{+0.77}_{-0.61}$ & $>-2.22$ & $>-4.04$ & $1.05^{+0.02}_{-0.03}$ & $0.05^{+0.02}_{-0.02}$ \\
MACSJ0429 & 
$5.14^{+2.03}_{-1.62}$ & $0.95^{+0.40}_{-0.28}$ & 
$6.30^{+2.74}_{-1.89}$ & $1.57^{+0.80}_{-0.55}$ & $>-1.96$ & $>-3.49$ & $0.96^{+0.02}_{-0.02}$ & $-0.04^{+0.02}_{-0.03}$ \\
MACSJ1206 & 
$4.31^{+1.43}_{-1.19}$ & $1.73^{+0.48}_{-0.40}$ & 
$5.51^{+2.32}_{-1.56}$ & $2.69^{+1.04}_{-0.98}$ & $>-2.40$ & $>-2.60$ & $1.12^{+0.02}_{-0.02}$ & $0.11^{+0.02}_{-0.01}$ \\
MACSJ0329 & 
$7.60^{+2.45}_{-1.95}$ & $0.88^{+0.23}_{-0.19}$ & 
$8.71^{+2.63}_{-2.19}$ & $1.50^{+0.50}_{-0.39}$ & $>-2.08$ & $>-2.17$ & $1.01^{+0.02}_{-0.02}$ & $0.008^{+0.022}_{-0.022}$ \\
RXJ1347 & 
$3.32^{+1.04}_{-0.88}$ & $3.59^{+1.06}_{-0.86}$ &
$4.15^{+1.56}_{-1.08}$ & $6.00^{+2.39}_{-2.08}$ & $>-2.49$ & $>-3.14$ &  $0.98^{+0.02}_{-0.02}$ & $-0.02^{+0.02}_{-0.02}$ \\
MACSJ1149* & 
$2.29^{+0.77}_{-0.60}$ & $2.53^{+0.67}_{-0.59}$ &
$2.94^{+1.14}_{-0.82}$ & $4.31^{+1.64}_{-1.46}$ & $>-1.13$ & $>-3.26$ & $0.92^{+0.02}_{-0.02}$ & $-0.08^{+0.03}_{-0.02}$ \\
MACSJ0717* & 
$1.94^{+0.43}_{-0.38}$ & $2.81^{+0.64}_{-0.58}$ &
$2.43^{+0.58}_{-0.49}$ & $4.87^{+2.07}_{-4.85}$ & $>-4.43$ & $>-4.79$ & $2.74^{+0.13}_{-0.13}$ & $1.01^{+0.05}_{-0.05}$ \\
MACSJ0647* & 
$4.44^{+2.12}_{-1.38}$ & $1.43^{+0.50}_{-0.41}$ & 
$5.63^{+2.63}_{-1.74}$ & $2.28^{+0.99}_{-0.82}$ & $>-1.90$ & $>-2.82$ & $1.01^{+0.03}_{-0.02}$ & $0.006^{+0.028}_{-0.024}$ \\
MACSJ0744 & 
$4.32^{+1.77}_{-1.23}$ & $1.62^{+0.53}_{-0.42}$ &
$5.38^{+2.48}_{-1.62}$ & $2.72^{+1.14}_{-0.96}$ & $>-1.86$ & $>-2.96$ & $0.98^{+0.02}_{-0.02}$ & $-0.02^{+0.02}_{-0.03}$ \\
\hline
\hline
\end{tabular}}
\end{minipage}
\end{table*}}}


\begin{figure*}[h]
\centering
\includegraphics[width=8.25cm]{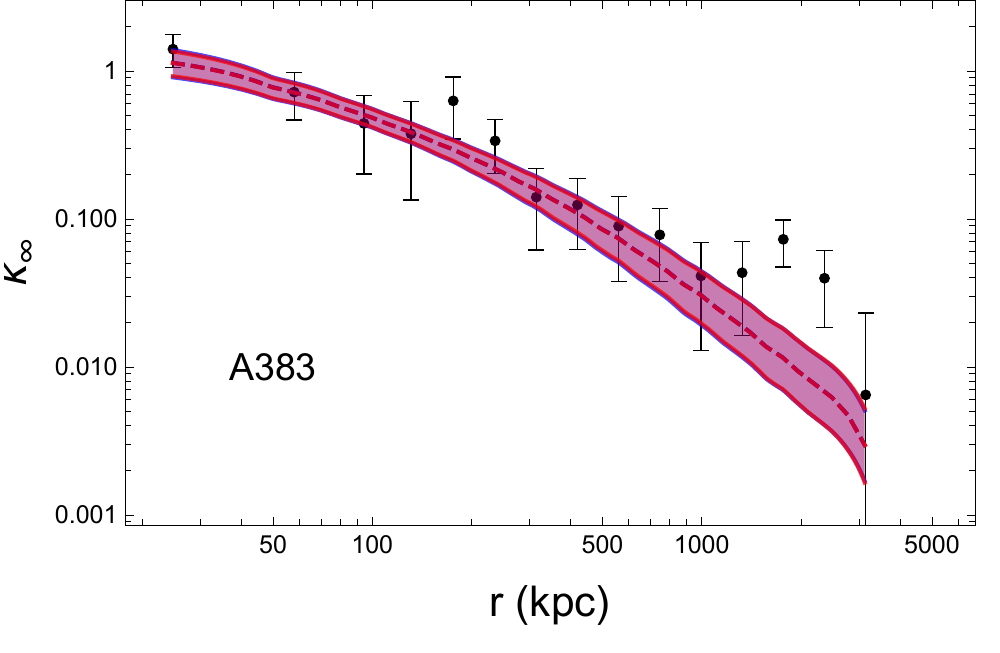}~~~
\includegraphics[width=8.25cm]{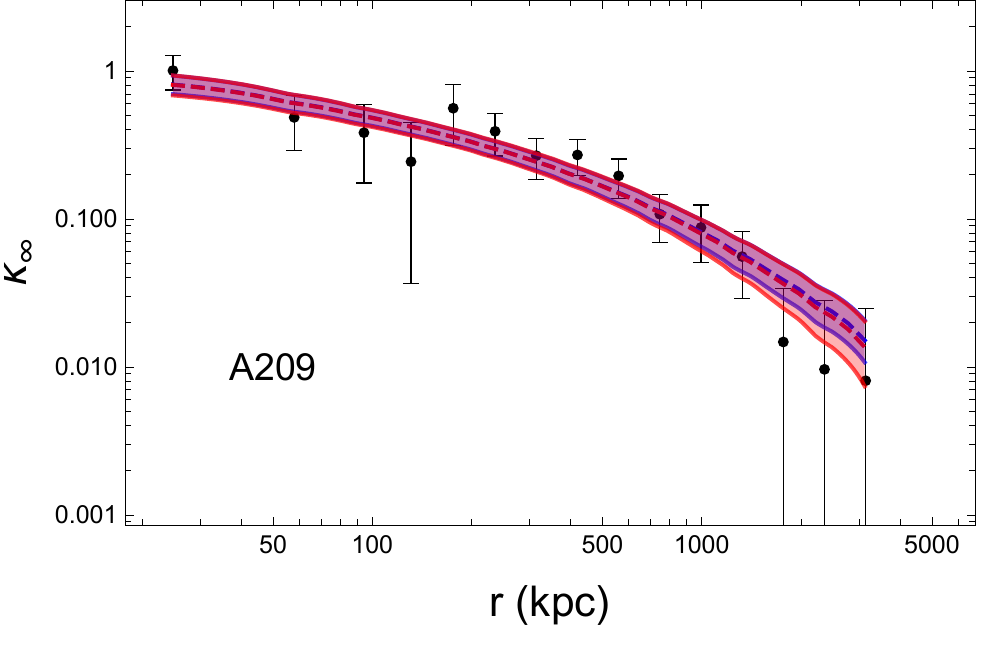}\\
~~~\\
\includegraphics[width=8.25cm]{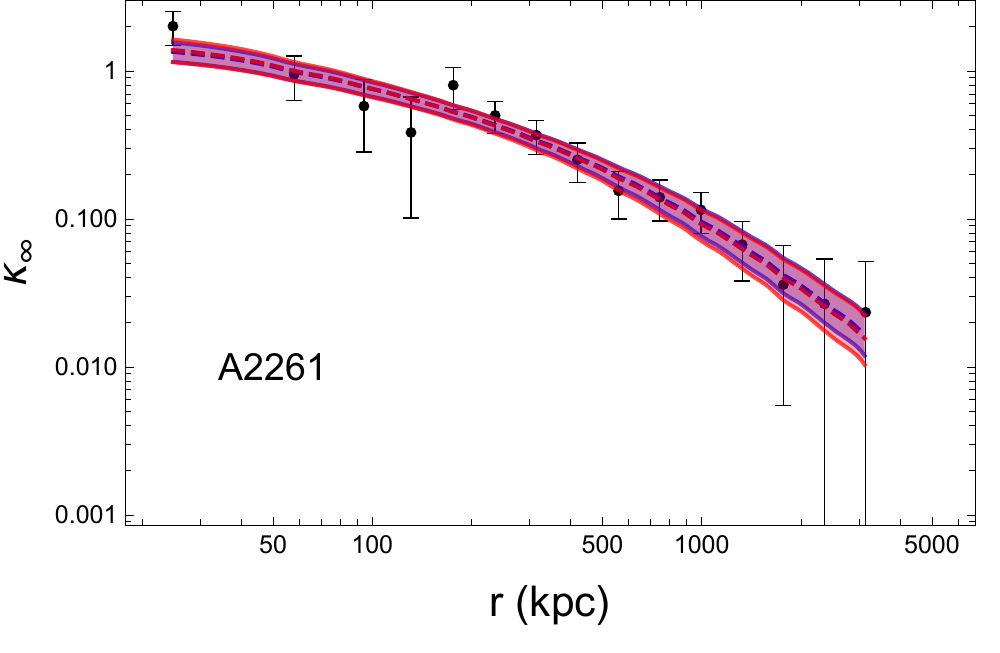}~~~
\includegraphics[width=8.25cm]{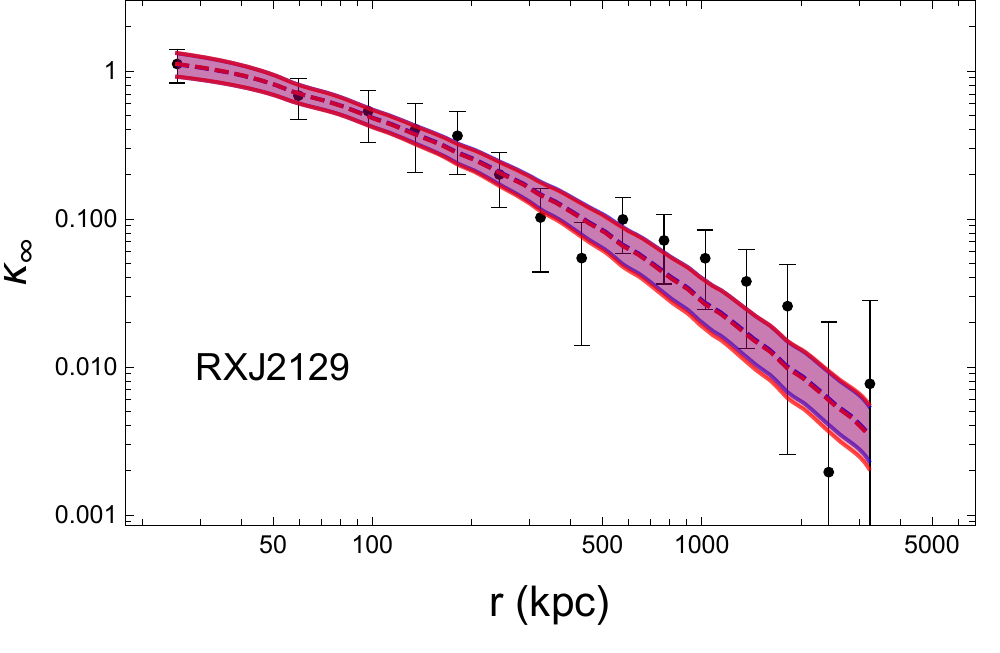}\\
~~~\\
\includegraphics[width=8.25cm]{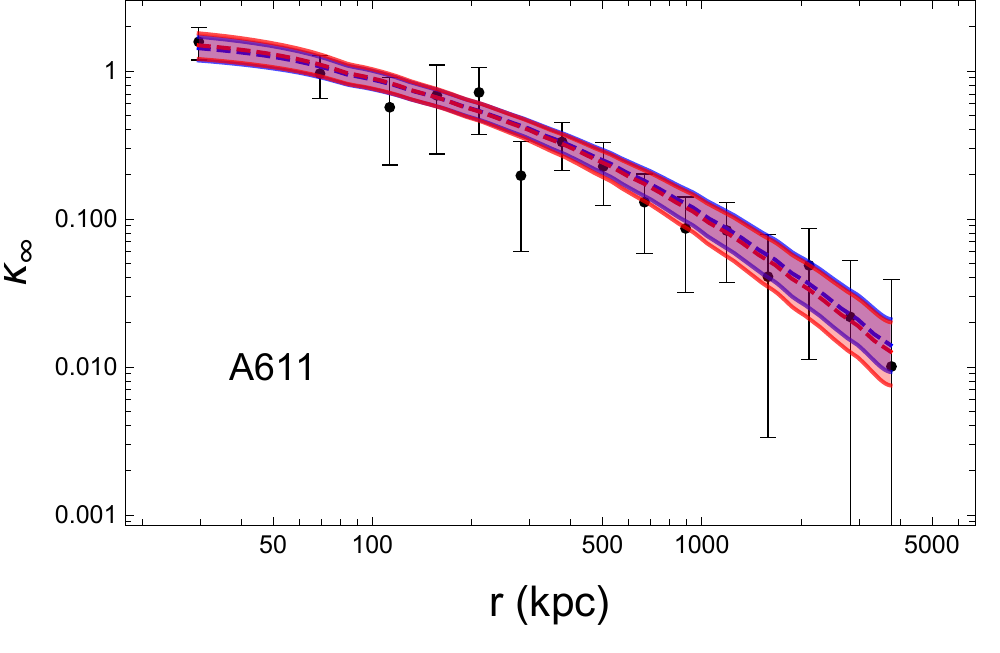}~~~
\includegraphics[width=8.25cm]{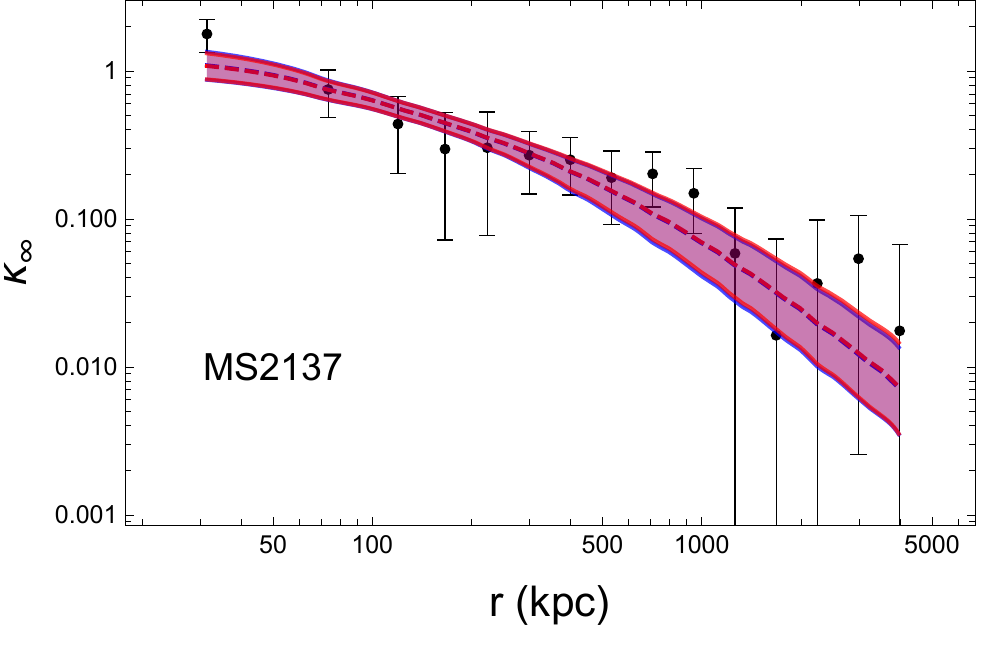}\\
~~~\\
\includegraphics[width=8.25cm]{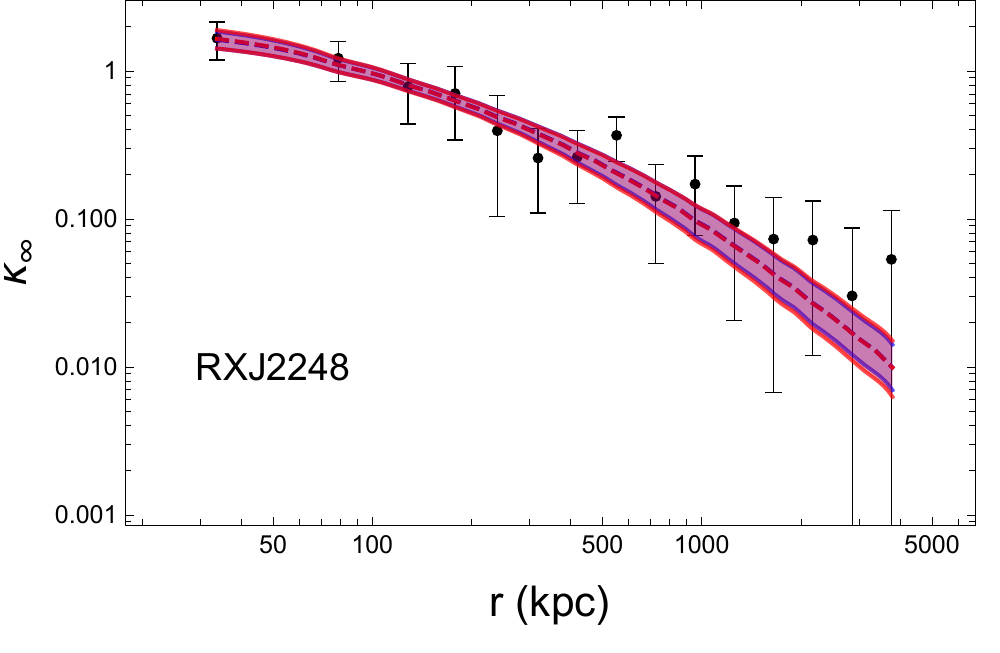}~~~
\includegraphics[width=8.25cm]{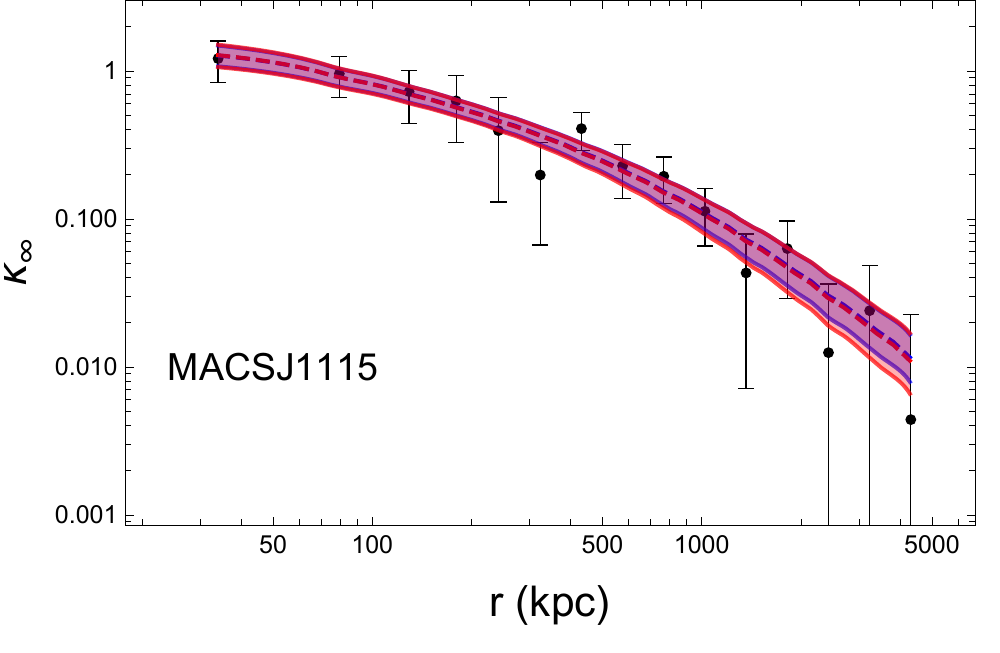}
\caption{Profiles: blue - GR; red - nonlocal.}\label{fig:k_profiles_1}
\end{figure*}

\begin{figure*}[h]
\centering
\includegraphics[width=8.25cm]{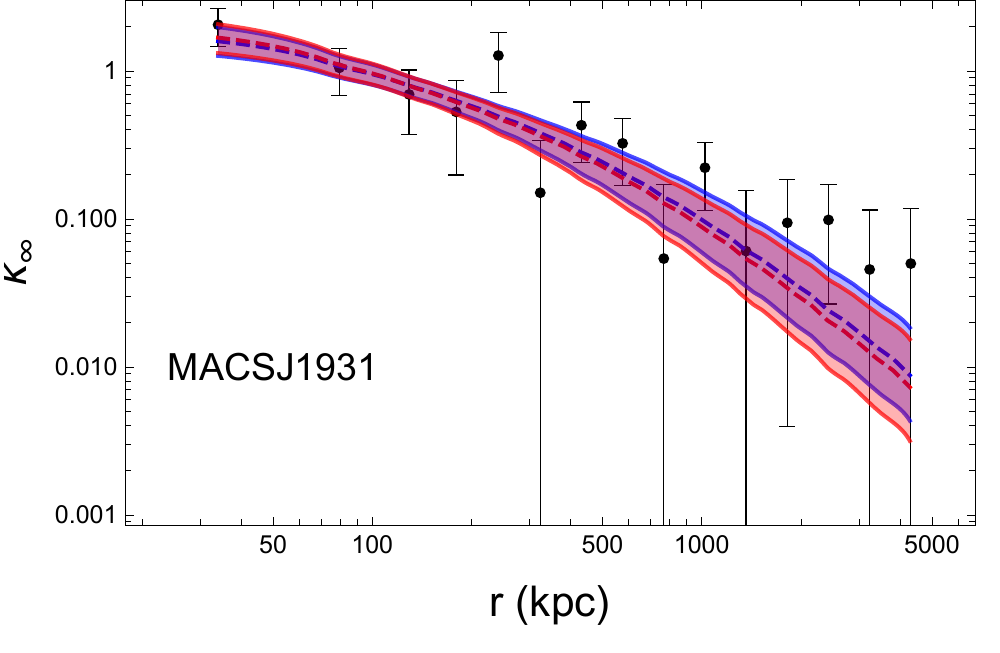}~~~
\includegraphics[width=8.25cm]{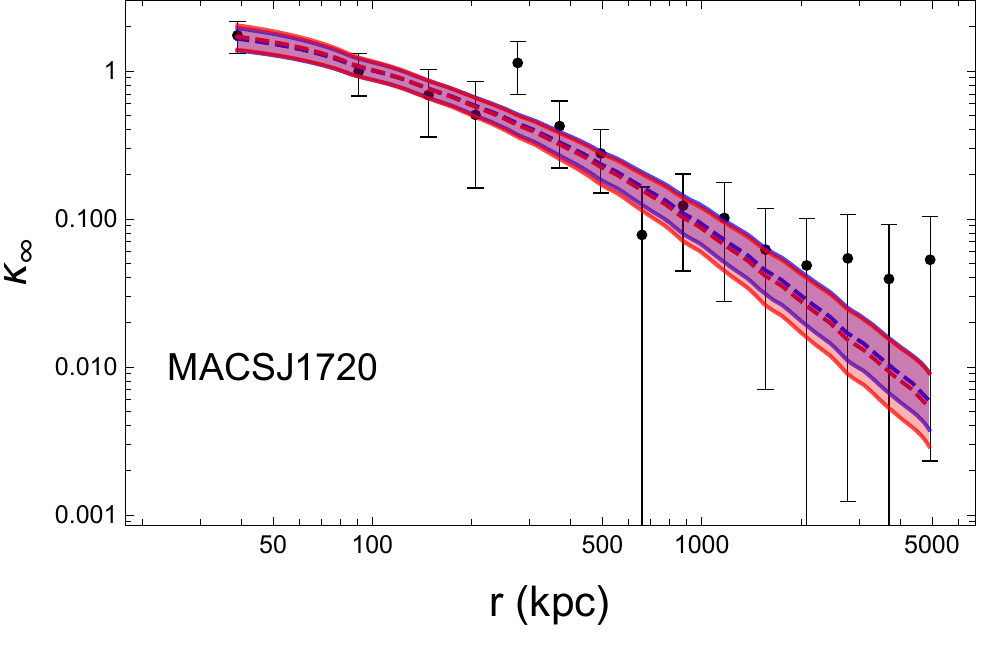}
~~~\\
\includegraphics[width=8.25cm]{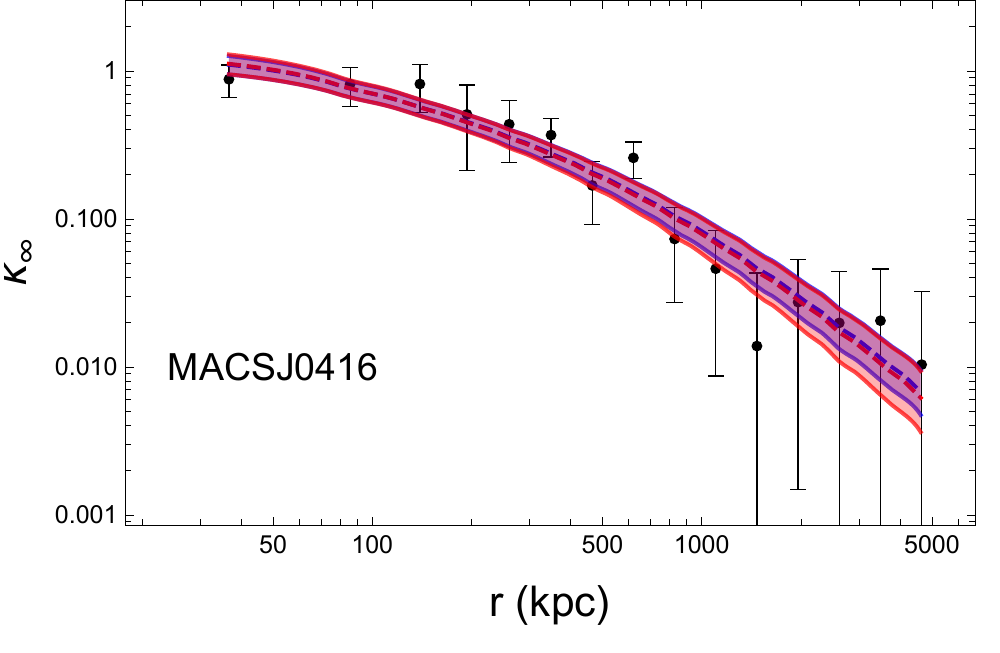}~~~
\includegraphics[width=8.25cm]{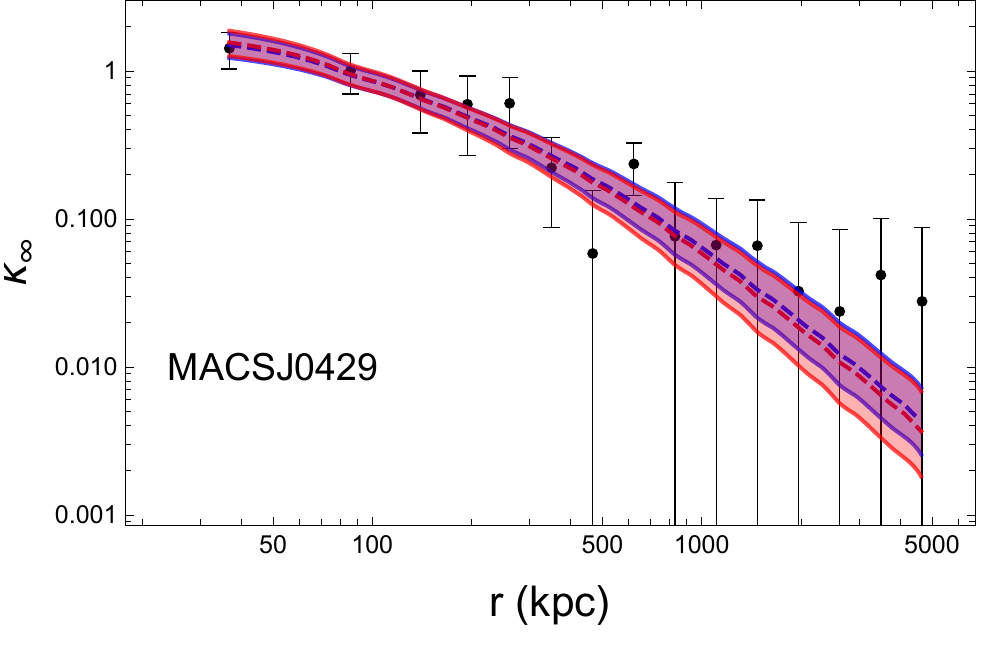}\\

\includegraphics[width=8.25cm]{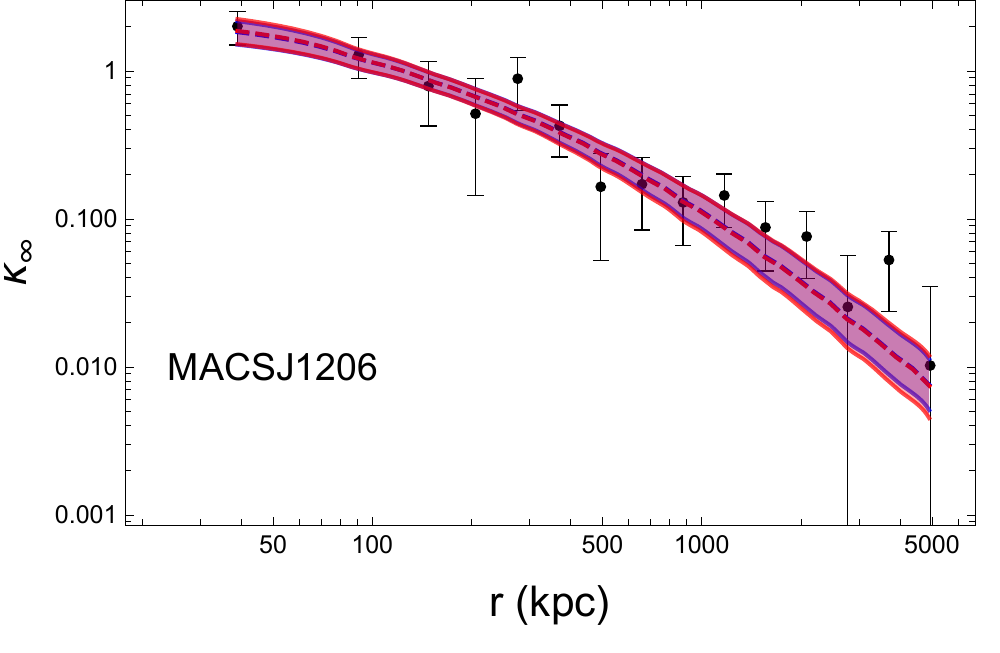}~~~
\includegraphics[width=8.25cm]{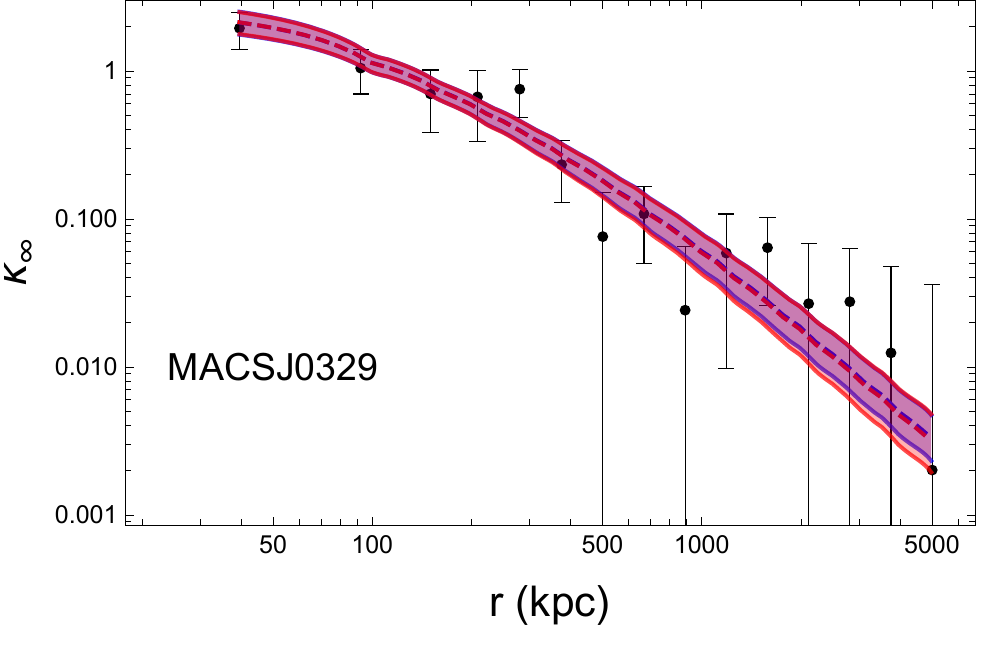}\\
~~~\\
\includegraphics[width=8.25cm]{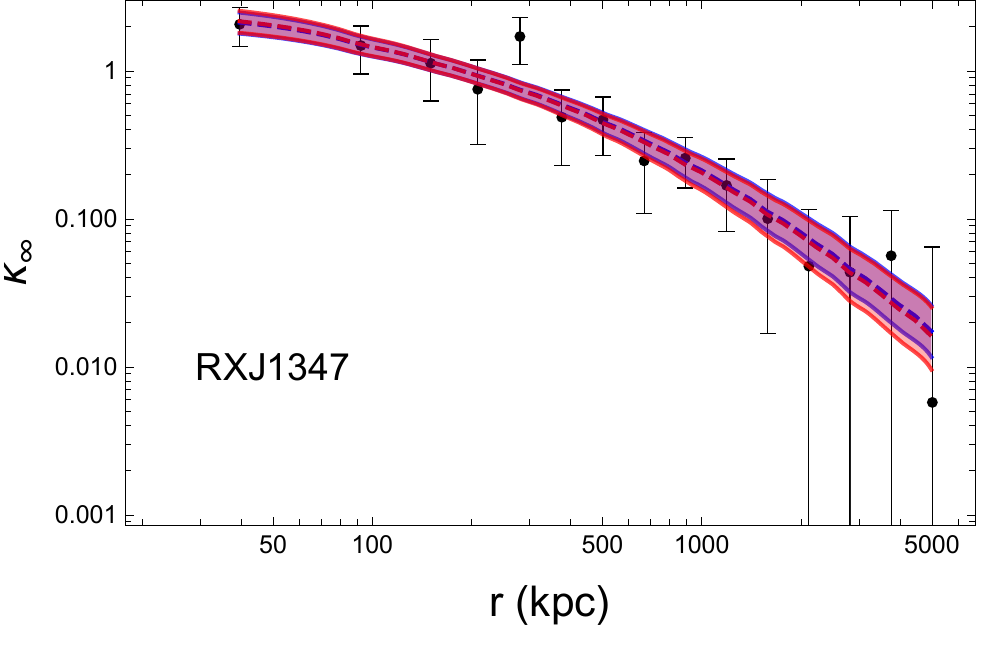}~~~
\includegraphics[width=8.25cm]{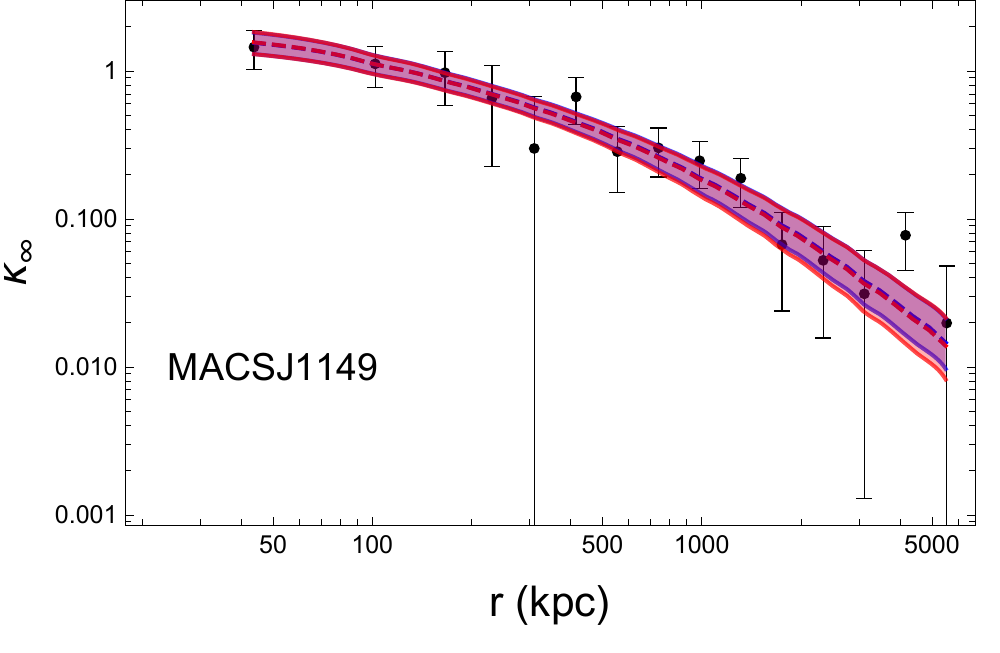}
\caption{Profiles: blue - GR; red - nonlocal.}\label{fig:k_profiles_2}
\end{figure*}

\begin{figure*}[h]
\centering
\includegraphics[width=8.25cm]{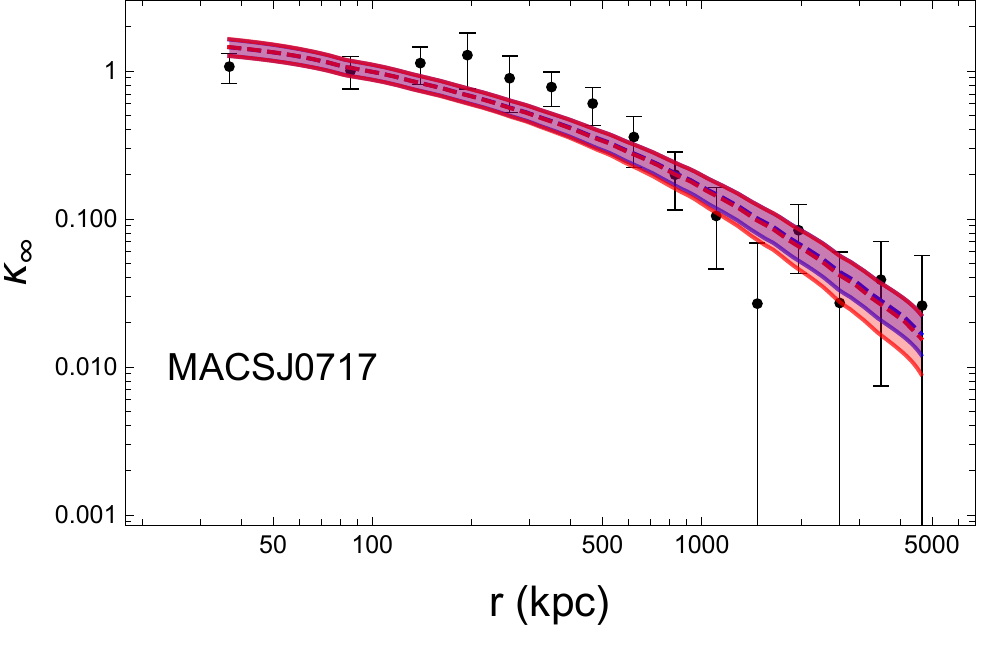}~~~
\includegraphics[width=8.25cm]{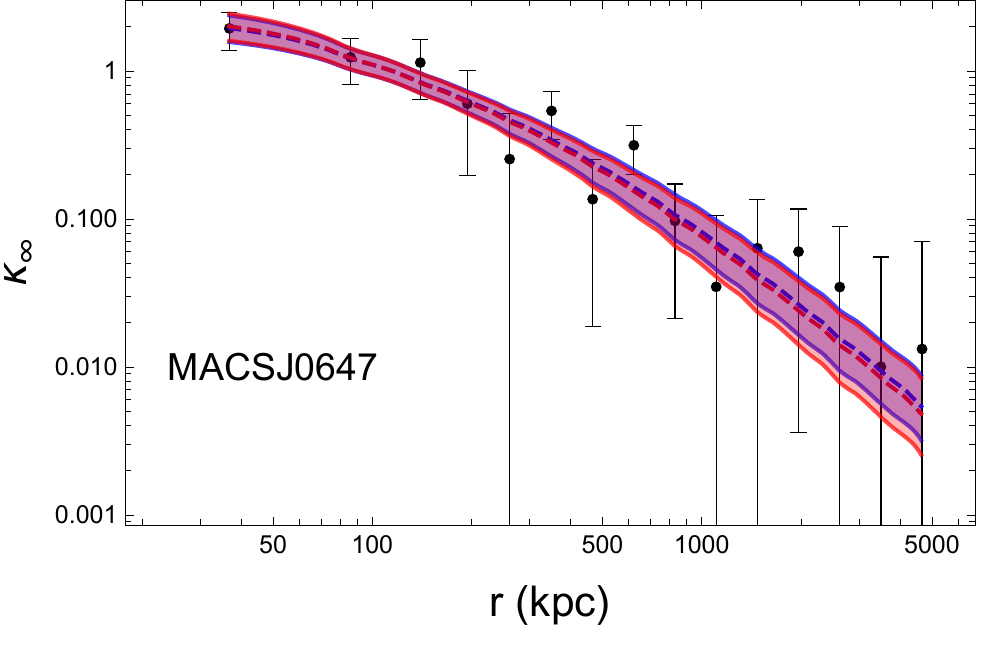}\\
~~~\\
\includegraphics[width=8.25cm]{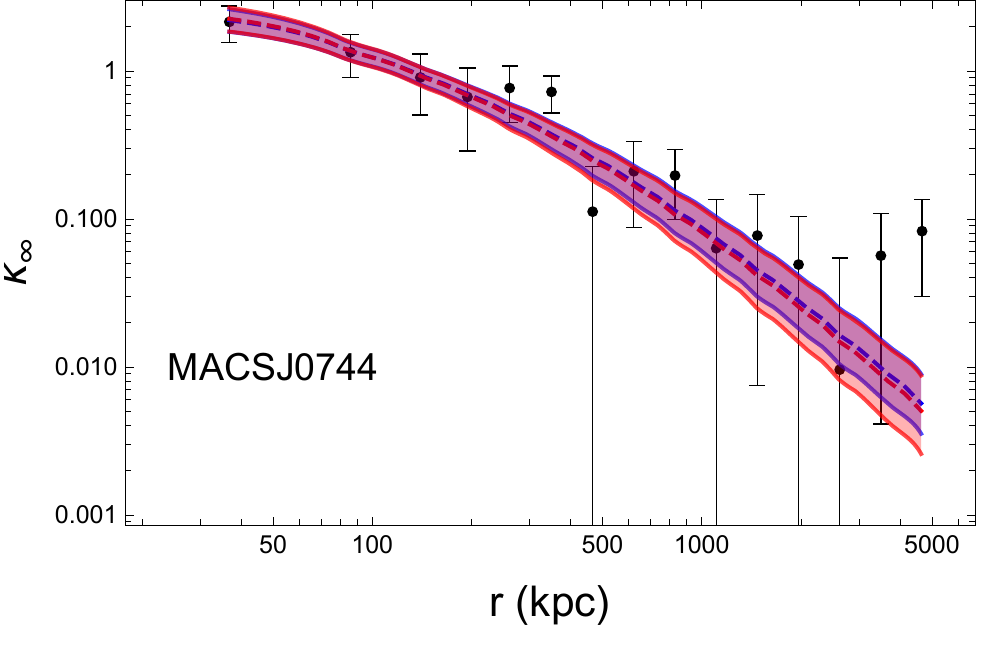}
\caption{Profiles: blue - GR; red - nonlocal.}\label{fig:k_profiles_3}
\end{figure*}

\bibliographystyle{apsrev4-1}
\bibliography{biblio}

\begin{thebibliography}{77}%
\makeatletter
\providecommand \@ifxundefined [1]{%
 \@ifx{#1\undefined}
}%
\providecommand \@ifnum [1]{%
 \ifnum #1\expandafter \@firstoftwo
 \else \expandafter \@secondoftwo
 \fi
}%
\providecommand \@ifx [1]{%
 \ifx #1\expandafter \@firstoftwo
 \else \expandafter \@secondoftwo
 \fi
}%
\providecommand \natexlab [1]{#1}%
\providecommand \enquote  [1]{``#1''}%
\providecommand \bibnamefont  [1]{#1}%
\providecommand \bibfnamefont [1]{#1}%
\providecommand \citenamefont [1]{#1}%
\providecommand \href@noop [0]{\@secondoftwo}%
\providecommand \href [0]{\begingroup \@sanitize@url \@href}%
\providecommand \@href[1]{\@@startlink{#1}\@@href}%
\providecommand \@@href[1]{\endgroup#1\@@endlink}%
\providecommand \@sanitize@url [0]{\catcode `\\12\catcode `\$12\catcode
  `\&12\catcode `\#12\catcode `\^12\catcode `\_12\catcode `\%12\relax}%
\providecommand \@@startlink[1]{}%
\providecommand \@@endlink[0]{}%
\providecommand \url  [0]{\begingroup\@sanitize@url \@url }%
\providecommand \@url [1]{\endgroup\@href {#1}{\urlprefix }}%
\providecommand \urlprefix  [0]{URL }%
\providecommand \Eprint [0]{\href }%
\providecommand \doibase [0]{http://dx.doi.org/}%
\providecommand \selectlanguage [0]{\@gobble}%
\providecommand \bibinfo  [0]{\@secondoftwo}%
\providecommand \bibfield  [0]{\@secondoftwo}%
\providecommand \translation [1]{[#1]}%
\providecommand \BibitemOpen [0]{}%
\providecommand \bibitemStop [0]{}%
\providecommand \bibitemNoStop [0]{.\EOS\space}%
\providecommand \EOS [0]{\spacefactor3000\relax}%
\providecommand \BibitemShut  [1]{\csname bibitem#1\endcsname}%
\let\auto@bib@innerbib\@empty
\bibitem [{\citenamefont {Aghanim}\ \emph {et~al.}(2020)\citenamefont {Aghanim}
  \emph {et~al.}}]{Planck:2018vyg}%
  \BibitemOpen
  \bibfield  {author} {\bibinfo {author} {\bibfnamefont {N.}~\bibnamefont
  {Aghanim}} \emph {et~al.} (\bibinfo {collaboration} {Planck}),\ }\href
  {\doibase 10.1051/0004-6361/201833910} {\bibfield  {journal} {\bibinfo
  {journal} {Astron. Astrophys.}\ }\textbf {\bibinfo {volume} {641}},\ \bibinfo
  {pages} {A6} (\bibinfo {year} {2020})},\ \bibinfo {note} {[Erratum:
  Astron.Astrophys. 652, C4 (2021)]},\ \Eprint
  {http://arxiv.org/abs/1807.06209} {arXiv:1807.06209 [astro-ph.CO]}
  \BibitemShut {NoStop}%
\bibitem [{\citenamefont {Alam}\ \emph {et~al.}(2021)\citenamefont {Alam} \emph
  {et~al.}}]{eBOSS:2020yzd}%
  \BibitemOpen
  \bibfield  {author} {\bibinfo {author} {\bibfnamefont {S.}~\bibnamefont
  {Alam}} \emph {et~al.} (\bibinfo {collaboration} {eBOSS}),\ }\href {\doibase
  10.1103/PhysRevD.103.083533} {\bibfield  {journal} {\bibinfo  {journal}
  {Phys. Rev. D}\ }\textbf {\bibinfo {volume} {103}},\ \bibinfo {pages}
  {083533} (\bibinfo {year} {2021})},\ \Eprint
  {http://arxiv.org/abs/2007.08991} {arXiv:2007.08991 [astro-ph.CO]}
  \BibitemShut {NoStop}%
\bibitem [{\citenamefont {Abbott}\ \emph {et~al.}(2022)\citenamefont {Abbott}
  \emph {et~al.}}]{DES:2021wwk}%
  \BibitemOpen
  \bibfield  {author} {\bibinfo {author} {\bibfnamefont {T.~M.~C.}\
  \bibnamefont {Abbott}} \emph {et~al.} (\bibinfo {collaboration} {DES}),\
  }\href {\doibase 10.1103/PhysRevD.105.023520} {\bibfield  {journal} {\bibinfo
   {journal} {Phys. Rev. D}\ }\textbf {\bibinfo {volume} {105}},\ \bibinfo
  {pages} {023520} (\bibinfo {year} {2022})},\ \Eprint
  {http://arxiv.org/abs/2105.13549} {arXiv:2105.13549 [astro-ph.CO]}
  \BibitemShut {NoStop}%
\bibitem [{\citenamefont {Bull}\ \emph {et~al.}(2016)\citenamefont {Bull} \emph
  {et~al.}}]{Bull:2015stt}%
  \BibitemOpen
  \bibfield  {author} {\bibinfo {author} {\bibfnamefont {P.}~\bibnamefont
  {Bull}} \emph {et~al.},\ }\href {\doibase 10.1016/j.dark.2016.02.001}
  {\bibfield  {journal} {\bibinfo  {journal} {Phys. Dark Univ.}\ }\textbf
  {\bibinfo {volume} {12}},\ \bibinfo {pages} {56} (\bibinfo {year} {2016})},\
  \Eprint {http://arxiv.org/abs/1512.05356} {arXiv:1512.05356 [astro-ph.CO]}
  \BibitemShut {NoStop}%
\bibitem [{\citenamefont {Bertone}\ \emph {et~al.}(2005)\citenamefont
  {Bertone}, \citenamefont {Hooper},\ and\ \citenamefont
  {Silk}}]{Bertone:2004pz}%
  \BibitemOpen
  \bibfield  {author} {\bibinfo {author} {\bibfnamefont {G.}~\bibnamefont
  {Bertone}}, \bibinfo {author} {\bibfnamefont {D.}~\bibnamefont {Hooper}}, \
  and\ \bibinfo {author} {\bibfnamefont {J.}~\bibnamefont {Silk}},\ }\href
  {\doibase 10.1016/j.physrep.2004.08.031} {\bibfield  {journal} {\bibinfo
  {journal} {Phys. Rept.}\ }\textbf {\bibinfo {volume} {405}},\ \bibinfo
  {pages} {279} (\bibinfo {year} {2005})},\ \Eprint
  {http://arxiv.org/abs/hep-ph/0404175} {arXiv:hep-ph/0404175 [hep-ph]}
  \BibitemShut {NoStop}%
\bibitem [{\citenamefont {Clifton}\ \emph {et~al.}(2012)\citenamefont
  {Clifton}, \citenamefont {Ferreira}, \citenamefont {Padilla},\ and\
  \citenamefont {Skordis}}]{Clifton:2011jh}%
  \BibitemOpen
  \bibfield  {author} {\bibinfo {author} {\bibfnamefont {T.}~\bibnamefont
  {Clifton}}, \bibinfo {author} {\bibfnamefont {P.~G.}\ \bibnamefont
  {Ferreira}}, \bibinfo {author} {\bibfnamefont {A.}~\bibnamefont {Padilla}}, \
  and\ \bibinfo {author} {\bibfnamefont {C.}~\bibnamefont {Skordis}},\ }\href
  {\doibase 10.1016/j.physrep.2012.01.001} {\bibfield  {journal} {\bibinfo
  {journal} {Phys. Rept.}\ }\textbf {\bibinfo {volume} {513}},\ \bibinfo
  {pages} {1} (\bibinfo {year} {2012})},\ \Eprint
  {http://arxiv.org/abs/1106.2476} {arXiv:1106.2476 [astro-ph.CO]} \BibitemShut
  {NoStop}%
\bibitem [{\citenamefont {Ishak}(2019)}]{Ishak:2018his}%
  \BibitemOpen
  \bibfield  {author} {\bibinfo {author} {\bibfnamefont {M.}~\bibnamefont
  {Ishak}},\ }\href {\doibase 10.1007/s41114-018-0017-4} {\bibfield  {journal}
  {\bibinfo  {journal} {Living Rev. Rel.}\ }\textbf {\bibinfo {volume} {22}},\
  \bibinfo {pages} {1} (\bibinfo {year} {2019})},\ \Eprint
  {http://arxiv.org/abs/1806.10122} {arXiv:1806.10122 [astro-ph.CO]}
  \BibitemShut {NoStop}%
\bibitem [{\citenamefont {Capozziello}(2002)}]{Capozziello:2002rd}%
  \BibitemOpen
  \bibfield  {author} {\bibinfo {author} {\bibfnamefont {S.}~\bibnamefont
  {Capozziello}},\ }\href {\doibase 10.1142/S0218271802002025} {\bibfield
  {journal} {\bibinfo  {journal} {Int. J. Mod. Phys. D}\ }\textbf {\bibinfo
  {volume} {11}},\ \bibinfo {pages} {483} (\bibinfo {year} {2002})},\ \Eprint
  {http://arxiv.org/abs/gr-qc/0201033} {arXiv:gr-qc/0201033} \BibitemShut
  {NoStop}%
\bibitem [{\citenamefont {Sotiriou}\ and\ \citenamefont
  {Faraoni}(2010)}]{RevModPhys.82.451}%
  \BibitemOpen
  \bibfield  {author} {\bibinfo {author} {\bibfnamefont {T.~P.}\ \bibnamefont
  {Sotiriou}}\ and\ \bibinfo {author} {\bibfnamefont {V.}~\bibnamefont
  {Faraoni}},\ }\href {\doibase 10.1103/RevModPhys.82.451} {\bibfield
  {journal} {\bibinfo  {journal} {Rev. Mod. Phys.}\ }\textbf {\bibinfo {volume}
  {82}},\ \bibinfo {pages} {451} (\bibinfo {year} {2010})}\BibitemShut
  {NoStop}%
\bibitem [{\citenamefont {Capozziello}\ and\ \citenamefont
  {De~Laurentis}(2012)}]{Capozziello:2012ie}%
  \BibitemOpen
  \bibfield  {author} {\bibinfo {author} {\bibfnamefont {S.}~\bibnamefont
  {Capozziello}}\ and\ \bibinfo {author} {\bibfnamefont {M.}~\bibnamefont
  {De~Laurentis}},\ }\href {\doibase 10.1002/andp.201200109} {\bibfield
  {journal} {\bibinfo  {journal} {Annalen Phys.}\ }\textbf {\bibinfo {volume}
  {524}},\ \bibinfo {pages} {545} (\bibinfo {year} {2012})}\BibitemShut
  {NoStop}%
\bibitem [{\citenamefont {Capozziello}\ and\ \citenamefont {{De
  Laurentis}}(2011)}]{CAPOZZIELLO2011167}%
  \BibitemOpen
  \bibfield  {author} {\bibinfo {author} {\bibfnamefont {S.}~\bibnamefont
  {Capozziello}}\ and\ \bibinfo {author} {\bibfnamefont {M.}~\bibnamefont {{De
  Laurentis}}},\ }\href {\doibase
  https://doi.org/10.1016/j.physrep.2011.09.003} {\bibfield  {journal}
  {\bibinfo  {journal} {Physics Reports}\ }\textbf {\bibinfo {volume} {509}},\
  \bibinfo {pages} {167} (\bibinfo {year} {2011})}\BibitemShut {NoStop}%
\bibitem [{\citenamefont {De~Felice}\ and\ \citenamefont
  {Tsujikawa}(2010)}]{ReviewDeFelice}%
  \BibitemOpen
  \bibfield  {author} {\bibinfo {author} {\bibfnamefont {A.}~\bibnamefont
  {De~Felice}}\ and\ \bibinfo {author} {\bibfnamefont {S.}~\bibnamefont
  {Tsujikawa}},\ }\href {\doibase 10.12942/lrr-2010-3} {\bibfield  {journal}
  {\bibinfo  {journal} {Living Reviews in Relativity}\ } (\bibinfo {year}
  {2010}),\ 10.12942/lrr-2010-3}\BibitemShut {NoStop}%
\bibitem [{\citenamefont {Nojiri}\ and\ \citenamefont
  {Odintsov}(2011)}]{Nojiri:2010wj}%
  \BibitemOpen
  \bibfield  {author} {\bibinfo {author} {\bibfnamefont {S.}~\bibnamefont
  {Nojiri}}\ and\ \bibinfo {author} {\bibfnamefont {S.~D.}\ \bibnamefont
  {Odintsov}},\ }\href {\doibase 10.1016/j.physrep.2011.04.001} {\bibfield
  {journal} {\bibinfo  {journal} {Phys. Rept.}\ }\textbf {\bibinfo {volume}
  {505}},\ \bibinfo {pages} {59} (\bibinfo {year} {2011})},\ \Eprint
  {http://arxiv.org/abs/1011.0544} {arXiv:1011.0544 [gr-qc]} \BibitemShut
  {NoStop}%
\bibitem [{\citenamefont {Nojiri}\ \emph
  {et~al.}(2017{\natexlab{a}})\citenamefont {Nojiri}, \citenamefont
  {Odintsov},\ and\ \citenamefont {Oikonomou}}]{Oikonomou}%
  \BibitemOpen
  \bibfield  {author} {\bibinfo {author} {\bibfnamefont {S.}~\bibnamefont
  {Nojiri}}, \bibinfo {author} {\bibfnamefont {S.~D.}\ \bibnamefont
  {Odintsov}}, \ and\ \bibinfo {author} {\bibfnamefont {V.~K.}\ \bibnamefont
  {Oikonomou}},\ }\href {\doibase 10.1016/j.physrep.2017.06.001} {\bibfield
  {journal} {\bibinfo  {journal} {Phys. Rept.}\ }\textbf {\bibinfo {volume}
  {692}},\ \bibinfo {pages} {1} (\bibinfo {year} {2017}{\natexlab{a}})},\
  \Eprint {http://arxiv.org/abs/1705.11098} {arXiv:1705.11098 [gr-qc]}
  \BibitemShut {NoStop}%
\bibitem [{\citenamefont {Copeland}\ \emph {et~al.}(2006)\citenamefont
  {Copeland}, \citenamefont {Sami},\ and\ \citenamefont
  {Tsujikawa}}]{doi:10.1142/S021827180600942X}%
  \BibitemOpen
  \bibfield  {author} {\bibinfo {author} {\bibfnamefont {E.~J.}\ \bibnamefont
  {Copeland}}, \bibinfo {author} {\bibfnamefont {M.}~\bibnamefont {Sami}}, \
  and\ \bibinfo {author} {\bibfnamefont {S.}~\bibnamefont {Tsujikawa}},\ }\href
  {\doibase 10.1142/S021827180600942X} {\bibfield  {journal} {\bibinfo
  {journal} {International Journal of Modern Physics D}\ }\textbf {\bibinfo
  {volume} {15}},\ \bibinfo {pages} {1753} (\bibinfo {year}
  {2006})}\BibitemShut {NoStop}%
\bibitem [{\citenamefont {Peebles}\ and\ \citenamefont
  {Ratra}(2003)}]{RevModPhys.75.559}%
  \BibitemOpen
  \bibfield  {author} {\bibinfo {author} {\bibfnamefont {P.~J.~E.}\
  \bibnamefont {Peebles}}\ and\ \bibinfo {author} {\bibfnamefont
  {B.}~\bibnamefont {Ratra}},\ }\href {\doibase 10.1103/RevModPhys.75.559}
  {\bibfield  {journal} {\bibinfo  {journal} {Rev. Mod. Phys.}\ }\textbf
  {\bibinfo {volume} {75}},\ \bibinfo {pages} {559} (\bibinfo {year}
  {2003})}\BibitemShut {NoStop}%
\bibitem [{\citenamefont {Amendola~\textit{et al.}}(2018)}]{Euclid}%
  \BibitemOpen
  \bibfield  {author} {\bibinfo {author} {\bibfnamefont {L.}~\bibnamefont
  {Amendola~\textit{et al.}}},\ }\href {\doibase 10.1007/s41114-017-0010-3}
  {\bibfield  {journal} {\bibinfo  {journal} {Living Reviews in Relativity}\ }
  (\bibinfo {year} {2018}),\ 10.1007/s41114-017-0010-3}\BibitemShut {NoStop}%
\bibitem [{\citenamefont {Huterer}\ and\ \citenamefont
  {Shafer}(2017)}]{Huterer_2017}%
  \BibitemOpen
  \bibfield  {author} {\bibinfo {author} {\bibfnamefont {D.}~\bibnamefont
  {Huterer}}\ and\ \bibinfo {author} {\bibfnamefont {D.~L.}\ \bibnamefont
  {Shafer}},\ }\href {\doibase 10.1088/1361-6633/aa997e} {\bibfield  {journal}
  {\bibinfo  {journal} {Reports on Progress in Physics}\ }\textbf {\bibinfo
  {volume} {81}},\ \bibinfo {pages} {016901} (\bibinfo {year}
  {2017})}\BibitemShut {NoStop}%
\bibitem [{\citenamefont {Cai}\ \emph {et~al.}(2016)\citenamefont {Cai},
  \citenamefont {Capozziello}, \citenamefont {De~Laurentis},\ and\
  \citenamefont {Saridakis}}]{Cai:2015emx}%
  \BibitemOpen
  \bibfield  {author} {\bibinfo {author} {\bibfnamefont {Y.-F.}\ \bibnamefont
  {Cai}}, \bibinfo {author} {\bibfnamefont {S.}~\bibnamefont {Capozziello}},
  \bibinfo {author} {\bibfnamefont {M.}~\bibnamefont {De~Laurentis}}, \ and\
  \bibinfo {author} {\bibfnamefont {E.~N.}\ \bibnamefont {Saridakis}},\ }\href
  {\doibase 10.1088/0034-4885/79/10/106901} {\bibfield  {journal} {\bibinfo
  {journal} {Rept. Prog. Phys.}\ }\textbf {\bibinfo {volume} {79}},\ \bibinfo
  {pages} {106901} (\bibinfo {year} {2016})},\ \Eprint
  {http://arxiv.org/abs/1511.07586} {arXiv:1511.07586 [gr-qc]} \BibitemShut
  {NoStop}%
\bibitem [{\citenamefont {Unzicker}\ and\ \citenamefont
  {Case}(2005)}]{unzicker2005translation}%
  \BibitemOpen
  \bibfield  {author} {\bibinfo {author} {\bibfnamefont {A.}~\bibnamefont
  {Unzicker}}\ and\ \bibinfo {author} {\bibfnamefont {T.}~\bibnamefont
  {Case}},\ }\href@noop {} {\enquote {\bibinfo {title} {Translation of
  einstein's attempt of a unified field theory with teleparallelism},}\ }
  (\bibinfo {year} {2005}),\ \Eprint {http://arxiv.org/abs/physics/0503046}
  {arXiv:physics/0503046 [physics.hist-ph]} \BibitemShut {NoStop}%
\bibitem [{\citenamefont {Bamba}\ \emph {et~al.}(2013)\citenamefont {Bamba},
  \citenamefont {Capozziello}, \citenamefont {De~Laurentis}, \citenamefont
  {Nojiri},\ and\ \citenamefont {S\'aez-G\'omez}}]{Bamba}%
  \BibitemOpen
  \bibfield  {author} {\bibinfo {author} {\bibfnamefont {K.}~\bibnamefont
  {Bamba}}, \bibinfo {author} {\bibfnamefont {S.}~\bibnamefont {Capozziello}},
  \bibinfo {author} {\bibfnamefont {M.}~\bibnamefont {De~Laurentis}}, \bibinfo
  {author} {\bibfnamefont {S.}~\bibnamefont {Nojiri}}, \ and\ \bibinfo {author}
  {\bibfnamefont {D.}~\bibnamefont {S\'aez-G\'omez}},\ }\href {\doibase
  10.1016/j.physletb.2013.10.022} {\bibfield  {journal} {\bibinfo  {journal}
  {Phys. Lett. B}\ }\textbf {\bibinfo {volume} {727}},\ \bibinfo {pages} {194}
  (\bibinfo {year} {2013})},\ \Eprint {http://arxiv.org/abs/1309.2698}
  {arXiv:1309.2698 [gr-qc]} \BibitemShut {NoStop}%
\bibitem [{\citenamefont {Capozziello}\ and\ \citenamefont
  {Bajardi}(2021)}]{Capozziello:ReviewNonLocal}%
  \BibitemOpen
  \bibfield  {author} {\bibinfo {author} {\bibfnamefont {S.}~\bibnamefont
  {Capozziello}}\ and\ \bibinfo {author} {\bibfnamefont {F.}~\bibnamefont
  {Bajardi}},\ }\href {https://doi.org/10.1142/S0218271822300099} {\bibfield
  {journal} {\bibinfo  {journal} {Int. J. Mod. Phys. D}\ } (\bibinfo {year}
  {2021})}\BibitemShut {NoStop}%
\bibitem [{\citenamefont {Buoninfante}\ \emph {et~al.}(2018)\citenamefont
  {Buoninfante}, \citenamefont {Koshelev}, \citenamefont {Lambiase},\ and\
  \citenamefont {Mazumdar}}]{Buoninfante2018}%
  \BibitemOpen
  \bibfield  {author} {\bibinfo {author} {\bibfnamefont {L.}~\bibnamefont
  {Buoninfante}}, \bibinfo {author} {\bibfnamefont {A.~S.}\ \bibnamefont
  {Koshelev}}, \bibinfo {author} {\bibfnamefont {G.}~\bibnamefont {Lambiase}},
  \ and\ \bibinfo {author} {\bibfnamefont {A.}~\bibnamefont {Mazumdar}},\
  }\href {\doibase 10.1088/1475-7516/2018/09/034} {\bibfield  {journal}
  {\bibinfo  {journal} {Journal of Cosmology and Astroparticle Physics}\
  }\textbf {\bibinfo {volume} {2018}},\ \bibinfo {pages} {034–034} (\bibinfo
  {year} {2018})}\BibitemShut {NoStop}%
\bibitem [{\citenamefont {Modesto}(2015)}]{Modesto:2013ioa}%
  \BibitemOpen
  \bibfield  {author} {\bibinfo {author} {\bibfnamefont {L.}~\bibnamefont
  {Modesto}},\ }in\ \href {\doibase 10.1142/9789814623995_0098} {\emph
  {\bibinfo {booktitle} {{13th Marcel Grossmann Meeting on Recent Developments
  in Theoretical and Experimental General Relativity, Astrophysics, and
  Relativistic Field Theories}}}}\ (\bibinfo {year} {2015})\ pp.\ \bibinfo
  {pages} {1128--1130},\ \Eprint {http://arxiv.org/abs/1302.6348}
  {arXiv:1302.6348 [hep-th]} \BibitemShut {NoStop}%
\bibitem [{\citenamefont {Briscese}\ \emph {et~al.}(2013)\citenamefont
  {Briscese}, \citenamefont {Marcian\`o}, \citenamefont {Modesto},\ and\
  \citenamefont {Saridakis}}]{PhysRevD.87.083507}%
  \BibitemOpen
  \bibfield  {author} {\bibinfo {author} {\bibfnamefont {F.}~\bibnamefont
  {Briscese}}, \bibinfo {author} {\bibfnamefont {A.}~\bibnamefont
  {Marcian\`o}}, \bibinfo {author} {\bibfnamefont {L.}~\bibnamefont {Modesto}},
  \ and\ \bibinfo {author} {\bibfnamefont {E.~N.}\ \bibnamefont {Saridakis}},\
  }\href {\doibase 10.1103/PhysRevD.87.083507} {\bibfield  {journal} {\bibinfo
  {journal} {Phys. Rev. D}\ }\textbf {\bibinfo {volume} {87}},\ \bibinfo
  {pages} {083507} (\bibinfo {year} {2013})}\BibitemShut {NoStop}%
\bibitem [{\citenamefont {Conroy}\ \emph {et~al.}(2015)\citenamefont {Conroy},
  \citenamefont {Koivisto}, \citenamefont {Mazumdar},\ and\ \citenamefont
  {Teimouri}}]{Conroy:2014eja}%
  \BibitemOpen
  \bibfield  {author} {\bibinfo {author} {\bibfnamefont {A.}~\bibnamefont
  {Conroy}}, \bibinfo {author} {\bibfnamefont {T.}~\bibnamefont {Koivisto}},
  \bibinfo {author} {\bibfnamefont {A.}~\bibnamefont {Mazumdar}}, \ and\
  \bibinfo {author} {\bibfnamefont {A.}~\bibnamefont {Teimouri}},\ }\href
  {\doibase 10.1088/0264-9381/32/1/015024} {\bibfield  {journal} {\bibinfo
  {journal} {Class. Quant. Grav.}\ }\textbf {\bibinfo {volume} {32}},\ \bibinfo
  {pages} {015024} (\bibinfo {year} {2015})},\ \Eprint
  {http://arxiv.org/abs/1406.4998} {arXiv:1406.4998 [hep-th]} \BibitemShut
  {NoStop}%
\bibitem [{\citenamefont {Barvinsky}(2015)}]{doi:10.1142/S0217732315400039}%
  \BibitemOpen
  \bibfield  {author} {\bibinfo {author} {\bibfnamefont {A.~O.}\ \bibnamefont
  {Barvinsky}},\ }\href {\doibase 10.1142/S0217732315400039} {\bibfield
  {journal} {\bibinfo  {journal} {Modern Physics Letters A}\ }\textbf {\bibinfo
  {volume} {30}},\ \bibinfo {pages} {1540003} (\bibinfo {year}
  {2015})}\BibitemShut {NoStop}%
\bibitem [{\citenamefont {Deser}\ and\ \citenamefont
  {Woodard}(2007)}]{Deser:2007jk}%
  \BibitemOpen
  \bibfield  {author} {\bibinfo {author} {\bibfnamefont {S.}~\bibnamefont
  {Deser}}\ and\ \bibinfo {author} {\bibfnamefont {R.~P.}\ \bibnamefont
  {Woodard}},\ }\href {\doibase 10.1103/PhysRevLett.99.111301} {\bibfield
  {journal} {\bibinfo  {journal} {Phys. Rev. Lett.}\ }\textbf {\bibinfo
  {volume} {99}},\ \bibinfo {pages} {111301} (\bibinfo {year} {2007})},\
  \Eprint {http://arxiv.org/abs/0706.2151} {arXiv:0706.2151 [astro-ph]}
  \BibitemShut {NoStop}%
\bibitem [{\citenamefont {Capozziello}\ and\ \citenamefont
  {Capriolo}(2021)}]{Capriolo1}%
  \BibitemOpen
  \bibfield  {author} {\bibinfo {author} {\bibfnamefont {S.}~\bibnamefont
  {Capozziello}}\ and\ \bibinfo {author} {\bibfnamefont {M.}~\bibnamefont
  {Capriolo}},\ }\href {\doibase 10.1088/1361-6382/ac1720} {\bibfield
  {journal} {\bibinfo  {journal} {Class. Quant. Grav.}\ }\textbf {\bibinfo
  {volume} {38}},\ \bibinfo {pages} {175008} (\bibinfo {year} {2021})},\
  \Eprint {http://arxiv.org/abs/2107.06972} {arXiv:2107.06972 [gr-qc]}
  \BibitemShut {NoStop}%
\bibitem [{\citenamefont {Capozziello}\ \emph {et~al.}(2020)\citenamefont
  {Capozziello}, \citenamefont {Capriolo},\ and\ \citenamefont
  {Nojiri}}]{Capriolo2}%
  \BibitemOpen
  \bibfield  {author} {\bibinfo {author} {\bibfnamefont {S.}~\bibnamefont
  {Capozziello}}, \bibinfo {author} {\bibfnamefont {M.}~\bibnamefont
  {Capriolo}}, \ and\ \bibinfo {author} {\bibfnamefont {S.}~\bibnamefont
  {Nojiri}},\ }\href {\doibase 10.1016/j.physletb.2020.135821} {\bibfield
  {journal} {\bibinfo  {journal} {Phys. Lett. B}\ }\textbf {\bibinfo {volume}
  {810}},\ \bibinfo {pages} {135821} (\bibinfo {year} {2020})},\ \Eprint
  {http://arxiv.org/abs/2009.12777} {arXiv:2009.12777 [gr-qc]} \BibitemShut
  {NoStop}%
\bibitem [{\citenamefont {Maggiore}\ and\ \citenamefont
  {Mancarella}(2014)}]{RRmodel}%
  \BibitemOpen
  \bibfield  {author} {\bibinfo {author} {\bibfnamefont {M.}~\bibnamefont
  {Maggiore}}\ and\ \bibinfo {author} {\bibfnamefont {M.}~\bibnamefont
  {Mancarella}},\ }\href {\doibase 10.1103/physrevd.90.023005} {\bibfield
  {journal} {\bibinfo  {journal} {Physical Review D}\ }\textbf {\bibinfo
  {volume} {90}} (\bibinfo {year} {2014}),\
  10.1103/physrevd.90.023005}\BibitemShut {NoStop}%
\bibitem [{\citenamefont {Maggiore}(2014)}]{RTmodel}%
  \BibitemOpen
  \bibfield  {author} {\bibinfo {author} {\bibfnamefont {M.}~\bibnamefont
  {Maggiore}},\ }\href {\doibase 10.1103/physrevd.89.043008} {\bibfield
  {journal} {\bibinfo  {journal} {Physical Review D}\ }\textbf {\bibinfo
  {volume} {89}} (\bibinfo {year} {2014}),\
  10.1103/physrevd.89.043008}\BibitemShut {NoStop}%
\bibitem [{\citenamefont {Dirian}\ \emph {et~al.}(2016)\citenamefont {Dirian},
  \citenamefont {Foffa}, \citenamefont {Kunz}, \citenamefont {Maggiore},\ and\
  \citenamefont {Pettorino}}]{Dirian:2016puz}%
  \BibitemOpen
  \bibfield  {author} {\bibinfo {author} {\bibfnamefont {Y.}~\bibnamefont
  {Dirian}}, \bibinfo {author} {\bibfnamefont {S.}~\bibnamefont {Foffa}},
  \bibinfo {author} {\bibfnamefont {M.}~\bibnamefont {Kunz}}, \bibinfo {author}
  {\bibfnamefont {M.}~\bibnamefont {Maggiore}}, \ and\ \bibinfo {author}
  {\bibfnamefont {V.}~\bibnamefont {Pettorino}},\ }\href {\doibase
  10.1088/1475-7516/2016/05/068} {\bibfield  {journal} {\bibinfo  {journal}
  {JCAP}\ }\textbf {\bibinfo {volume} {05}},\ \bibinfo {pages} {068} (\bibinfo
  {year} {2016})},\ \Eprint {http://arxiv.org/abs/1602.03558} {arXiv:1602.03558
  [astro-ph.CO]} \BibitemShut {NoStop}%
\bibitem [{\citenamefont {Belgacem}\ \emph {et~al.}(2018)\citenamefont
  {Belgacem}, \citenamefont {Dirian}, \citenamefont {Foffa},\ and\
  \citenamefont {Maggiore}}]{RRtest2018}%
  \BibitemOpen
  \bibfield  {author} {\bibinfo {author} {\bibfnamefont {E.}~\bibnamefont
  {Belgacem}}, \bibinfo {author} {\bibfnamefont {Y.}~\bibnamefont {Dirian}},
  \bibinfo {author} {\bibfnamefont {S.}~\bibnamefont {Foffa}}, \ and\ \bibinfo
  {author} {\bibfnamefont {M.}~\bibnamefont {Maggiore}},\ }\href {\doibase
  10.1088/1475-7516/2018/03/002} {\bibfield  {journal} {\bibinfo  {journal}
  {Journal of Cosmology and Astroparticle Physics}\ }\textbf {\bibinfo {volume}
  {2018}},\ \bibinfo {pages} {002–002} (\bibinfo {year} {2018})}\BibitemShut
  {NoStop}%
\bibitem [{\citenamefont {Nesseris}\ and\ \citenamefont
  {Tsujikawa}(2014)}]{Nesseris:2014mea}%
  \BibitemOpen
  \bibfield  {author} {\bibinfo {author} {\bibfnamefont {S.}~\bibnamefont
  {Nesseris}}\ and\ \bibinfo {author} {\bibfnamefont {S.}~\bibnamefont
  {Tsujikawa}},\ }\href {\doibase 10.1103/PhysRevD.90.024070} {\bibfield
  {journal} {\bibinfo  {journal} {Phys. Rev. D}\ }\textbf {\bibinfo {volume}
  {90}},\ \bibinfo {pages} {024070} (\bibinfo {year} {2014})},\ \Eprint
  {http://arxiv.org/abs/1402.4613} {arXiv:1402.4613 [astro-ph.CO]} \BibitemShut
  {NoStop}%
\bibitem [{\citenamefont {Acunzo}\ \emph {et~al.}(2022)\citenamefont {Acunzo},
  \citenamefont {Bajardi},\ and\ \citenamefont {Capozziello}}]{Acunzo}%
  \BibitemOpen
  \bibfield  {author} {\bibinfo {author} {\bibfnamefont {A.}~\bibnamefont
  {Acunzo}}, \bibinfo {author} {\bibfnamefont {F.}~\bibnamefont {Bajardi}}, \
  and\ \bibinfo {author} {\bibfnamefont {S.}~\bibnamefont {Capozziello}},\
  }\href {\doibase 10.1016/j.physletb.2022.136907} {\bibfield  {journal}
  {\bibinfo  {journal} {Phys. Lett. B}\ }\textbf {\bibinfo {volume} {826}},\
  \bibinfo {pages} {136907} (\bibinfo {year} {2022})},\ \Eprint
  {http://arxiv.org/abs/2111.07285} {arXiv:2111.07285 [gr-qc]} \BibitemShut
  {NoStop}%
\bibitem [{\citenamefont {Nojiri}\ and\ \citenamefont
  {Odintsov}(2008)}]{Nojiri:2007uq}%
  \BibitemOpen
  \bibfield  {author} {\bibinfo {author} {\bibfnamefont {S.}~\bibnamefont
  {Nojiri}}\ and\ \bibinfo {author} {\bibfnamefont {S.~D.}\ \bibnamefont
  {Odintsov}},\ }\href {\doibase 10.1016/j.physletb.2007.12.001} {\bibfield
  {journal} {\bibinfo  {journal} {Phys. Lett. B}\ }\textbf {\bibinfo {volume}
  {659}},\ \bibinfo {pages} {821} (\bibinfo {year} {2008})},\ \Eprint
  {http://arxiv.org/abs/0708.0924} {arXiv:0708.0924 [hep-th]} \BibitemShut
  {NoStop}%
\bibitem [{\citenamefont {Salzano}\ \emph {et~al.}(2017)\citenamefont
  {Salzano}, \citenamefont {Mota}, \citenamefont {Capozziello},\ and\
  \citenamefont {Donahue}}]{Salzano:2017qac}%
  \BibitemOpen
  \bibfield  {author} {\bibinfo {author} {\bibfnamefont {V.}~\bibnamefont
  {Salzano}}, \bibinfo {author} {\bibfnamefont {D.~F.}\ \bibnamefont {Mota}},
  \bibinfo {author} {\bibfnamefont {S.}~\bibnamefont {Capozziello}}, \ and\
  \bibinfo {author} {\bibfnamefont {M.}~\bibnamefont {Donahue}},\ }\href
  {\doibase 10.1103/PhysRevD.95.044038} {\bibfield  {journal} {\bibinfo
  {journal} {Phys. Rev. D}\ }\textbf {\bibinfo {volume} {95}},\ \bibinfo
  {pages} {044038} (\bibinfo {year} {2017})},\ \Eprint
  {http://arxiv.org/abs/1701.03517} {arXiv:1701.03517 [astro-ph.CO]}
  \BibitemShut {NoStop}%
\bibitem [{\citenamefont {Salzano}\ \emph {et~al.}(2016)\citenamefont
  {Salzano}, \citenamefont {Mota}, \citenamefont {Dabrowski},\ and\
  \citenamefont {Capozziello}}]{Salzano:2016udu}%
  \BibitemOpen
  \bibfield  {author} {\bibinfo {author} {\bibfnamefont {V.}~\bibnamefont
  {Salzano}}, \bibinfo {author} {\bibfnamefont {D.~F.}\ \bibnamefont {Mota}},
  \bibinfo {author} {\bibfnamefont {M.~P.}\ \bibnamefont {Dabrowski}}, \ and\
  \bibinfo {author} {\bibfnamefont {S.}~\bibnamefont {Capozziello}},\ }\href
  {\doibase 10.1088/1475-7516/2016/10/033} {\bibfield  {journal} {\bibinfo
  {journal} {JCAP}\ }\textbf {\bibinfo {volume} {10}},\ \bibinfo {pages} {033}
  (\bibinfo {year} {2016})},\ \Eprint {http://arxiv.org/abs/1607.02606}
  {arXiv:1607.02606 [astro-ph.CO]} \BibitemShut {NoStop}%
\bibitem [{\citenamefont {{Laudato}}\ \emph {et~al.}(2022)\citenamefont
  {{Laudato}}, \citenamefont {{Salzano}},\ and\ \citenamefont
  {{Umetsu}}}]{Laudato:2021mnm}%
  \BibitemOpen
  \bibfield  {author} {\bibinfo {author} {\bibfnamefont {E.}~\bibnamefont
  {{Laudato}}}, \bibinfo {author} {\bibfnamefont {V.}~\bibnamefont
  {{Salzano}}}, \ and\ \bibinfo {author} {\bibfnamefont {K.}~\bibnamefont
  {{Umetsu}}},\ }\href {\doibase 10.1093/mnras/stac180} {\bibfield  {journal}
  {\bibinfo  {journal} {Monthly Notices of the Royal Astronomical Society}\
  }\textbf {\bibinfo {volume} {511}},\ \bibinfo {pages} {1878} (\bibinfo {year}
  {2022})},\ \Eprint {http://arxiv.org/abs/2110.11019} {arXiv:2110.11019
  [astro-ph.CO]} \BibitemShut {NoStop}%
\bibitem [{\citenamefont {Postman~\textit{et al.}}(2012)}]{CLASH2012}%
  \BibitemOpen
  \bibfield  {author} {\bibinfo {author} {\bibfnamefont {M.}~\bibnamefont
  {Postman~\textit{et al.}}},\ }\href {\doibase 10.1088/0067-0049/199/2/25}
  {\bibfield  {journal} {\bibinfo  {journal} {The Astrophysical Journal
  Supplement Series}\ }\textbf {\bibinfo {volume} {199}},\ \bibinfo {pages}
  {25} (\bibinfo {year} {2012})}\BibitemShut {NoStop}%
\bibitem [{\citenamefont {Dialektopoulos}\ \emph {et~al.}(2019)\citenamefont
  {Dialektopoulos}, \citenamefont {Borka}, \citenamefont {Capozziello},
  \citenamefont {Borka~Jovanovi\'c},\ and\ \citenamefont
  {Jovanovi\'c}}]{Dialektopoulos:2018iph}%
  \BibitemOpen
  \bibfield  {author} {\bibinfo {author} {\bibfnamefont {K.~F.}\ \bibnamefont
  {Dialektopoulos}}, \bibinfo {author} {\bibfnamefont {D.}~\bibnamefont
  {Borka}}, \bibinfo {author} {\bibfnamefont {S.}~\bibnamefont {Capozziello}},
  \bibinfo {author} {\bibfnamefont {V.}~\bibnamefont {Borka~Jovanovi\'c}}, \
  and\ \bibinfo {author} {\bibfnamefont {P.}~\bibnamefont {Jovanovi\'c}},\
  }\href {\doibase 10.1103/PhysRevD.99.044053} {\bibfield  {journal} {\bibinfo
  {journal} {Phys. Rev. D}\ }\textbf {\bibinfo {volume} {99}},\ \bibinfo
  {pages} {044053} (\bibinfo {year} {2019})},\ \Eprint
  {http://arxiv.org/abs/1812.09289} {arXiv:1812.09289 [astro-ph.GA]}
  \BibitemShut {NoStop}%
\bibitem [{\citenamefont {Borka}\ \emph {et~al.}(2021)\citenamefont {Borka},
  \citenamefont {Jovanovi\'c}, \citenamefont {Capozziello}, \citenamefont
  {Zakharov},\ and\ \citenamefont {Jovanovi\'c}}]{Borka:2021omc}%
  \BibitemOpen
  \bibfield  {author} {\bibinfo {author} {\bibfnamefont {D.}~\bibnamefont
  {Borka}}, \bibinfo {author} {\bibfnamefont {V.~B.}\ \bibnamefont
  {Jovanovi\'c}}, \bibinfo {author} {\bibfnamefont {S.}~\bibnamefont
  {Capozziello}}, \bibinfo {author} {\bibfnamefont {A.~F.}\ \bibnamefont
  {Zakharov}}, \ and\ \bibinfo {author} {\bibfnamefont {P.}~\bibnamefont
  {Jovanovi\'c}},\ }\href {\doibase 10.3390/universe7110407} {\bibfield
  {journal} {\bibinfo  {journal} {Universe}\ }\textbf {\bibinfo {volume} {7}},\
  \bibinfo {pages} {407} (\bibinfo {year} {2021})},\ \Eprint
  {http://arxiv.org/abs/2111.00578} {arXiv:2111.00578 [gr-qc]} \BibitemShut
  {NoStop}%
\bibitem [{\citenamefont {Bahamonde}\ \emph {et~al.}(2017)\citenamefont
  {Bahamonde}, \citenamefont {Capozziello},\ and\ \citenamefont
  {Dialektopoulos}}]{Bahamonde:2017sdo}%
  \BibitemOpen
  \bibfield  {author} {\bibinfo {author} {\bibfnamefont {S.}~\bibnamefont
  {Bahamonde}}, \bibinfo {author} {\bibfnamefont {S.}~\bibnamefont
  {Capozziello}}, \ and\ \bibinfo {author} {\bibfnamefont {K.~F.}\ \bibnamefont
  {Dialektopoulos}},\ }\href {\doibase 10.1140/epjc/s10052-017-5283-x}
  {\bibfield  {journal} {\bibinfo  {journal} {Eur. Phys. J. C}\ }\textbf
  {\bibinfo {volume} {77}},\ \bibinfo {pages} {722} (\bibinfo {year} {2017})},\
  \Eprint {http://arxiv.org/abs/1708.06310} {arXiv:1708.06310 [gr-qc]}
  \BibitemShut {NoStop}%
\bibitem [{\citenamefont {Deffayet}\ and\ \citenamefont
  {Woodard}(2009)}]{WoodardDistFun}%
  \BibitemOpen
  \bibfield  {author} {\bibinfo {author} {\bibfnamefont {C.}~\bibnamefont
  {Deffayet}}\ and\ \bibinfo {author} {\bibfnamefont {R.}~\bibnamefont
  {Woodard}},\ }\href {\doibase 10.1088/1475-7516/2009/08/023} {\bibfield
  {journal} {\bibinfo  {journal} {Journal of Cosmology and Astroparticle
  Physics}\ }\textbf {\bibinfo {volume} {2009}},\ \bibinfo {pages} {023–023}
  (\bibinfo {year} {2009})}\BibitemShut {NoStop}%
\bibitem [{\citenamefont {Nersisyan}\ \emph {et~al.}(2017)\citenamefont
  {Nersisyan}, \citenamefont {Cid},\ and\ \citenamefont
  {Amendola}}]{NersisyanConstraints2017}%
  \BibitemOpen
  \bibfield  {author} {\bibinfo {author} {\bibfnamefont {H.}~\bibnamefont
  {Nersisyan}}, \bibinfo {author} {\bibfnamefont {A.~F.}\ \bibnamefont {Cid}},
  \ and\ \bibinfo {author} {\bibfnamefont {L.}~\bibnamefont {Amendola}},\
  }\href {\doibase 10.1088/1475-7516/2017/04/046} {\bibfield  {journal}
  {\bibinfo  {journal} {Journal of Cosmology and Astroparticle Physics}\
  }\textbf {\bibinfo {volume} {2017}},\ \bibinfo {pages} {046–046} (\bibinfo
  {year} {2017})}\BibitemShut {NoStop}%
\bibitem [{\citenamefont {Park}(2018)}]{Park2018}%
  \BibitemOpen
  \bibfield  {author} {\bibinfo {author} {\bibfnamefont {S.}~\bibnamefont
  {Park}},\ }\href {\doibase 10.1103/physrevd.97.044006} {\bibfield  {journal}
  {\bibinfo  {journal} {Physical Review D}\ }\textbf {\bibinfo {volume} {97}}
  (\bibinfo {year} {2018}),\ 10.1103/physrevd.97.044006}\BibitemShut {NoStop}%
\bibitem [{\citenamefont {Amendola}\ \emph {et~al.}(2019)\citenamefont
  {Amendola}, \citenamefont {Dirian}, \citenamefont {Nersisyan},\ and\
  \citenamefont {Park}}]{AmendolaIdrianConstraints}%
  \BibitemOpen
  \bibfield  {author} {\bibinfo {author} {\bibfnamefont {L.}~\bibnamefont
  {Amendola}}, \bibinfo {author} {\bibfnamefont {Y.}~\bibnamefont {Dirian}},
  \bibinfo {author} {\bibfnamefont {H.}~\bibnamefont {Nersisyan}}, \ and\
  \bibinfo {author} {\bibfnamefont {S.}~\bibnamefont {Park}},\ }\href {\doibase
  10.1088/1475-7516/2019/03/045} {\bibfield  {journal} {\bibinfo  {journal}
  {Journal of Cosmology and Astroparticle Physics}\ }\textbf {\bibinfo {volume}
  {2019}},\ \bibinfo {pages} {045–045} (\bibinfo {year} {2019})}\BibitemShut
  {NoStop}%
\bibitem [{\citenamefont {Deser}\ and\ \citenamefont
  {Woodard}(2013)}]{DeserScreening}%
  \BibitemOpen
  \bibfield  {author} {\bibinfo {author} {\bibfnamefont {S.}~\bibnamefont
  {Deser}}\ and\ \bibinfo {author} {\bibfnamefont {R.}~\bibnamefont
  {Woodard}},\ }\href {\doibase 10.1088/1475-7516/2013/11/036} {\bibfield
  {journal} {\bibinfo  {journal} {Journal of Cosmology and Astroparticle
  Physics}\ }\textbf {\bibinfo {volume} {2013}},\ \bibinfo {pages} {036}
  (\bibinfo {year} {2013})}\BibitemShut {NoStop}%
\bibitem [{\citenamefont {Belgacem}\ \emph {et~al.}(2019)\citenamefont
  {Belgacem}, \citenamefont {Finke}, \citenamefont {Frassino},\ and\
  \citenamefont {Maggiore}}]{LunarLaserRanging}%
  \BibitemOpen
  \bibfield  {author} {\bibinfo {author} {\bibfnamefont {E.}~\bibnamefont
  {Belgacem}}, \bibinfo {author} {\bibfnamefont {A.}~\bibnamefont {Finke}},
  \bibinfo {author} {\bibfnamefont {A.}~\bibnamefont {Frassino}}, \ and\
  \bibinfo {author} {\bibfnamefont {M.}~\bibnamefont {Maggiore}},\ }\href
  {\doibase 10.1088/1475-7516/2019/02/035} {\bibfield  {journal} {\bibinfo
  {journal} {Journal of Cosmology and Astroparticle Physics}\ }\textbf
  {\bibinfo {volume} {2019}},\ \bibinfo {pages} {035–035} (\bibinfo {year}
  {2019})}\BibitemShut {NoStop}%
\bibitem [{\citenamefont {Vainshtein}(1972)}]{Vainshtein:1972sx}%
  \BibitemOpen
  \bibfield  {author} {\bibinfo {author} {\bibfnamefont {A.~I.}\ \bibnamefont
  {Vainshtein}},\ }\href {\doibase 10.1016/0370-2693(72)90147-5} {\bibfield
  {journal} {\bibinfo  {journal} {Phys. Lett. B}\ }\textbf {\bibinfo {volume}
  {39}},\ \bibinfo {pages} {393} (\bibinfo {year} {1972})}\BibitemShut
  {NoStop}%
\bibitem [{\citenamefont {Capozziello}\ and\ \citenamefont
  {Tsujikawa}(2008)}]{Capozziello:2007eu}%
  \BibitemOpen
  \bibfield  {author} {\bibinfo {author} {\bibfnamefont {S.}~\bibnamefont
  {Capozziello}}\ and\ \bibinfo {author} {\bibfnamefont {S.}~\bibnamefont
  {Tsujikawa}},\ }\href {\doibase 10.1103/PhysRevD.77.107501} {\bibfield
  {journal} {\bibinfo  {journal} {Phys. Rev. D}\ }\textbf {\bibinfo {volume}
  {77}},\ \bibinfo {pages} {107501} (\bibinfo {year} {2008})},\ \Eprint
  {http://arxiv.org/abs/0712.2268} {arXiv:0712.2268 [gr-qc]} \BibitemShut
  {NoStop}%
\bibitem [{\citenamefont {Capozziello}\ \emph {et~al.}(1996)\citenamefont
  {Capozziello}, \citenamefont {De~Ritis}, \citenamefont {Rubano},\ and\
  \citenamefont {Scudellaro}}]{Capozziello:1996bi}%
  \BibitemOpen
  \bibfield  {author} {\bibinfo {author} {\bibfnamefont {S.}~\bibnamefont
  {Capozziello}}, \bibinfo {author} {\bibfnamefont {R.}~\bibnamefont
  {De~Ritis}}, \bibinfo {author} {\bibfnamefont {C.}~\bibnamefont {Rubano}}, \
  and\ \bibinfo {author} {\bibfnamefont {P.}~\bibnamefont {Scudellaro}},\
  }\href {\doibase 10.1007/BF02742992} {\bibfield  {journal} {\bibinfo
  {journal} {Riv. Nuovo Cim.}\ }\textbf {\bibinfo {volume} {19N4}},\ \bibinfo
  {pages} {1} (\bibinfo {year} {1996})}\BibitemShut {NoStop}%
\bibitem [{\citenamefont {Dialektopoulos}\ and\ \citenamefont
  {Capozziello}(2018)}]{Dialektopoulos:2018qoe}%
  \BibitemOpen
  \bibfield  {author} {\bibinfo {author} {\bibfnamefont {K.~F.}\ \bibnamefont
  {Dialektopoulos}}\ and\ \bibinfo {author} {\bibfnamefont {S.}~\bibnamefont
  {Capozziello}},\ }\href {\doibase 10.1142/S0219887818400078} {\bibfield
  {journal} {\bibinfo  {journal} {Int. J. Geom. Meth. Mod. Phys.}\ }\textbf
  {\bibinfo {volume} {15}},\ \bibinfo {pages} {1840007} (\bibinfo {year}
  {2018})},\ \Eprint {http://arxiv.org/abs/1808.03484} {arXiv:1808.03484
  [gr-qc]} \BibitemShut {NoStop}%
\bibitem [{\citenamefont {Tsamparlis}\ and\ \citenamefont
  {Paliathanasis}(2018)}]{sym10070233}%
  \BibitemOpen
  \bibfield  {author} {\bibinfo {author} {\bibfnamefont {M.}~\bibnamefont
  {Tsamparlis}}\ and\ \bibinfo {author} {\bibfnamefont {A.}~\bibnamefont
  {Paliathanasis}},\ }\href {https://www.mdpi.com/2073-8994/10/7/233}
  {\bibfield  {journal} {\bibinfo  {journal} {Symmetry}\ }\textbf {\bibinfo
  {volume} {10}} (\bibinfo {year} {2018})}\BibitemShut {NoStop}%
\bibitem [{\citenamefont {Bahamonde}\ \emph {et~al.}(2019)\citenamefont
  {Bahamonde}, \citenamefont {Bamba},\ and\ \citenamefont
  {Camci}}]{Bahamonde2019}%
  \BibitemOpen
  \bibfield  {author} {\bibinfo {author} {\bibfnamefont {S.}~\bibnamefont
  {Bahamonde}}, \bibinfo {author} {\bibfnamefont {K.}~\bibnamefont {Bamba}}, \
  and\ \bibinfo {author} {\bibfnamefont {U.}~\bibnamefont {Camci}},\ }\href
  {\doibase 10.1088/1475-7516/2019/02/016} {\bibfield  {journal} {\bibinfo
  {journal} {Journal of Cosmology and Astroparticle Physics}\ }\textbf
  {\bibinfo {volume} {2019}},\ \bibinfo {pages} {016–016} (\bibinfo {year}
  {2019})}\BibitemShut {NoStop}%
\bibitem [{\citenamefont {Dimakis}\ \emph {et~al.}(2017)\citenamefont
  {Dimakis}, \citenamefont {Giacomini}, \citenamefont {Jamal}, \citenamefont
  {Leon},\ and\ \citenamefont {Paliathanasis}}]{PhysRevD.95.064031}%
  \BibitemOpen
  \bibfield  {author} {\bibinfo {author} {\bibfnamefont {N.}~\bibnamefont
  {Dimakis}}, \bibinfo {author} {\bibfnamefont {A.}~\bibnamefont {Giacomini}},
  \bibinfo {author} {\bibfnamefont {S.}~\bibnamefont {Jamal}}, \bibinfo
  {author} {\bibfnamefont {G.}~\bibnamefont {Leon}}, \ and\ \bibinfo {author}
  {\bibfnamefont {A.}~\bibnamefont {Paliathanasis}},\ }\href {\doibase
  10.1103/PhysRevD.95.064031} {\bibfield  {journal} {\bibinfo  {journal} {Phys.
  Rev. D}\ }\textbf {\bibinfo {volume} {95}},\ \bibinfo {pages} {064031}
  (\bibinfo {year} {2017})}\BibitemShut {NoStop}%
\bibitem [{\citenamefont {Nojiri}\ \emph
  {et~al.}(2017{\natexlab{b}})\citenamefont {Nojiri}, \citenamefont
  {Odintsov},\ and\ \citenamefont {Oikonomou}}]{Nojiri2017}%
  \BibitemOpen
  \bibfield  {author} {\bibinfo {author} {\bibfnamefont {S.}~\bibnamefont
  {Nojiri}}, \bibinfo {author} {\bibfnamefont {S.}~\bibnamefont {Odintsov}}, \
  and\ \bibinfo {author} {\bibfnamefont {V.}~\bibnamefont {Oikonomou}},\ }\href
  {\doibase 10.1016/j.physrep.2017.06.001} {\bibfield  {journal} {\bibinfo
  {journal} {Physics Reports}\ }\textbf {\bibinfo {volume} {692}},\ \bibinfo
  {pages} {1–104} (\bibinfo {year} {2017}{\natexlab{b}})}\BibitemShut
  {NoStop}%
\bibitem [{\citenamefont {Wetterich}(1998)}]{Wetterich1998}%
  \BibitemOpen
  \bibfield  {author} {\bibinfo {author} {\bibfnamefont {C.}~\bibnamefont
  {Wetterich}},\ }\href {\doibase 10.1023/a:1018837319976} {\bibfield
  {journal} {\bibinfo  {journal} {General Relativity and Gravitation}\ }\textbf
  {\bibinfo {volume} {30}},\ \bibinfo {pages} {159–172} (\bibinfo {year}
  {1998})}\BibitemShut {NoStop}%
\bibitem [{\citenamefont {Modesto}(2012)}]{Modesto1}%
  \BibitemOpen
  \bibfield  {author} {\bibinfo {author} {\bibfnamefont {L.}~\bibnamefont
  {Modesto}},\ }\href {\doibase 10.1103/PhysRevD.86.044005} {\bibfield
  {journal} {\bibinfo  {journal} {Phys. Rev. D}\ }\textbf {\bibinfo {volume}
  {86}},\ \bibinfo {pages} {044005} (\bibinfo {year} {2012})},\ \Eprint
  {http://arxiv.org/abs/1107.2403} {arXiv:1107.2403 [hep-th]} \BibitemShut
  {NoStop}%
\bibitem [{\citenamefont {Capozziello}\ and\ \citenamefont
  {Saez-Gomez}(2012)}]{Diego}%
  \BibitemOpen
  \bibfield  {author} {\bibinfo {author} {\bibfnamefont {S.}~\bibnamefont
  {Capozziello}}\ and\ \bibinfo {author} {\bibfnamefont {D.}~\bibnamefont
  {Saez-Gomez}},\ }\href {\doibase 10.1002/andp.201100244} {\bibfield
  {journal} {\bibinfo  {journal} {Annalen Phys.}\ }\textbf {\bibinfo {volume}
  {524}},\ \bibinfo {pages} {279} (\bibinfo {year} {2012})},\ \Eprint
  {http://arxiv.org/abs/1107.0948} {arXiv:1107.0948 [gr-qc]} \BibitemShut
  {NoStop}%
\bibitem [{\citenamefont {Capozziello}\ \emph {et~al.}(2007)\citenamefont
  {Capozziello}, \citenamefont {Cardone},\ and\ \citenamefont
  {Troisi}}]{Cardo}%
  \BibitemOpen
  \bibfield  {author} {\bibinfo {author} {\bibfnamefont {S.}~\bibnamefont
  {Capozziello}}, \bibinfo {author} {\bibfnamefont {V.~F.}\ \bibnamefont
  {Cardone}}, \ and\ \bibinfo {author} {\bibfnamefont {A.}~\bibnamefont
  {Troisi}},\ }\href {\doibase 10.1111/j.1365-2966.2007.11401.x} {\bibfield
  {journal} {\bibinfo  {journal} {Mon. Not. Roy. Astron. Soc.}\ }\textbf
  {\bibinfo {volume} {375}},\ \bibinfo {pages} {1423} (\bibinfo {year}
  {2007})},\ \Eprint {http://arxiv.org/abs/astro-ph/0603522}
  {arXiv:astro-ph/0603522} \BibitemShut {NoStop}%
\bibitem [{\citenamefont {Navarro}\ \emph {et~al.}(1996)\citenamefont
  {Navarro}, \citenamefont {Frenk},\ and\ \citenamefont {White}}]{NFW1996}%
  \BibitemOpen
  \bibfield  {author} {\bibinfo {author} {\bibfnamefont {J.~F.}\ \bibnamefont
  {Navarro}}, \bibinfo {author} {\bibfnamefont {C.~S.}\ \bibnamefont {Frenk}},
  \ and\ \bibinfo {author} {\bibfnamefont {S.~D.~M.}\ \bibnamefont {White}},\
  }\href {\doibase 10.1086/177173} {\bibfield  {journal} {\bibinfo  {journal}
  {The Astrophysical Journal}\ }\textbf {\bibinfo {volume} {462}},\ \bibinfo
  {pages} {563} (\bibinfo {year} {1996})}\BibitemShut {NoStop}%
\bibitem [{\citenamefont {Child}\ \emph {et~al.}(2018)\citenamefont {Child},
  \citenamefont {Habib}, \citenamefont {Heitmann}, \citenamefont {Frontiere},
  \citenamefont {Finkel}, \citenamefont {Pope},\ and\ \citenamefont
  {Morozov}}]{Child2018}%
  \BibitemOpen
  \bibfield  {author} {\bibinfo {author} {\bibfnamefont {H.~L.}\ \bibnamefont
  {Child}}, \bibinfo {author} {\bibfnamefont {S.}~\bibnamefont {Habib}},
  \bibinfo {author} {\bibfnamefont {K.}~\bibnamefont {Heitmann}}, \bibinfo
  {author} {\bibfnamefont {N.}~\bibnamefont {Frontiere}}, \bibinfo {author}
  {\bibfnamefont {H.}~\bibnamefont {Finkel}}, \bibinfo {author} {\bibfnamefont
  {A.}~\bibnamefont {Pope}}, \ and\ \bibinfo {author} {\bibfnamefont
  {V.}~\bibnamefont {Morozov}},\ }\href {\doibase 10.3847/1538-4357/aabf95}
  {\bibfield  {journal} {\bibinfo  {journal} {Astrophys. J.}\ }\textbf
  {\bibinfo {volume} {859}},\ \bibinfo {pages} {55} (\bibinfo {year} {2018})},\
  \Eprint {http://arxiv.org/abs/1804.10199} {arXiv:1804.10199 [astro-ph.CO]}
  \BibitemShut {NoStop}%
\bibitem [{\citenamefont {Umetsu}\ \emph {et~al.}(2016)\citenamefont {Umetsu},
  \citenamefont {Zitrin}, \citenamefont {Gruen}, \citenamefont {Merten},
  \citenamefont {Donahue},\ and\ \citenamefont {Postman}}]{Umetsu:2015baa}%
  \BibitemOpen
  \bibfield  {author} {\bibinfo {author} {\bibfnamefont {K.}~\bibnamefont
  {Umetsu}}, \bibinfo {author} {\bibfnamefont {A.}~\bibnamefont {Zitrin}},
  \bibinfo {author} {\bibfnamefont {D.}~\bibnamefont {Gruen}}, \bibinfo
  {author} {\bibfnamefont {J.}~\bibnamefont {Merten}}, \bibinfo {author}
  {\bibfnamefont {M.}~\bibnamefont {Donahue}}, \ and\ \bibinfo {author}
  {\bibfnamefont {M.}~\bibnamefont {Postman}},\ }\href {\doibase
  10.3847/0004-637X/821/2/116} {\bibfield  {journal} {\bibinfo  {journal}
  {Astrophys. J.}\ }\textbf {\bibinfo {volume} {821}},\ \bibinfo {pages} {116}
  (\bibinfo {year} {2016})},\ \Eprint {http://arxiv.org/abs/1507.04385}
  {arXiv:1507.04385 [astro-ph.CO]} \BibitemShut {NoStop}%
\bibitem [{\citenamefont {Bartelmann}\ and\ \citenamefont
  {Schneider}(2001)}]{LensingReview}%
  \BibitemOpen
  \bibfield  {author} {\bibinfo {author} {\bibfnamefont {M.}~\bibnamefont
  {Bartelmann}}\ and\ \bibinfo {author} {\bibfnamefont {P.}~\bibnamefont
  {Schneider}},\ }\href {\doibase 10.1016/s0370-1573(00)00082-x} {\bibfield
  {journal} {\bibinfo  {journal} {Physics Reports}\ }\textbf {\bibinfo {volume}
  {340}},\ \bibinfo {pages} {291–472} (\bibinfo {year} {2001})}\BibitemShut
  {NoStop}%
\bibitem [{\citenamefont {{Umetsu}}(2020)}]{Umetsu2020}%
  \BibitemOpen
  \bibfield  {author} {\bibinfo {author} {\bibfnamefont {K.}~\bibnamefont
  {{Umetsu}}},\ }\href {\doibase 10.1007/s00159-020-00129-w} {\bibfield
  {journal} {\bibinfo  {journal} {Astron. Astrophys. Rev.}\ }\textbf {\bibinfo
  {volume} {28}},\ \bibinfo {eid} {7} (\bibinfo {year} {2020})},\ \Eprint
  {http://arxiv.org/abs/2007.00506} {arXiv:2007.00506 [astro-ph.CO]}
  \BibitemShut {NoStop}%
\bibitem [{\citenamefont {{Umetsu}}\ \emph {et~al.}(2014)\citenamefont
  {{Umetsu}}, \citenamefont {{Medezinski}}, \citenamefont {{Nonino}},
  \citenamefont {{Merten}}, \citenamefont {{Postman}}, \citenamefont
  {{Meneghetti}}, \citenamefont {{Donahue}}, \citenamefont {{Czakon}},
  \citenamefont {{Molino}}, \citenamefont {{Seitz}}, \citenamefont {{Gruen}},
  \citenamefont {{Lemze}}, \citenamefont {{Balestra}}, \citenamefont
  {{Ben{\'\i}tez}}, \citenamefont {{Biviano}}, \citenamefont {{Broadhurst}},
  \citenamefont {{Ford}}, \citenamefont {{Grillo}}, \citenamefont
  {{Koekemoer}}, \citenamefont {{Melchior}}, \citenamefont {{Mercurio}},
  \citenamefont {{Moustakas}}, \citenamefont {{Rosati}},\ and\ \citenamefont
  {{Zitrin}}}]{Umetsu2014}%
  \BibitemOpen
  \bibfield  {author} {\bibinfo {author} {\bibfnamefont {K.}~\bibnamefont
  {{Umetsu}}}, \bibinfo {author} {\bibfnamefont {E.}~\bibnamefont
  {{Medezinski}}}, \bibinfo {author} {\bibfnamefont {M.}~\bibnamefont
  {{Nonino}}}, \bibinfo {author} {\bibfnamefont {J.}~\bibnamefont {{Merten}}},
  \bibinfo {author} {\bibfnamefont {M.}~\bibnamefont {{Postman}}}, \bibinfo
  {author} {\bibfnamefont {M.}~\bibnamefont {{Meneghetti}}}, \bibinfo {author}
  {\bibfnamefont {M.}~\bibnamefont {{Donahue}}}, \bibinfo {author}
  {\bibfnamefont {N.}~\bibnamefont {{Czakon}}}, \bibinfo {author}
  {\bibfnamefont {A.}~\bibnamefont {{Molino}}}, \bibinfo {author}
  {\bibfnamefont {S.}~\bibnamefont {{Seitz}}}, \bibinfo {author} {\bibfnamefont
  {D.}~\bibnamefont {{Gruen}}}, \bibinfo {author} {\bibfnamefont
  {D.}~\bibnamefont {{Lemze}}}, \bibinfo {author} {\bibfnamefont
  {I.}~\bibnamefont {{Balestra}}}, \bibinfo {author} {\bibfnamefont
  {N.}~\bibnamefont {{Ben{\'\i}tez}}}, \bibinfo {author} {\bibfnamefont
  {A.}~\bibnamefont {{Biviano}}}, \bibinfo {author} {\bibfnamefont
  {T.}~\bibnamefont {{Broadhurst}}}, \bibinfo {author} {\bibfnamefont
  {H.}~\bibnamefont {{Ford}}}, \bibinfo {author} {\bibfnamefont
  {C.}~\bibnamefont {{Grillo}}}, \bibinfo {author} {\bibfnamefont
  {A.}~\bibnamefont {{Koekemoer}}}, \bibinfo {author} {\bibfnamefont
  {P.}~\bibnamefont {{Melchior}}}, \bibinfo {author} {\bibfnamefont
  {A.}~\bibnamefont {{Mercurio}}}, \bibinfo {author} {\bibfnamefont
  {J.}~\bibnamefont {{Moustakas}}}, \bibinfo {author} {\bibfnamefont
  {P.}~\bibnamefont {{Rosati}}}, \ and\ \bibinfo {author} {\bibfnamefont
  {A.}~\bibnamefont {{Zitrin}}},\ }\href {\doibase 10.1088/0004-637X/795/2/163}
  {\bibfield  {journal} {\bibinfo  {journal} {Astrophys. J.}\ }\textbf
  {\bibinfo {volume} {795}},\ \bibinfo {eid} {163} (\bibinfo {year} {2014})},\
  \Eprint {http://arxiv.org/abs/1404.1375} {arXiv:1404.1375 [astro-ph.CO]}
  \BibitemShut {NoStop}%
\bibitem [{\citenamefont {Meneghetti}\ \emph {et~al.}(2014)\citenamefont
  {Meneghetti} \emph {et~al.}}]{Meneghetti:2014xna}%
  \BibitemOpen
  \bibfield  {author} {\bibinfo {author} {\bibfnamefont {M.}~\bibnamefont
  {Meneghetti}} \emph {et~al.},\ }\href {\doibase 10.1088/0004-637X/797/1/34}
  {\bibfield  {journal} {\bibinfo  {journal} {Astrophys. J.}\ }\textbf
  {\bibinfo {volume} {797}},\ \bibinfo {pages} {34} (\bibinfo {year} {2014})},\
  \Eprint {http://arxiv.org/abs/1404.1384} {arXiv:1404.1384 [astro-ph.CO]}
  \BibitemShut {NoStop}%
\bibitem [{\citenamefont {{Zitrin}}\ \emph {et~al.}(2015)\citenamefont
  {{Zitrin}}, \citenamefont {{Fabris}}, \citenamefont {{Merten}}, \citenamefont
  {{Melchior}}, \citenamefont {{Meneghetti}}, \citenamefont {{Koekemoer}},
  \citenamefont {{Coe}}, \citenamefont {{Maturi}}, \citenamefont
  {{Bartelmann}}, \citenamefont {{Postman}}, \citenamefont {{Umetsu}},
  \citenamefont {{Seidel}}, \citenamefont {{Sendra}}, \citenamefont
  {{Broadhurst}}, \citenamefont {{Balestra}}, \citenamefont {{Biviano}},
  \citenamefont {{Grillo}}, \citenamefont {{Mercurio}}, \citenamefont
  {{Nonino}}, \citenamefont {{Rosati}}, \citenamefont {{Bradley}},
  \citenamefont {{Carrasco}}, \citenamefont {{Donahue}}, \citenamefont
  {{Ford}}, \citenamefont {{Frye}},\ and\ \citenamefont
  {{Moustakas}}}]{Zitrin2015}%
  \BibitemOpen
  \bibfield  {author} {\bibinfo {author} {\bibfnamefont {A.}~\bibnamefont
  {{Zitrin}}}, \bibinfo {author} {\bibfnamefont {A.}~\bibnamefont {{Fabris}}},
  \bibinfo {author} {\bibfnamefont {J.}~\bibnamefont {{Merten}}}, \bibinfo
  {author} {\bibfnamefont {P.}~\bibnamefont {{Melchior}}}, \bibinfo {author}
  {\bibfnamefont {M.}~\bibnamefont {{Meneghetti}}}, \bibinfo {author}
  {\bibfnamefont {A.}~\bibnamefont {{Koekemoer}}}, \bibinfo {author}
  {\bibfnamefont {D.}~\bibnamefont {{Coe}}}, \bibinfo {author} {\bibfnamefont
  {M.}~\bibnamefont {{Maturi}}}, \bibinfo {author} {\bibfnamefont
  {M.}~\bibnamefont {{Bartelmann}}}, \bibinfo {author} {\bibfnamefont
  {M.}~\bibnamefont {{Postman}}}, \bibinfo {author} {\bibfnamefont
  {K.}~\bibnamefont {{Umetsu}}}, \bibinfo {author} {\bibfnamefont
  {G.}~\bibnamefont {{Seidel}}}, \bibinfo {author} {\bibfnamefont
  {I.}~\bibnamefont {{Sendra}}}, \bibinfo {author} {\bibfnamefont
  {T.}~\bibnamefont {{Broadhurst}}}, \bibinfo {author} {\bibfnamefont
  {I.}~\bibnamefont {{Balestra}}}, \bibinfo {author} {\bibfnamefont
  {A.}~\bibnamefont {{Biviano}}}, \bibinfo {author} {\bibfnamefont
  {C.}~\bibnamefont {{Grillo}}}, \bibinfo {author} {\bibfnamefont
  {A.}~\bibnamefont {{Mercurio}}}, \bibinfo {author} {\bibfnamefont
  {M.}~\bibnamefont {{Nonino}}}, \bibinfo {author} {\bibfnamefont
  {P.}~\bibnamefont {{Rosati}}}, \bibinfo {author} {\bibfnamefont
  {L.}~\bibnamefont {{Bradley}}}, \bibinfo {author} {\bibfnamefont
  {M.}~\bibnamefont {{Carrasco}}}, \bibinfo {author} {\bibfnamefont
  {M.}~\bibnamefont {{Donahue}}}, \bibinfo {author} {\bibfnamefont
  {H.}~\bibnamefont {{Ford}}}, \bibinfo {author} {\bibfnamefont {B.~L.}\
  \bibnamefont {{Frye}}}, \ and\ \bibinfo {author} {\bibfnamefont
  {J.}~\bibnamefont {{Moustakas}}},\ }\href {\doibase
  10.1088/0004-637X/801/1/44} {\bibfield  {journal} {\bibinfo  {journal}
  {Astrophys. J.}\ }\textbf {\bibinfo {volume} {801}},\ \bibinfo {eid} {44}
  (\bibinfo {year} {2015})},\ \Eprint {http://arxiv.org/abs/1411.1414}
  {arXiv:1411.1414 [astro-ph.CO]} \BibitemShut {NoStop}%
\bibitem [{\citenamefont {Dunkley}\ \emph {et~al.}(2005)\citenamefont
  {Dunkley}, \citenamefont {Bucher}, \citenamefont {Ferreira}, \citenamefont
  {Moodley},\ and\ \citenamefont {Skordis}}]{10.1111/j.1365-2966.2004.08464.x}%
  \BibitemOpen
  \bibfield  {author} {\bibinfo {author} {\bibfnamefont {J.}~\bibnamefont
  {Dunkley}}, \bibinfo {author} {\bibfnamefont {M.}~\bibnamefont {Bucher}},
  \bibinfo {author} {\bibfnamefont {P.~G.}\ \bibnamefont {Ferreira}}, \bibinfo
  {author} {\bibfnamefont {K.}~\bibnamefont {Moodley}}, \ and\ \bibinfo
  {author} {\bibfnamefont {C.}~\bibnamefont {Skordis}},\ }\href {\doibase
  10.1111/j.1365-2966.2004.08464.x} {\bibfield  {journal} {\bibinfo  {journal}
  {Monthly Notices of the Royal Astronomical Society}\ }\textbf {\bibinfo
  {volume} {356}},\ \bibinfo {pages} {925} (\bibinfo {year}
  {2005})}\BibitemShut {NoStop}%
\bibitem [{\citenamefont {Mukherjee}\ \emph {et~al.}(2006)\citenamefont
  {Mukherjee}, \citenamefont {Parkinson},\ and\ \citenamefont
  {Liddle}}]{Mukherjee_2006}%
  \BibitemOpen
  \bibfield  {author} {\bibinfo {author} {\bibfnamefont {P.}~\bibnamefont
  {Mukherjee}}, \bibinfo {author} {\bibfnamefont {D.}~\bibnamefont
  {Parkinson}}, \ and\ \bibinfo {author} {\bibfnamefont {A.~R.}\ \bibnamefont
  {Liddle}},\ }\href {\doibase 10.1086/501068} {\bibfield  {journal} {\bibinfo
  {journal} {The Astrophysical Journal}\ }\textbf {\bibinfo {volume} {638}},\
  \bibinfo {pages} {L51} (\bibinfo {year} {2006})}\BibitemShut {NoStop}%
\bibitem [{\citenamefont {Nesseris}\ and\ \citenamefont
  {Garc{\'{\i}}a-Bellido}(2013)}]{Nesseris_2013}%
  \BibitemOpen
  \bibfield  {author} {\bibinfo {author} {\bibfnamefont {S.}~\bibnamefont
  {Nesseris}}\ and\ \bibinfo {author} {\bibfnamefont {J.}~\bibnamefont
  {Garc{\'{\i}}a-Bellido}},\ }\href {\doibase 10.1088/1475-7516/2013/08/036}
  {\bibfield  {journal} {\bibinfo  {journal} {Journal of Cosmology and
  Astroparticle Physics}\ }\textbf {\bibinfo {volume} {2013}},\ \bibinfo
  {pages} {036} (\bibinfo {year} {2013})}\BibitemShut {NoStop}%
\bibitem [{\citenamefont {Jeffreys}(1998)}]{JeffreysScale}%
  \BibitemOpen
  \bibfield  {author} {\bibinfo {author} {\bibfnamefont {H.}~\bibnamefont
  {Jeffreys}},\ }\href@noop {} {\emph {\bibinfo {title} {{The Theory of
  Probability}}}}\ (\bibinfo  {publisher} {{Oxford University Press}},\
  \bibinfo {year} {1998})\BibitemShut {NoStop}%
\bibitem [{\citenamefont {Merten}\ \emph {et~al.}(2015)\citenamefont {Merten}
  \emph {et~al.}}]{Merten2015}%
  \BibitemOpen
  \bibfield  {author} {\bibinfo {author} {\bibfnamefont {J.}~\bibnamefont
  {Merten}} \emph {et~al.},\ }\href {\doibase 10.1088/0004-637X/806/1/4}
  {\bibfield  {journal} {\bibinfo  {journal} {Astrophys. J.}\ }\textbf
  {\bibinfo {volume} {806}},\ \bibinfo {pages} {4} (\bibinfo {year} {2015})},\
  \Eprint {http://arxiv.org/abs/1404.1376} {arXiv:1404.1376 [astro-ph.CO]}
  \BibitemShut {NoStop}%
\bibitem [{\citenamefont {Correa}\ \emph {et~al.}(2015)\citenamefont {Correa},
  \citenamefont {Wyithe}, \citenamefont {Schaye},\ and\ \citenamefont
  {Duffy}}]{Correa:2015dva}%
  \BibitemOpen
  \bibfield  {author} {\bibinfo {author} {\bibfnamefont {C.~A.}\ \bibnamefont
  {Correa}}, \bibinfo {author} {\bibfnamefont {J.~S.~B.}\ \bibnamefont
  {Wyithe}}, \bibinfo {author} {\bibfnamefont {J.}~\bibnamefont {Schaye}}, \
  and\ \bibinfo {author} {\bibfnamefont {A.~R.}\ \bibnamefont {Duffy}},\ }\href
  {\doibase 10.1093/mnras/stv1363} {\bibfield  {journal} {\bibinfo  {journal}
  {Mon. Not. Roy. Astron. Soc.}\ }\textbf {\bibinfo {volume} {452}},\ \bibinfo
  {pages} {1217} (\bibinfo {year} {2015})},\ \Eprint
  {http://arxiv.org/abs/1502.00391} {arXiv:1502.00391 [astro-ph.CO]}
  \BibitemShut {NoStop}%
\bibitem [{\citenamefont {Diemer}\ and\ \citenamefont
  {Joyce}(2019)}]{Diemer:2018vmz}%
  \BibitemOpen
  \bibfield  {author} {\bibinfo {author} {\bibfnamefont {B.}~\bibnamefont
  {Diemer}}\ and\ \bibinfo {author} {\bibfnamefont {M.}~\bibnamefont {Joyce}},\
  }\href {\doibase 10.3847/1538-4357/aafad6} {\bibfield  {journal} {\bibinfo
  {journal} {Astrophys. J.}\ }\textbf {\bibinfo {volume} {871}},\ \bibinfo
  {pages} {168} (\bibinfo {year} {2019})},\ \Eprint
  {http://arxiv.org/abs/1809.07326} {arXiv:1809.07326 [astro-ph.CO]}
  \BibitemShut {NoStop}%
\end{thebibliography}%

\end{document}